\documentstyle[aps,eqsecnum,epsfig]{revtex}
\def\vec{\bf} 
\begin{document}
\title{Dynamical Renormalization Group Approach to Quantum Kinetics
in Scalar and Gauge Theories\footnote{To appear in Physical Review D}}
\author{{\bf 
D. Boyanovsky,$^{1,2}$
H.J. de Vega,$^{2,1}$
and
S.-Y. Wang$^{1}$}}
\address
{${}^{1}$Department of Physics and Astronomy, University of
Pittsburgh, Pittsburgh, PA. 15260, U.S.A.\\
${}^{2}$LPTHE, Universit\'e Pierre et Marie Curie (Paris VI) et Denis Diderot 
(Paris VII), Tour 16, 1er. \'etage, 4, Place Jussieu, 75252 Paris, Cedex 05, 
France}
\date{\today}
\maketitle

\begin{abstract}
We derive quantum kinetic equations from a quantum field 
theory implementing a diagrammatic perturbative expansion 
improved by a resummation via the dynamical renormalization group. 
The method begins by obtaining the equation of motion of the 
distribution function in perturbation theory. 
The solution of this equation of motion 
reveals secular terms that grow in time, the dynamical renormalization
group resums these secular terms in real time and leads directly to
the quantum kinetic equation. This method allows to include
consistently medium effects via resummations akin to hard thermal
loops but away from equilibrium. A close relationship between this
approach and the renormalization group in Euclidean field theory is established. 
In particular, coarse graining, stationary solutions,
relaxation time approximation and relaxation rates have a natural parallel as 
irrelevant operators, fixed points, linearization and stability
exponents in the Euclidean renormalization group, respectively. 
We used this method to study the 
relaxation in a cool gas of pions and sigma mesons in the
$O(4)$ chiral linear sigma model. 
We obtain in relaxation time approximation
the pion and sigma meson relaxation rates. 
We also find that in large momentum limit 
emission and absorption of massless pions result
in threshold infrared divergence in sigma meson relaxation rate
and lead to a crossover behavior in relaxation. 
We then study the relaxation of charged quasiparticles in 
scalar electrodynamics (SQED). 
We begin with a {\em gauge invariant} description of the 
distribution function and implement the hard thermal loop 
resummation for longitudinal and transverse photons 
as well as for the scalars. While longitudinal,
Debye screened photons lead to purely exponential relaxation, 
transverse photons, only dynamically screened by 
Landau damping lead to anomalous (non-exponential) relaxation,
thus leading to a crossover between two different relaxational regimes. 
We emphasize that infrared divergent damping rates
are indicative of non-exponential relaxation and the dynamical
renormalization group reveals the correct relaxation directly in real time. 
Furthermore the relaxational time scales for charged quasiparticles are 
similar to those found in QCD in a self-consistent HTL resummation.  
Finally we also show that this method provides a natural framework 
to interpret and resolve the issue of pinch singularities out of equilibrium 
and establish a direct correspondence between pinch singularities and 
secular terms in time-dependent perturbation theory. 
We argue that this method is particularly well suited to study quantum kinetics 
and transport in gauge theories.   
\end{abstract}


\section{Introduction}

The search for the quark-gluon plasma (QGP)
at the Relativistic Heavy Ion Collider (RHIC) 
and the forthcoming Large Hadron Collider (LHC) 
has the potential of providing clear evidence for the formation of a
deconfined plasma of quarks and gluons and hopefully to study 
the chiral phase transition. Perhaps this is the only opportunity to
study phase transitions that are conjectured 
to occur in particle physics with earth-bound accelerators and an
intense theoretical effort has developed  
parallel to the experimental program that seeks to understand the 
signatures of the QGP and the chiral phase transition~\cite{qgp,books}. 
An important part of the program is to assess whether the plasma, once formed,
achieves a state of thermodynamic equilibrium and if so on what time
scales. This is an important question since current estimates suggest
that at the energies and luminosities to be achieved at RHIC,  
the spatial and temporal scales for the existence of the QGP are of
the order of $20~\mbox{fm}$~\cite{qgp}. The description of the
space-time evolution in an ultrarelativistic heavy ion collision
requires the understanding of phenomena on different time and spatial
scales. Ideally, such a description should begin from the parton distribution 
functions of the colliding nuclei as the initial state and evolve this
state in time using QCD to obtain the kinetic 
and chemical equilibration of partons, the emergence of hydrodynamics
and the hadronization and freeze-out stages~\cite{geiger}.   
An important part of the program to study the space-time evolution
from first principles seeks to establish a consistent 
kinetic description of transport phenomena in a dense partonic
environment. Such kinetic description has the potential 
of providing  a detailed understanding of collective flow, observables
(hadronic and electromagnetic) such as multiparticle 
distributions, charmonium suppression, freeze out of hadrons and other
important experimental signatures that will lead to an unambiguous
determination of whether a QGP has been formed and the observables of
phase transitions. This premise justifies an important theoretical
effort to obtain such a kinetic description from first
principles. During the last few years there have been important
advances in this program, from  derivations of kinetic and transport equations 
from first principles in QCD~\cite{geiger,heinz,mrow,bass} and scalar
field theories~\cite{lawrie,boyanrk,frenkel,rau} to numerical codes
that describe the space-time evolution in 
terms of partonic cascades~\cite{geiger} that include screening
corrections in the scattering cross sections~\cite{wang,eskola}
and more recently nonequilibrium
dynamics has been studied via lattice simulations~\cite{smit,venu,raja,mueller3}

The kinetic description to study hot and/or dense quantum field theory
systems is also of fundamental importance in the understanding of the
emergence of  hydrodynamics  in the long-wavelength limit
of a quantum field theory~\cite{Jeon:1996zm} and more recently a 
transport approach has been advocated as a description of the
collective dynamics of soft degrees of freedom in hot
QCD~\cite{Arnold:1999xk,Bodeker:1999ud,Blaizot:1999xk,Litim:1999id,basagoiti}.
The typical approach to derive transport equations begins by
introducing a Wigner transform of a particular nonequilibrium Green's
functions at two  different space-time
points~\cite{geiger,heinz,mrow,Blaizot:1999xk,daniel} (a gauge
covariant Wigner transform in the case of gauge theories) and often
requires a quasiparticle approximation~\cite{mrow,daniel}. The
rationale behind a Wigner transform of a nonequilibrium Green's
function is the assumption of a wide separation between the
microscopic (fast) and relaxational (slow) time scales, typically
justified in a weakly coupled theory. A recent derivation of transport
equations for a hot QCD plasma along these lines has recently been
reported in~\cite{Blaizot:1999xk}, however  the collisional terms
obtained in the quasiparticle and relaxation time approximations turn
out to be infrared divergent.    

Thus, the importance of a fundamental understanding of transport in
quantum field theory from first principles, with the direct
application to the experimental aspects of the search for the QGP
justifies the study of transport phenomena from many different
perspectives. In this article we present a novel method to obtain
quantum kinetic equations directly from the underlying quantum field
theory implementing a dynamical renormalization group  
resummation. Such approach has been recently introduced to study the
relaxation of mean-fields of hard charged scalars 
in a gauge theory~\cite{boyrgir}. This method allowed to obtain
directly in Ref.~\cite{boyrgir} the anomalous relaxation of hard
charged excitations in an Abelian gauge theory~\cite{iancu}, 
providing an interpretation of 
infrared divergent damping rates~\cite{robinfra} in terms of
non-exponential relaxation and pointed to a shortcoming in the
interpretation of quasiparticle relaxation in terms of complex poles
in the propagator. Infrared divergences associated with the
emission and absorption of long-wavelength gauge bosons are ubiquitous in
gauge theories. Thus, this novel approach is particularly suitable to study
transport phenomena in gauge theories.  

\vspace{3mm}
\noindent{\bf Goals and Strategy}: 
The  goals of this article are to provide a novel and alternative
derivation of quantum kinetic equations directly 
from the microscopic quantum field theory in real time and apply this
program to several relevant cases of interest. We consider
 scalar theories describing pions and sigma mesons and
gauge theories. This approach allows to include consistently medium effects,
such as nonequilibrium generalizations of the hard thermal loop
resummation, describes anomalous relaxation and reveals the proper
time scales for relaxation directly in real time. There are several
advantages that this program offers as compared to other approaches to
transport phenomena:  
\begin{itemize}
\item{(i) It allows to study the crossover between different relaxational
behavior in real time. This is relevant in the case of resonances
where the medium may enhance threshold effects. } 

\item{(ii) It describes non-exponential relaxation in a clear manner and
treats threshold effects consistently, 
providing a real-time interpretation of infrared divergent damping
rates in gauge theories,} 

\item{(iii) It provides a systematic field-theoretical method to include
higher order corrections and  
allows to incorporate self-consistently medium effects such as for
example a resummation of hard  
thermal loops~\cite{htl,rob2,rob3} that are necessary to determine the
relevant degrees of freedom and their microscopic 
time scales.}

\item{(iv) It resolves the issue of pinch singularities that often
appear in calculations of physical quantities 
out of equilibrium. }

\end{itemize}
The strategy to be followed is a generalization of the methods
introduced in Ref.~\cite{boyrgir} but adapted to the description 
of quantum kinetics. 
The starting point is the identification of the distribution function
of the quasiparticles which could require a resummation of medium
effects (the equivalent of hard thermal
loops~\cite{htl,rob2,rob3}). The equation of motion for this
distribution function is solved in a perturbative 
expansion in terms of nonequilibrium Feynman diagrams. The
perturbative solution in real time displays secular terms, 
i.e., terms that grow in time and invalidate the perturbative
expansion beyond a particular time scale (recognized 
{\em a posteriori} to be the relaxational time scale). The dynamical
renormalization group implements a systematic 
resummation of these secular terms and the resulting renormalization
group equation is the quantum kinetic equation.  

The validity of this approach hinges upon the basic assumption of a
wide separation between the microscopic and the relaxational time
scales. Such an assumption underlies every approach to a kinetic
description and is generally justified in weakly coupled
theories. Unlike other approaches in terms of a truncation of the
equations of motion for the Wigner distribution function, the main
ingredient in the approach presented here is a 
perturbative diagrammatic evaluation of the time evolution of the
proper distribution function in real time~\cite{boyanrk} improved via a
renormalization group resummation of the secular divergences. 

An important bonus of this approach is that it illuminates  the origin
and provides a natural resolution of pinch 
singularities~\cite{landsmann,altherr} found in
perturbation theory out of equilibrium. The perturbative real-time
approach combined with the 
renormalization group resummation reveal clearly that these are
indicative of the nonequilibrium evolution of the 
distribution functions. In this framework, pinch singularities are the
manifestation of secular terms.  

The article is organized as follows: In Sec.~II we summarize the
main ingredients of nonequilibrium field theory 
to establish the perturbative framework. In Sec.~III we study the
familiar case of a scalar field theory, including in addition the
nonequilibrium resummation akin to the hard thermal loops to 
account for the effective masses in the medium and  
therefore the relevant microscopic time scales. In Sec.~IV we
discuss in detail the main features of the dynamical 
renormalization group approach to quantum kinetics, compare it to the
more familiar renormalization group of Euclidean 
quantum field theory and provide an easy-to-follow recipe to obtain
quantum kinetic equations. In Sec.~V we apply 
these techniques to obtain the kinetic equations for cool pions and
sigma mesons in the $O(4)$ linear sigma model 
in the chiral limit. In relaxation time approximation we obtain
the relaxation rates for pions and sigma mesons. 
This case allows us to highlight the power of this approach to study
threshold effects on the relaxation of resonances, 
in particular the crossover between two different relaxational regimes
as a function of the momentum of the resonance. 
This aspect becomes phenomenological important in view of the recent
studies by Hatsuda and collaborators~\cite{hatsuda}  
that reveal a dropping of the sigma mass near the chiral phase
transition and an enhancement of threshold effects with 
potential observational consequences in heavy ion collisions.

In Sec.~VI we study the relaxation of charged quasiparticles in the full range 
of momenta in SQED. This theory has the same hard thermal loop
structure at lowest order as QED and QCD~\cite{rebhan,thoma,boyhtl,thoma2} 
and shares many features of these theories such as the lack of magnetic 
screening mass. In particular, in this Abelian case we provide a {\em
gauge invariant} description of the quasiparticle 
distribution function, thus bypassing the complications associated
with the gauge covariant Wigner transforms of the 
charged field Green's function. The hard thermal loop
resummation~\cite{htl,rob2,rob3} is included in the scalar as 
well as in the gauge boson spectral densities. We find that the
exchange of HTL resummed longitudinal photons leads 
to exponential relaxation but the exchange of dynamically screened
transverse photons leads to anomalous relaxation, 
thus leading to a crossover behavior in the relaxation of the
distribution function as a function of the momentum of 
the charged particle. The real-time description of relaxation
advocated in this article bypasses the ambiguities associated with an
infrared divergent damping rate~\cite{Blaizot:1999xk,thoma}. In Sec.~VII 
we discuss the issue of pinch 
singularities found in calculations in nonequilibrium field theory
and establish the equivalence between these and 
secular terms in the perturbative expansion, these singularities are
thus resolved via the resummation provided by 
the dynamical renormalization group.  

We summarize our results and discuss further implications and future
directions in the conclusions.

\section{Real-time Nonequilibrium Techniques}

The field theoretical methods to describe nonequilibrium processes
have been studied at length in the literature to which the reader should refer  
for a more detailed presentation~\cite{landsmann,schwinger,kt,keldysh,chou,disip,tadpole,lebellac}.
Here we only highlight those aspects and details that are necessary for our purposes.

The basic ingredient is the time
evolution of density matrix prepared initially at time $t=t_0$, 
which leads to the generating functional of nonequilibrium 
Green's functions in terms of a path integral defined on a contour 
in the complex time plane. 

The contour has two branches running forward and backward in the real time 
axis corresponding to the unitary evolution operator forward in time that 
pre-multiplies the density
matrix at $t_0$ and the hermitian conjugate that post-multiplies 
it and determines evolution backwards in time. 
The initial density matrix determines the boundary conditions on the propagators. 

This is a standard formulation of nonequilibrium 
quantum field theory known as 
Schwinger-Keldysh or closed-time-path (CTP)~\cite{landsmann,schwinger,kt,keldysh,chou,disip,tadpole,lebellac}.
Fields defined on the forward and backward branches 
are labeled respectively with ``$+$'' and ``$-$'' superscripts and 
are  treated independently.  
Introducing sources on the CTP contour, one can easily construct the 
nonequilibrium generating functional, which generates nonequilibrium 
Green's functions through functional derivatives with respect to sources much in the same manner as the
usual formulation of amplitudes in terms of path integrals.
 
The path integral along the CTP contour is in 
terms of the effective Lagrangian defined by
\begin{equation}
{\cal{L}}_{\text{noneq}}[{\Psi}^+,{\Psi}^-]=
{\cal{L}}[{\Psi}^+]-{\cal L}[{\Psi}^-]~,\label{noneqlag}
\end{equation}
where ${\cal L}[\Psi]$ denotes the corresponding Lagrangian in usual
field theory and $\Psi$ denotes any generic (bosonic or fermionic) field.
The advantage of the path integral representation with the above 
nonequilibrium, effective Lagrangian is that it is straightforward to 
construct diagrammatically a perturbative expansion of the nonequilibrium 
Green's functions in terms of modified nonequilibrium Feynman rules.
These nonequilibrium Feynman rules are as follows.

\begin{itemize}

\item{(i) The number of vertices is doubled: Those associated with fields on the 
``$+$'' branch are the usual interaction vertices, 
while those associated with fields on the ``$-$'' branch have the opposite 
sign. }

\item{(ii) There are four propagators corresponding to the possible contractions of fields among the 
two branches.  
Besides the usual time-ordered (Feynman) propagators which are associated
with fields on the ``$+$'' branch, there are anti-time ordered propagators
associated with fields on the ``$-$'' branch and 
the Wightman functions associated with fields on different branches.} 
 
\item{(iii) The combinatoric factors of the Feynman diagrams are the same as those in the usual 
calculation of S-matrix elements in field theory.}

\end{itemize}

For a scalar (bosonic) field $\Phi(x)$,
the spatial Fourier transform of the nonequilibrium propagators 
are defined by (the extension to the case of a gauge or fermionic field
is straightforward)
\begin{mathletters}
\begin{eqnarray}
&&G_{\vec k}^{>}(t,t')=i \int d^3x  \, e^{-i{\vec k}\cdot{\vec x}} \,
\Big\langle \Phi({{\vec x}},t) \Phi({{\vec 0}},t') \Big\rangle~, 
\label{greater}\\
&&G_{\vec k}^{<}(t,t')=i \int d^3x \, e^{-i{\vec k}\cdot{\vec x}} \,
\Big\langle \Phi({{\vec 0}},t') \Phi({{\vec x}},t) \Big\rangle ~,
\label{gsmaller}\\
&&G_{\vec k}^{++}(t,t')=G_{\vec k}^{>}(t,t')\theta(t-t')+G_{\vec k}^{<}(t,t')
\theta(t'-t)\, ,                                   \label{gplusplus}\\
&&G_{\vec k}^{--}(t,t')=G_{\vec k}^{>}(t,t')\theta(t'-t)+G_{\vec k}^{<}(t,t')
\theta(t-t')~, \label{gminusminus}\\
&&G_{\vec k}^{+-}(t,t')=G_{\vec k}^{<}(t,t') ~,\label{gplusminus} \\
&&G_{\vec k}^{-+}(t,t')=G_{\vec k}^{>}(t,t') ~, \label{gminusplus}
\end{eqnarray}\label{noneqpropagator}
\end{mathletters}
where $\langle\cdots\rangle$ denotes the 
expectation value with respect to
the initial density matrix. 
From the definitions of the 
nonequilibrium propagators eqs.~(\ref{noneqpropagator}), it is clear
that they satisfy the identity:
\begin{equation}
G_{\vec k}^{++}(t,t')+G_{\vec k}^{--}(t,t')-
G_{\vec k}^{+-}(t,t')-G_{\vec k}^{-+}(t,t')=0~.
\end{equation}
The retarded and advanced propagators are defined as
\begin{eqnarray*}
G_{{\rm R},\vec k}(t,t')&=&G_{\vec k}^{++}(t,t')-G_{\vec k}^{+-}(t,t')
=\left[G_{\vec k}^>(t,t')-G_{\vec k}^<(t,t')\right]\theta(t-t')~,\\
G_{{\rm A},\vec k}(t,t')&=&G_{\vec k}^{++}(t,t')-G_{\vec k}^{-+}(t,t')
=\left[G_{\vec k}^<(t,t')-G_{\vec k}^>(t,t')\right]\theta(t'-t)~,
\end{eqnarray*}
which are useful in the discussion of the pinch singularities
discussed in a later section (see Sec.~\ref{section:pinch}).

It now remains to specify the initial state. If we were considering
the situation in {\em equilibrium} the 
natural initial density matrix would describe a {\em thermal} initial
state for the free particles at temperature $ T $. The density matrix of
this initial state is $ \hat{\rho}=\exp(-H_0/T) $, where $ H_0 $ is the free
Hamiltonian of the system, and the time evolution is with the full
interacting Hamiltonian. This is tantamount 
to switching on the interaction at $t=t_0$. If the full Hamiltonian
does not commute with $ H_0 $ the density matrix {\bf evolves out of
equilibrium} for $ t> t_0 $. This choice of the thermal initial state
for the free particles determines the usual  
Kubo-Martin-Schwinger (KMS) conditions on the Green's functions:
\begin{equation}
G_{\vec k}^<(t,t')=G_{\vec k}^>(t-i\beta,t')~.\label{kms}
\end{equation}
Perturbative expansions are carried out with the following real-time equilibrium
free quasiparticle Green's functions:
\begin{mathletters}
\begin{eqnarray}
G_{\vec k}^{>}(t,t')&=&\frac{i}{2\omega_{\vec k}}
\left\{[1+n_B(\omega_{\vec k})]\;
e^{-i\omega_{\vec k}(t-t')}+n_B(\omega_{\vec k})\;e^{i\omega_{\vec k}(t- t')} 
\right\}~~,
\label{ggreat}\\
G_{\vec k}^{<}(t,t')&=&\frac{i}{2\omega_{\vec k}}\left\{
n_B(\omega_{\vec k})~e^{-i\omega_{\vec k}(t- t')}
+[1+n_B(\omega_{\vec k})]\;e^{i\omega_{\vec k}(t-t')}
\right\}~~,
\label{gsmall}\\
\omega_{\vec k}&=&\sqrt{k^2+m^2}~,\quad\quad\quad
n_B(\omega)=[\exp(\beta\omega)-1]^{-1}~,\label{bosefactor}
\end{eqnarray}\label{ffpropagator}
\end{mathletters}
where (here and henceforth) $k=|{\vec k}|$, and $m$ is the 
mass of the field and $n_B(\omega)$ is the equilibrium 
Bose-Einstein distribution function.

In a hot and/or dense medium the definition of the quasiparticles whose
distribution function we want to study  
may require a resummation scheme such as for example that of hard
thermal loops generalized to nonequilibrium 
situations. In these cases, the Hamiltonian is rearranged in such a
way that part of the interaction is self-consistently 
included in the part of the Hamiltonian that commutes with the
quasiparticle number operator, call it for convenience $ H_0 $ and
specific counterterms are included in the interacting part $ H_I $ to
avoid double counting.  

As we are interested in obtaining an equation of
evolution for a quasiparticle distribution function, the most natural
initial state corresponds 
to a density matrix that is diagonal in the basis of free
quasiparticles, i.e. that commutes with 
$H_0$. This initial density matrix is then evolved in time with the
full Hamiltonian, and if the interaction 
does not commute with $H_0$ the distribution function of these
quasiparticles  will evolve in time.  

The distribution function $ n_{\vec k}(t_0) $  
is the expectation value of the operator that counts  these
quasiparticles in the 
initial density matrix. Under the assumption that the initial density
matrix is diagonal in the basis of this 
quasiparticle number,  perturbative expansions are carried out 
with the following nonequilibrium free quasiparticle Green's 
functions:
\begin{mathletters}
\begin{eqnarray}
G_{\vec k}^{>}(t,t')&=&\frac{i}{2\omega_{\vec k}}\left\{[1+n_{\vec k}(t_0)]\;
e^{-i\omega_{\vec k}(t-t')}+n_{\vec k}(t_0)\; e^{i\omega_{\vec k}(t- t')} 
\right\}~,\label{fqp:ggreat}\\
G_{\vec k}^{<}(t,t')&=&\frac{i}{2\omega_{\vec k}}\left\{
n_{\vec k}(t_0)\; e^{-i\omega_{\vec k}(t- t')}
+[1+n_{\vec k}(t_0)]\;e^{i\omega_{\vec k}(t-t')}\right\}~~,
\label{fqp:gsmall}
\end{eqnarray}\label{fqppropagator}
\end{mathletters}
where $ \omega_{\vec k} $ is the dispersion relation for the free
quasiparticle. In this picture the width of the quasiparticles arises
from their interaction and is related to the relaxation rate of the
distribution function in relaxation time approximation.  
This point will become more clear in the
sections that follow where we implement this program in detail. 

Finally, it is easy to check that the (bosonic) 
free quasiparticle Green's functions, 
eqs.~(\ref{fqppropagator}) and (\ref{ffpropagator}), satisfy
\begin{equation}
G_{\vec k}^{>}(t,t') = G_{\vec k}^{<}(t',t)~, \label{gfrelation}
\end{equation}
which will be useful in our following calculations.

\section{Self-interacting Scalar Theory}\label{section:scalar}

We begin our investigation with a self-interacting scalar theory. 
The Lagrangian density is 
given by
\begin{equation}
{\cal{L}}[\Phi] = \frac{1}{2} \left(\partial_{\mu}\Phi\right)^2 - 
\frac{1}{2} m^2_{0}\Phi^2-\frac{\lambda}{4!}\Phi^4 ~,
\label{scalar:barelagrangian}
\end{equation}
where $m_0$ is the bare mass.

As mentioned in the introduction, the first step towards understanding
the kinetic regime is the identification of the {\em microscopic} time
scales in the problem. In a medium, the bare particles are dressed by
the interactions becoming quasiparticles. One is interested in
describing the relaxation of these quasiparticles. Thus the important
microscopic time scales are those associated with the quasiparticles
and not the bare particles. If a kinetic equation is obtained in
some perturbative scheme, such a scheme should be in terms of the
quasiparticles which already implies a resummation of the
perturbative expansion. This is precisely the rationale behind the
resummation of the hard thermal loops in finite temperature field
theory~\cite{htl,rob2,rob3} and also behind the self-consistent 
treatment~\cite{lawrie,boyanrk}.

In a scalar field theory in {\em equilibrium} such a self-consistent resummation 
can be implemented by writing in the Lagrangian
\begin{equation}
m^2_0= m^2_{\rm eff}+\delta m^2~,\label{scalar:counterterm}
\end{equation}
where $m_{\rm eff}$ is the renormalized and {\em temperature dependent}
quasiparticle thermal effective mass 
which enters in the propagators, and $\delta m^2$
is a counterterm which will cancel a subset of Feynman diagrams in the
perturbative expansion and is considered part of the interaction
Lagrangian. As shown in Ref.~\cite{parwani} for the scalar field theory case, 
this method implements a resummation akin to the hard thermal loops in
a gauge theory~\cite{htl,rob2,rob3}. Parwani showed~\cite{parwani} that this
resummation is effectively implemented by solving the following 
self-consistent gap equation for $m^2_{\rm eff}$~\cite{boyanrk,parwani,dolan}
\begin{equation}
m^2_{\rm eff}= m^2_0 +\frac{\lambda}{2} 
\langle\Phi^2\rangle~,\quad
\langle\Phi^2\rangle = \int \frac{d^3q}{(2\pi)^3}
\frac{1+2\,n_B(\omega_{\vec k})}{2\omega_{\vec k}}~,
\label{scalar:gapeq}
\end{equation}
with $\omega_{\vec k} = \sqrt{k^2+m^2_{\rm eff}}$. 
The divergences (quadratic and logarithmic in terms of a spatial momentum 
cutoff) in the zero-temperature part of eq.~(\ref{scalar:gapeq}) can
be absorbed into a renormalization of the bare mass by a subtraction at
some renormalization scale. 
A convenient choice corresponds to a renormalization scale at $T=0$ and 
$m(T=0)=m$ is the zero-temperature mass.

For $T \gg m_{\rm eff}$, the solution of the gap equation is given
by~\cite{parwani,dolan} 
\begin{equation}
m^2_{\rm eff}= m^2+{\lambda \over 2}\left[\ \frac{T^2}{12}-
\frac{m_{\rm eff}\,T}{4 \pi} + {\cal{O}} \left(m^2_{\rm eff} \ln 
\left[\frac{m_{\rm eff}}{T}\right]\right)\right]~.
\label{scalar:thermalmass1}
\end{equation}
In particular, for $ T\gg \sqrt{\lambda} \; T\gg  m $, we can neglect
the zero-temperature mass $ m $ and obtain
\begin{equation}
m^2_{\rm eff}=\frac{\lambda \; T^2}{24} +
{\cal{O}}\left(\lambda^{3/2}T^2\right)~.  
\label{scalar:thermalmass2}
\end{equation}
In the massless case, $ m_{\rm eff} $ serves as an infrared cutoff for the loop 
integrals~\cite{parwani,elmfors}.
The leading term of eq.~(\ref{scalar:thermalmass2}) provides the
correct microscopic time scale at large temperature. 

We note that this renormalized and
temperature dependent mass determines the important time scales in the
medium but is {\bf not} the position of the quasiparticle pole (or,
strictly speaking, resonance). 

When the temperature is much larger than the renormalized zero
temperature mass, the hard thermal loop resummation is needed to
incorporate the physically relevant time and length scales in the
perturbative expansion.
For a hard quasiparticle $ k\sim T $, 
while for a soft quasiparticle $ k\lesssim \sqrt{\lambda}\;T $, hence
the longest microscopic time  scale of the system is
$ t_{\rm micro} \sim 1/\sqrt{\lambda}\; T\sim 1/m_{\rm eff}$. 

\subsection{Quantum kinetic equation}

In this subsection we obtain the evolution equations for the distribution
functions of quasiparticles. For this we consider an initial state out
of equilibrium described by a density matrix that is diagonal in 
the basis of the free quasiparticles, but with nonequilibrium 
distribution functions. If the medium is hot these quasiparticles
will have an effective mass $m_{\rm eff}$ which
will result from medium effects, much in the same manner as the
temperature dependent thermal mass in the equilibrium 
situation described above. This mass will be very different from the
bare mass $m_0$ in the absence of medium 
effects and must be taken into account for the correct assessment of
the microscopic time scales. Thus, we write the 
Hamiltonian in terms of the in medium dressed mass $m_{\rm eff}$
and a counterterm $ \delta m^2 = m^2_0-m^2_{\rm eff} $ 
which will be treated as part of the perturbation and required to
cancel the mass shifts consistently in perturbation 
theory. This is the nonequilibrium generalization of the resummation
described above in the equilibrium case. We emphasize 
that $ m^2_{\rm eff} $ depends on the initial distribution of quasiparticles.
This observation  will become important later 
when we discuss the time evolution of the distribution functions and
therefore of the effective mass.   

We write the Hamiltonian of the theory as
\begin{mathletters}
\begin{eqnarray}
&&H = H_0 + H_{\text{int}}~,\label{scalar:hsplit} \\
&&H_0 = \frac{1}{2} \int d^3x \left[\Pi^2 +({\vec \nabla}\Phi)^2 +
m^2_{\rm eff}\; \Phi^2\right] ~, \label{scalar:hfree}\\
&&H_{\text{int}} = \int d^3x \left[
\frac{\lambda}{4!}\;\Phi^4+\frac{1}{2}\delta m^2\;\Phi^2\right]~,
\label{scalar:hint}
\end{eqnarray}
\end{mathletters}
where $\Pi({\vec x},t)=\dot\Phi({\vec x},t)$ is the canonical momentum, 
and the mass counterterm has been absorbed in the interaction. 
Here and henceforth, a dot denotes derivative with respect to time. 
The free part of the Hamiltonian $H_0$ describes free
quasiparticles of renormalized finite-temperature mass $m_{\rm eff}$ and is
diagonal and Gaussian in terms of free quasiparticle creation and
annihilation of operators $ a^{\dagger}({\vec k})$ and $a({\vec k})$. 

With this definition, the lifetime of the quasiparticles
will be a consequence of interactions. In this manner, the
nonequilibrium equivalent of the hard
thermal loops (in the sense that the distribution functions are non-thermal) 
 which in this theory amount to local terms, have 
been absorbed in the definition of the effective mass. This guarantees
that the microscopic time scales are explicit in the quasiparticle
Hamiltonian. 

As discussed in the previous section, we consider that the initial
density matrix at time $t=t_0$, is diagonal in the basis 
of free quasiparticles, but with out of equilibrium 
initial distribution functions $n_{\vec k}(t_0)$. 
The Heisenberg field operators at time $t$ are now written as
\begin{mathletters}                                         
\begin{eqnarray}
\Phi({\vec x},t)&=& \int \frac{d^3 k}{(2\pi)^{3/2}}\; \Phi({\vec k},t) \;
e^{i {\vec k} \cdot {\vec x}}~, \quad \quad \Phi({\vec k},t)= 
\frac{1}{\sqrt{2\omega_{\vec k}}} \left[a({\vec k},t)
+a^{\dagger}({-\vec k},t) \right]~,\\
\Pi({\vec x},t)&=&\int \frac{d^3 k}{(2\pi)^{3/2}}\; \Pi({\vec k},t) \; 
e^{i {\vec k} \cdot {\vec x}} ~, \quad \quad \Pi({\vec k},t) = i
\sqrt{\frac{\omega_{\vec k}}{2}}
\left[a^{\dagger}({-\vec k},t)-a({\vec k},t)\right]~,
\label{initialexpansion}
\end{eqnarray}
\end{mathletters}
where $ a^{\dagger}({\vec k},t) $ and $ a({\vec k},t) $ are, respectively,
creation and annihilation operators at time $t$ and 
$ \omega_{\vec k}=\sqrt{k^2+m^2_{\rm eff}} $.
The expectation value of quasiparticle number operators $n_{\vec k}(t)$ 
can be expressed in terms of the field $\Phi({\vec k},t)$ and the conjugate
momentum $\Pi({\vec k},t)$ as follows
\begin{eqnarray}
n_{\vec k}(t) &=& \left\langle a^{\dagger}({\vec k},t)\,a({\vec k},t)
\right\rangle \nonumber\\ 
&=& {1 \over 2\omega_{\vec k} }\biggl\{
\left\langle \Pi({\vec k},t)\,\Pi({-\vec k},t)\right\rangle + 
\omega^2_{\vec k}\,
\left\langle\Phi({\vec k},t)\,\Phi({-\vec k},t)\right\rangle\nonumber\\
&& +\,i\omega_{\vec k}\Bigl[\left\langle\Phi({\vec k},t)\,\Pi({-\vec k},t)
\right\rangle - \left\langle\Pi({\vec k},t)\,\Phi({-\vec
k},t)\right\rangle\Bigr]\biggr\}~, 
\label{ocupation}
\end{eqnarray}
where the bracket $\langle \cdots \rangle$ means an average over the
Gaussian density matrix defined by the initial distribution functions
$n_{\vec k}(t_0)$. The time-dependent distribution (\ref{ocupation}) is
interpreted as the quasiparticle distribution function.

The interaction Hamiltonian in momentum space is given by
\begin{equation}
H_{\text{int}} =\frac{\lambda}{4!}\frac{1}{(2\pi)^3}\int 
\prod_{i=1}^{4}d^3 q_i\,\Phi({{\vec q}_i},t)\;
\delta^3({\vec q}_1 +{\vec q}_2 +{\vec q}_3 +{\vec q}_4)
+ \frac{\delta m^2}{2}
\int d^3 q\,\Phi({\vec q},t) \; \Phi({-\vec q},t)~.\label{intTlarge}
\end{equation}
Taking the derivative of $n_{\vec k} (t)$ with respect to time and using the
Heisenberg field equations, we find
\begin{eqnarray}
{\dot{n}}_{\vec k} (t) &=&-\frac{1}{2\omega_{\vec k}} 
\Bigg[\frac{\lambda}{6} \,
\Big\langle [\Phi^3({\vec k},t)]\,\Pi({-\vec k},t)+
\Pi({\vec k},t) \,[\Phi^3({\vec k},t)]\Big\rangle  \nonumber \\
&& +\, \delta m^2 \, \Big\langle \Phi({\vec k},t) \, 
\Pi(-{\vec k},t) +\Pi({\vec k},t) \,
\Phi({-\vec k},t)\Big\rangle \Bigg]~, \label{nkdot}
\end{eqnarray}
where we use the compact notation:
\begin{equation}
[\Phi^3({\vec k},t)] \equiv\frac{1}{(2\pi)^3} \int
d^3 q_1 \;  d^3 q_2 \;  d^3 q_3\,\Phi({\vec q}_1,t) \; \Phi({\vec q}_2,t) \; 
\Phi({\vec q}_3,t)\;\delta^3\left({\vec k}-{\vec q}_1-{\vec q}_2-{\vec
q}_3\right)~. 
\end{equation}

In a perturbative expansion care is needed to handle the canonical
momentum [$\Pi({\vec k},t) = \dot{\Phi}({\vec k},t)$]
and the scalar field at the same time because of Schwinger terms. 
This ambiguity is avoided by noticing that
\begin{eqnarray}
\left\langle \Pi({\vec k},t)\, [\Phi^3({-\vec k},t)] \right\rangle&=&\text{Tr}
\left[\hat{\rho}(t_0)\, \Pi_{\vec k}(t)\,[\Phi^3({-\vec k},t)]  \right]\nonumber\\
&\equiv& \lim_{ t \rightarrow t'} {\partial \over \partial t'}
\text{Tr}\left[[\Phi^3({-\vec k},t)]^+ \,\hat{\rho}(t_0)\,\Phi^-({\vec k},t')
\right]\nonumber\\
&=&\left.{\partial \over \partial t'}\left\langle 
[\Phi^3({-\vec k},t)]^+ \,\Phi^-({\vec k},t')\right\rangle\right|_{t'=t}~,
\end{eqnarray}
where we used the cyclic property of the trace and the ``$\pm$''
superscripts for the fields refer to field insertions obtained as
variational derivatives with respect to sources in the forward ($+$) time
branch and backward ($-$) time branch in the nonequilibrium
generating functional.

We now use the canonical commutation relation between $\Pi$ and $\Phi$
and define the mass counterterm $\delta m^2=\lambda\Delta/6$ 
to write the above expression as
\begin{eqnarray}
{\dot{n}}_{\vec k} (t) & = & 
-\frac{\lambda}{12\omega_{\vec k}}\Bigg\{
\frac{\partial}{\partial t'} \bigg[2 \Big\langle
[\Phi^3({\vec k},t)]^+\,\Phi^-({-\vec k},t')\Big\rangle
+\Delta\Bigl[\Bigl\langle\Phi^+({\vec k},t)\Phi^-({-\vec k},t')\Big\rangle
 +\Big\langle\Phi^+({\vec k},t')\Phi^-({-\vec k},t)\Bigr\rangle\Bigr]
\bigg]_{t=t'} \nonumber \\
&& + 3i \int \frac{d^3 q}{(2\pi)^3}\Big\langle 
\Phi^+({\vec q},t)\Phi^-({-\vec q},t)\Big\rangle\Bigg\}~.\label{nkdot1}
\end{eqnarray}
The right-hand side of eq.~(\ref{nkdot1}) can be obtained 
perturbatively in weak coupling expansion in $\lambda$. 
Such a perturbative expansion is in terms of the nonequilibrium vertices and 
Green's functions eqs.~(\ref{noneqpropagator}) with the basic 
Green's functions given by eqs.~(\ref{fqppropagator}).
At order $ {\cal O}(\lambda) $ the right hand side of (\ref{nkdot1}) vanishes
identically. This is a consequence of the fact that the initial
density matrix is diagonal in the basis of free quasiparticles. 

Figs.~1a-1c display the contributions up to two loops to the
kinetic equation (\ref{nkdot1}). The tadpole diagrams, depicted in 
Figs.~1a and 1b as well as the last term in eq.~(\ref{nkdot1})
are canceled by the proper choice of $\Delta$. 

An important point to notice is that these Green's functions include the proper
microscopic scales as the contribution of the hard thermal loops have
been incorporated by summing the tadpole diagrams. The propagators
entering in the calculations are the resummed propagators. The terms
with $\Delta$ are required to cancel the tadpoles to all orders.

Thus, from the formidable
expression (\ref{nkdot1}) only the first term remains after $ \Delta $
is properly chosen in order to cancel the tadpole diagrams. This
requirement guarantees that the  mass in the propagators is the
effective mass that includes the microscopic time scales. Hence,
we find that the final form of the kinetic equation is given by 
\begin{equation}
{\dot{n}}_{\vec k} (t) = -\frac{\lambda}{6\omega_{\vec k}} \frac{\partial} 
{\partial t'} \left[ \Big\langle [\Phi^3({\vec k},t)]^+\,\Phi^-({-\vec k},t')
\Big\rangle \right]_{t= t'}~,\label{ndotequa}
\end{equation}
with the understanding that no tadpole diagrams contribute to the
above equations as they are automatically canceled by the terms
containing $\Delta$ in eq.~(\ref{nkdot1}).

To lowest order the condition that the tadpoles are canceled 
leads to the following condition on $\Delta$
\begin{equation}
\Delta = -3\int \frac{d^3q}{(2\pi)^3} \frac{1+2n_{\vec q}(t_0)}{2\,\omega_{\vec q}}~,
\end{equation}
therefore the effective 
mass is the solution to the self-consistent gap equation
\begin{equation}
m^2_{\rm eff} = m^2_0 + \frac{\lambda}{2} \int \frac{d^3q}{(2\pi)^3}
\frac{1+2\,n_{\vec q}(t_0)}{2\omega_{\vec q}}~,\quad   
\omega_{\vec q} = \sqrt{q^2+m^2_{\rm eff}} \label{selfconmass}~.
\end{equation} 
We see that the requirement that the term proportional to $\Delta$ in
the kinetic equation cancels the tadpole contributions is equivalent
to the hard thermal loop resummation in the equilibrium
case~\cite{parwani} and makes explicit that $m^2_{\rm eff}$ is a functional
of the {\em initial} nonequilibrium distribution functions.  

As will be discussed in detail below, such an expansion will be
meaningful for times 
$t\ll t_{\text{rel}}=|n_{\vec k}(t)/{\dot{n}}_{\vec k} (t)|$,
where $t_{\text{rel}}$ is the relaxational time scale for the nonequilibrium
distribution function. For small enough coupling we expect that
$t_{\text{rel}}$ will be large enough such that there is a wide separation
between the microscopic and the relaxational time scales that will
warrant such an approximation (see discussion below). 

To two-loop order, the time evolution of the distribution function
that follows from eq.(\ref{ndotequa}) is given by
\begin{eqnarray}
{\dot{n}}_{\vec k}(t) &=& \frac{\lambda^2}{3}\frac{1}{2\omega_{\vec k}}\,
\int \frac{d^3 q_1}{(2 \pi)^3 2\omega_{{\vec q}_1}}
\frac{d^3 q_2}{(2 \pi)^3 2\omega_{{\vec q}_2}}
\frac{d^3 q_3}{(2 \pi)^3 2\omega_{{\vec q}_3}}
\int^t_{t_0}dt''~(2\pi)^3 
\delta^3({\vec k}-{\vec q}_1-{\vec q}_2-{\vec q}_3)\nonumber\\
&& \times \bigg[{\cal N}_1(t_0)
\cos[(\omega_{\vec k}+\omega_{{\vec q}_1}+\omega_{{\vec q}_2}+
\omega_{{\vec q}_3})(t-t'')] 
+ 3{\cal N}_2(t_0)\cos[(\omega_{\vec k}+\omega_{{\vec q}_1}+ 
\omega_{{\vec q}_2}-\omega_{{\vec q}_3})(t-t'')] \nonumber \\
&& +\,3{\cal N}_3(t_0)
\cos[(\omega_{\vec k}-\omega_{{\vec q}_1}-\omega_{{\vec q}_2}
+\omega_{{\vec q}_3})(t-t'')]
+{\cal N}_4(t_0)
\cos[(\omega_{\vec k}-\omega_{{\vec q}_1}-\omega_{{\vec q}_2}
-\omega_{{\vec q}_3}(t-t'')]\bigg],
\label{scalar:perturbkineq}
\end{eqnarray}
where
\begin{mathletters}
\begin{eqnarray}
{\cal N}_1(t) &=& [1+ n_{\vec k}(t)][1+ n_{{\vec q}_1}(t)]
[1+ n_{{\vec q}_2}(t)][1+ n_{{\vec q}_3}(t)]- 
n_{\vec k}(t) \; n_{{\vec q}_1}(t) \; n_{{\vec q}_2}(t) \; n_{{\vec q}_3}(t)~,
\label{scalar:ocupanumber1}\\ 
{\cal N}_2(t) &=& [1+ n_{\vec k}(t)][1+ n_{{\vec q}_1} (t)]
[1+ n_{{\vec q}_2}(t)] \; n_{{\vec q}_3} (t)- 
n_{\vec k}(t) \;  n_{{\vec q}_1}(t) \;  n_{{\vec q}_2}(t) \;
[1+n_{{\vec q}_3}(t)]~, \\ 
{\cal N}_3(t) &=& [1+ n_{\vec k} (t)] \; n_{{\vec q}_1}(t)  \;
n_{{\vec q}_2}(t) \; [1+ n_{{\vec q}_3} (t)] - 
n_{\vec k}(t) \;[1+n_{{\vec q}_1}(t)][1+ n_{{\vec q}_2}(t)] \;n_{{\vec
q}_3}(t)~,\\ {\cal N}_4(t) &=& [1+ n_{\vec k} (t) ] \; n_{{\vec q}_1}(t) \;
n_{{\vec q}_2}(t)\;n_{{\vec q}_3}(t)- n_{\vec k}(t) \; [1+ n_{{\vec
q}_1}(t)] [1+ n_{{\vec q}_2}(t)][1+n_{{\vec q}_3}(t)]~ \;. 
\label{scalar:ocupanumber}
\end{eqnarray}
\end{mathletters}
The kinetic equation (\ref{scalar:perturbkineq}) is retarded and causal.
The different contributions have a physical interpretation in terms of the
`gain minus loss' processes in the plasma. The first term describes the
creation of four particles minus the destruction of four particles in
the plasma, the second and fourth terms describe the creation of three
particles and destruction of one minus destruction of three and
creation of one, the third term is the {\em scattering} of two
particles off two particles and is the usual Boltzmann term.

Since the propagators entering in 
the perturbative expansion of the kinetic equation are in terms of the
distribution functions at the initial time $ t_0 $, the time integration can 
be done straightforwardly leading to the following equation:
\begin{equation}
{\dot{n}}_{\vec k}(t) = \frac{\lambda^2}{3}\int d\omega\, 
{\cal R}[\omega,{\vec k};{\cal N}_i(t_0)]\, 
\frac{\sin[(\omega-\omega_{\vec k})(t-t_0)]}{\pi(\omega-\omega_{\vec k})}~,
\label{ke:scalar}
\end{equation}
where ${\cal R}[\omega,{\vec k};{\cal N}_i(t_0)]$  is given by
\begin{eqnarray}
{\cal R}[\omega,{\vec k};{\cal N}_i(t_0)]&=&  \frac{\pi}{2\omega_{\vec k}} \,
\int \frac{d^3 q_1}{(2 \pi)^3 2\omega_{{\vec q}_1}}
\frac{d^3 q_2}{(2 \pi)^3 2\omega_{{\vec q}_2}}
\frac{d^3 q_3}{(2 \pi)^3 2\omega_{{\vec q}_3}}
(2\pi)^3 \delta^3({\vec k}-{\vec q}_1-{\vec q}_2-{\vec q}_3)\nonumber\\
&& \times \bigg[\delta (\omega+\omega_{{\vec q}_1}+\omega_{{\vec q}_2}+
\omega_{{\vec q}_3})\,{\cal N}_1(t_0) + 3\,\delta(\omega+\omega_{{\vec q}_1}+ 
\omega_{{\vec q}_2}-\omega_{{\vec q}_3})\,
{\cal N}_2(t_0) \nonumber \\
&& +\,3\delta(\omega-\omega_{{\vec q}_1}-\omega_{{\vec q}_2}
+\omega_{{\vec q}_3})\,
{\cal N}_3(t_0)+\delta(\omega-\omega_{{\vec q}_1}-\omega_{{\vec q}_2}
-\omega_{{\vec q}_3})\,{\cal N}_4(t_0)\bigg]~. \label{bigR}
\end{eqnarray}

We are now ready to solve the kinetic equation derived above.
Since $ {\cal R}[\omega,{\vec k};{\cal N}_i(t_0)] $ is fixed at initial
time $ t_0 $, eq.(\ref{ke:scalar}) can be solved by direct integration
over $ t $, thus leading to
\begin{equation}
n_{\vec k}(t)=n_{\vec k}(t_0)+\frac{\lambda^2}{3}\int d\omega\,{\cal
R}[\omega,{\vec k};{\cal
N}_i(t_0)]\frac{1-\cos[(\omega-\omega_{\vec k})(t-t_0)]}  
{\pi(\omega-\omega_{\vec k})^2}~. \label{integrated}
\end{equation}  
This expression gives the time evolution of the quasiparticle distribution
function to lowest order in perturbation theory, but only for early times. 
To make this statement more precise consider the limit $t \gg t_0$ in the 
expression between brackets in (\ref{integrated})
which  can be recognized from Fermi's Golden Rule of
elementary time-dependent perturbation theory
\begin{equation}
\lim_{t-t_0\to\infty}\frac{1-\cos[(\omega-\omega_{\vec k})(t-t_0)]}
{\pi(\omega-\omega_{\vec k})^2} = (t-t_0) \; \delta(\omega-\omega_{\vec k})~.
\label{FGR}  
\end{equation}
A more detailed evaluation of the long time limit is 
obtained by using the following expression~\cite{boyrgir}
\begin{equation}
\int_{-a}^{\infty} { dy \over y^2} \left( 1 - \cos yt \right) \; p(y)
 \buildrel{t \to \infty}\over= \pi \; t \; p(0) + {\cal P}
\int_{-a}^{\infty} { dy \over y^2} \; [ p(y) -  p(0)] 
+  {\cal O}\left( {1 \over t } \right)~, 
\label{FGR2}
\end{equation}
where $a$ is a fixed positive number, $p(y)$ is a smooth function for 
$-a\le y < \infty$ and is regular at $y=0$. 
Thus, {\em provided} that 
$ {\cal R}[\omega,{\vec k};{\cal N}_i(t_0)]$ 
is finite at $\omega= \omega_{\vec k}$, we find  
$n_{\vec k}(t)$ is given by 
\begin{eqnarray}
n_{\vec k}(t)&=& n_{\vec k}(t_0)+\frac{\lambda^2}{3}\,
{\cal R}[\omega_{\vec k},{\vec k};{\cal N}_i(t_0)]\,(t-t_0) +
\,\text{non-secular terms}~.
\label{scalar:perturb}
\end{eqnarray} 
The term that grows linearly with time is a {\bf secular term}, and by
{\em non-secular terms} in (\ref{scalar:perturb}) we refer to terms
that are bound at all times. The approximation above, replacing the
oscillatory terms with resonant denominators by 
$ t\,\delta(\omega-\omega_{\vec k})$ is the same as that invoked
in ordinary time-dependent perturbation theory leading to Fermi's Golden Rule.

Clearly, the presence of secular terms in time restricts the validity
of the perturbative expansion to a time interval $ t-t_0  \ll t_{\rm rel} $ with 
\begin{equation}
t_{\rm rel}({\vec k}) \approx \frac{3n_{\vec k}(t_0)}{\lambda^2 \;
{\cal R}[\omega_{\vec k},{\vec k};{\cal N}_i(t_0)]}~.
\label{timeinter} 
\end{equation}
Since the time scales in the integral in eq.(\ref{integrated}) are
of the order of or shorter than $t_{\rm micro} \sim 1/m_{\rm eff}$ the
asymptotic form given by (\ref{scalar:perturb}) is valid for $t-t_0
\gg t_{\rm micro}$.  
Therefore for weak coupling there is a regime of {\em intermediate
asymptotics} in time 
\begin{equation}
t_{\rm micro} \ll t-t_0 \ll t_{\rm rel}({\vec k}) \label{interasym}
\end{equation}
such that (i) the corrections to the distribution function is
dominated by the secular term, and (ii) perturbation theory is 
{\em valid}.  

We note two important features of this analysis: 
\begin{itemize}
\item{ (i) In the intermediate asymptotic regime (\ref{interasym}) the
only {\em explicit} dependence on the initial time $t_0$ is in the
secular term, 
since ${\cal R}[\omega_{\vec k},{\vec k};{\cal N}_i(t_0)]$ depends on
$t_0$ only implicitly through the initial distribution 
functions. These observations will become important for the analysis
that follows below.}  
\item {(ii) ${\cal R}[\omega_{\vec k},{\vec k};{\cal N}_i(t_0)]$ given by
eq.~(\ref{bigR}) with
(\ref{scalar:ocupanumber1})-(\ref{scalar:ocupanumber}) evaluated at
$ t_0 $ {\bf vanishes} if the initial distribution functions are the equilibrium 
ones as a result of the on-shell delta functions and the equilibrium
relation $1+n_B(\omega_{\vec q}) =\exp(\beta\omega_{\vec q}) n_B(\omega_{\vec q})$, 
in this case there are no secular terms in
the perturbative expansion.} 
\end{itemize}

To highlight the significance of the second point above in a manner
that will allow us to establish contact with the issue of pinch
singularities in a later section, we note that  
the secular term in eq.~(\ref{scalar:perturb}) corresponds to 
the net change of quasiparticles distribution function in 
the time interval $t-t_0$. To see this more explicitly,  let us rewrite   
\begin{equation}
\frac{\lambda^2}{3}{\cal R}[\omega_{\vec k},{\vec k};{\cal N}_i(t_0)]=
\frac{-i}{2\omega_{\vec k}}
\Big[[1+n_{\vec k}(t_0)]\,\Sigma^{<}(\omega_{\vec k},{\vec k};t_0)-
n_{\vec k}(t_0)\,\Sigma^{>}(\omega_{\vec k},{\vec k};t_0)\Big]~,
\label{secularpart}
\end{equation}
where $\Sigma^{>}(\omega_{\vec k},{\vec k};t_0)-
\Sigma^{<}(\omega_{\vec k},{\vec k};t_0)\equiv
2\,i\,{\rm Im}\Sigma_R(\omega_{\vec k},{\vec k};t_0)$
with $\Sigma_R(\omega_{\vec k},{\vec k};t_0)$ 
being the the on-shell
{\em retarded} scalar self-energy~\cite{boyanrk} calculated
to two-loop order in terms of the initial distribution functions 
$n_{\vec k}(t_0)$. 
Indeed, the first and the second terms in (\ref{secularpart}),
respectively, correspond to the `gain' and the `loss' parts 
in the usual Boltzmann collision term. 
Hence one can easily recognize that 
$\lambda^2{\cal R}[\omega_{\vec k},{\vec k};{\cal N}_i(t_0)]/3$
is the {\em net production rate of quasiparticles per unit time}\footnote{
See Sec.~4.4 in Ref.~\cite{lebellac}, especially pp.~83-84.}.
Moreover, the absence of secular term for a system in thermal equilibrium 
[for which $n_{\vec k}(t_0)= n_B(\omega_{\vec k})$]  
is a consequence of the KMS condition 
for the self-energy in thermal equilibrium:
\begin{equation}
\Sigma^{>}(\omega_{\vec k},{\vec k})=e^{\beta\omega_{\vec k}}
\Sigma^{<}(\omega_{\vec k},{\vec k})~.
\end{equation}

\subsection{Dynamical renormalization group: 
resummation of secular terms}

The dynamical renormalization group is a systematic 
generalization of multiple scale analysis and sums the secular
terms, thus improving the perturbative expansion~\cite{goldenfeld,kuni}. 
It was originally introduced to improve the asymptotic 
behavior of solutions of differential
equations~\cite{goldenfeld,kuni} to study pattern formation 
in condensed matter systems and has since
been adapted to studying the nonequilibrium evolution 
of mean-fields in quantum field theory~\cite{salgado} 
and the time evolution of quantum systems~\cite{qm}. 

For discussions of the dynamical renormalization group 
in other contexts, including applications to problems in quantum mechanics and
quantum field theory, see Refs.~\cite{goldenfeld,kuni,salgado,qm}. 

In this section we implement the dynamical renormalization group
resummation of secular divergences to improve the perturbative
expansion following the formulation presented in Ref.~\cite{boyrgir}.   

This is achieved by introducing the renormalized initial distribution
functions $n_{\vec p}(\tau)$,
which are related to the bare initial distribution function 
$n_{\vec p}(t_0)$ via a renormalization constant 
${\cal Z}_{\vec p}(\tau,t_0)$ by 
\begin{equation}
n_{\vec p}(t_0)= {\cal Z}_{\vec p}(\tau,t_0) \;
n_{\vec p}(\tau)~,\quad{\cal Z}_{\vec p}(\tau,t_0)=1 + \frac{\lambda^2}{3}\; 
z^{(1)}_{\vec p}(\tau,t_0)+\cdots~,
\label{scalar:renormal}
\end{equation}
where $\tau$ is an arbitrary renormalization scale and 
$ z^{(1)}_{\vec p}(\tau,t_0) $ 
will be chosen to cancel the secular term at a time scale $ \tau $. 
Substitute eq.~(\ref{scalar:renormal}) into eq.~(\ref{scalar:perturb}), 
to $ {\cal O}(\lambda^2) $ we obtain 
\begin{equation}
n_{\vec k}(t)=n_{\vec k}(\tau)+\frac{\lambda^2}{3} \; 
\left\{z^{(1)}_{\vec k}(\tau,t_0)n_{\vec k}(\tau)+(t-t_0)\;{\cal
R}[\omega_{\vec k},{\vec k};{\cal N}_i(\tau)]\;\right\}+{\cal O}(\lambda^4)~.
\end{equation}
To this order, the choice
\begin{equation}
z^{(1)}_{\vec k}(\tau,t_0)=-(\tau-t_0)\;{\cal R}[\omega_{\vec k},{\vec k};{\cal
N}_i(\tau)]/ n_{\vec k}(\tau)  
\end{equation} 
leads to
\begin{equation}
n_{\vec k}(t)=n_{\vec k}(\tau)+\frac{\lambda^2}{3}\,(t-\tau)\;
{\cal R}[\omega_{\vec k},{\vec k};{\cal N}_i(\tau)] +{\cal O}(\lambda^4)~.
\label{scalar:rensol}
\end{equation}
Whereas the original perturbative solution was only valid for times
such that the contribution from the secular term remains very small 
compared to the initial distribution function at time $t_0$, the
renormalized solution eq.~(\ref{scalar:rensol}) is valid for time 
intervals $t-\tau$ such that the secular term remains small, 
thus by choosing $\tau$ arbitrarily close to $t$ we have
improved the perturbative expansion.

To find the dependence of $n_{\vec k}(\tau)$ on $\tau$, we make use
of the fact that $n_{\vec k}(t)$ does not depend on the {\em arbitrary} 
scale $\tau$: a change in the renormalization 
point $\tau$ is compensated by a change in the renormalized distribution 
function. This leads to the 
{\em dynamical renormalization group equation} to lowest order  
\begin{equation}
\frac{d}{d\tau} n_{\vec k}(\tau)-\frac{\lambda^2}{3}\,
{\cal R}[\omega_{\vec k},{\vec k};{\cal N}_i(\tau)]=0~.
\label{scalar:drge}
\end{equation}

This renormalization of the distribution function  also affects the
effective mass of the quasiparticles since 
$m^2_{\rm eff}$ is determined from the self-consistent equation
(\ref{selfconmass}) which in turn is a consequence of 
the tadpole cancelation consistently in perturbation theory. Since the
effective mass is a functional of the 
distribution function it will be renormalized consistently. This is
physically correct since the in-medium effective 
masses will change under the time evolution of the distribution functions. 

Choosing the arbitrary scale $\tau$ to coincide with the time $t$ in 
eq.~(\ref{scalar:drge}), we obtain 
the {\em resummed} kinetic equation:
\begin{eqnarray}
{\dot{n}}_{\vec k}(t) &=& \frac{\lambda^2}{3}\frac{\pi}{2\omega_{\vec k}}
\int\frac{d^3 q_1}{(2 \pi)^3 2\omega_{{\vec q}_1}}
\frac{d^3 q_2}{(2 \pi)^3 2\omega_{{\vec q}_2}}
\frac{d^3 q_3}{(2 \pi)^3 2\omega_{{\vec q}_3}}
(2\pi)^3 \delta^3 ({\vec k}-{\vec q}_1 -{\vec q}_2- 
{\vec q}_3)\nonumber \\
&& 
\times \Big[ \delta( \omega_{\vec k} + \omega_{{\vec q}_1} 
+ \omega_{{\vec q}_2}+
\omega_{{\vec q}_3}) {\cal N}_1(t)+ 3\,\delta(\omega_{\vec k} 
+\omega_{{\vec q}_1}+
\omega_{{\vec q}_2}-\omega_{{\vec q}_3}){\cal N}_2(t)\nonumber \\
&&
+\,3\,\delta(\omega_{\vec k} -\omega_{{\vec q}_1}-\omega_{{\vec q}_2}
+\omega_{{\vec q}_3}){\cal N}_3(t)
+\delta(\omega_{\vec k} -\omega_{{\vec q}_1}-\omega_{{\vec q}_2}
-\omega_{{\vec q}_3}){\cal N}_4(t)\Big]~, \label{resucinetica}
\end{eqnarray}
\noindent where the $ {\cal N}_i(t) $ are given in
eqs.(\ref{scalar:ocupanumber1})-(\ref{scalar:ocupanumber}). To 
avoid cluttering of notation in the above expression we have not made
explicit the fact that the frequencies $ \omega_{\vec q} $ {\em depend
on time} through the time dependence of $m_{\rm eff}$ which is in turn
determined by the time dependence of the distribution function. 
Indeed, the renormalization group resummation leads at once to the
conclusion that the cancelation of tadpole terms by a proper choice of
$\Delta$ requires that at every time $ t $ the effective mass is the
solution of the {\em time-dependent} gap equation 

\begin{equation}
m^2_{\rm eff}(t) = m^2_0 + \frac{\lambda}{2}  \int \frac{d^3q}{(2\pi)^3}
\frac{1+2n_{\vec q}(t)}{2\omega_{\vec q}(t)}~,\quad  
\omega_{\vec q}(t) = \sqrt{q^2+m^2_{\rm eff}(t)} \label{selfconmassoft}~,
\end{equation} 
where $n_{\vec q}(t)$ is the solution of the kinetic
eq.(\ref{resucinetica}). Thus, the quantum kinetic equation that
includes a nonequilibrium generalization of the hard thermal 
loop resummation in this scalar theory is given by
(\ref{resucinetica}) with the frequencies  
$ \omega_{\vec q} \rightarrow \omega_{\vec q}(t) $ given as self-consistent solutions
of the time-dependent gap equation (\ref{selfconmassoft}) and of the kinetic equation
(\ref{resucinetica}).  

The quantum kinetic equation (\ref{resucinetica})
is therefore {\bf more general} than the familiar Boltzmann equation
for a scalar field theory in that it includes the 
proper in medium modifications of the quasiparticle masses. This
approach provides an alternative derivation of the 
self-consistent method proposed in Ref.~\cite{lawrie}. 

It is now evident that the dynamical renormalization group systematically 
resums the secular terms and the corresponding dynamical renormalization group 
equation extracts the {\em slow evolution} of the nonequilibrium system.  

For small departures from equilibrium the time scales for relaxation
can be obtained by linearizing the kinetic 
equation (\ref{resucinetica}) around the equilibrium solution at
$ t=t_0 $. This is the relaxation time approximation which assumes that
the distribution function for a fixed mode of momentum $ {\vec k} $ 
is perturbed slightly off equilibrium 
such that $ n_{\vec k}(t_0)=n_B(\omega_{\vec k})+\delta n_{\vec k}(t_0) $,
while all the other modes remain in equilibrium, i.e., 
$ n_{\vec k+q}(t_0)=n_B(\omega_{\vec k+q}) $ for $ {\vec q}\neq{\vec 0} $.

Recognizing that only the on-shell delta function that multiplies the
scattering term ${\cal N}_3(t)$ in (\ref{resucinetica}) is
fulfilled, we find that the linearized kinetic equation
(\ref{resucinetica}) reads 
\begin{equation}
\delta\dot{n}_{\vec k}(t) = -\gamma({\vec k})\,
\delta n_{\vec k}(t)~,\label{linearboltzt}
\end{equation}
where $\gamma({\vec k})$ is the scalar relaxation rate
\begin{eqnarray}
\gamma({\vec k})&=&\frac{\lambda^2\pi}{2\, \omega_{\vec k}}
\int \frac{d^3 q_1}{(2 \pi)^3 2\omega_{{\vec q}_1}}
\frac{d^3 q_2}{(2 \pi)^3 2\omega_{{\vec q}_2}}
\frac{d^3 q_3}{(2 \pi)^3 2\omega_{{\vec q}_3}}
(2\pi)^3 \delta^3 ({\vec k}-{\vec q}_1 -{\vec q}_2- {\vec q}_3) 
\delta(\omega_{\vec k} -\omega_{{\vec q}_1}-\omega_{{\vec q}_2}+
 \omega_{{\vec q}_3})\nonumber \\
&& \times \left\{ [1+n_B(\omega_{{\vec q}_1})]
\,[1+ n_B(\omega_{{\vec q}_2})]\, n_B(\omega_{{\vec q}_3})-
n_B(\omega_{{\vec q}_1})\,n_B(\omega_{{\vec q}_2})
\,[1+ n_B(\omega_{{\vec q}_3})] \right\}~~.
\end{eqnarray}
Solving eq.~(\ref{linearboltzt}) with the initial condition
$ \delta n_{\vec k}(t=t_0)=\delta n_{\vec k}(t_0) $, we find that the
quasiparticle distribution function in the linearized approximation
evolves in time in the following manner 
\begin{equation}
\delta n_{\vec k}(t)=\delta n_{\vec k}(t_0)\; e^{-\gamma({\vec k})(t-t_0)}~.
\label{expsol}
\end{equation}
The linearized approximation gives the time scales for relaxation for
situations close to equilibrium. In the case of soft momentum
($ T \gg m_{\rm eff}\gg k $) and high temperature $ \lambda T^2\gg m^2 $ we
obtain~\cite{boyanrk}  
\begin{equation}
t_{\rm rel}(k \approx 0)= [\gamma({k\approx 0})]^{-1}\approx
\frac{32 \sqrt{24\pi}}{\lambda^{3/2}~T}~. \label{trelax}
\end{equation}

For very weak coupling (as we have assumed), the relaxational time scale
is much larger that the microscopic one $t_{\rm micro}\sim 1/m_{\rm eff} 
\approx 1/\sqrt{\lambda}~T$, since
\begin{equation}
\frac{t_{\rm rel}}{t_{\rm micro}}
\sim\frac{1}{\lambda}\gg 1~.
\label{timescale}
\end{equation}
This verifies the assumption of separation of microscopic and 
relaxation scales in the weak coupling limit. 

\section{Comparison to the usual renormalization group and general strategy}

In order to relate this approach to obtain kinetic 
equations using a {\em dynamical renormalization group} 
to more familiar situations we now discuss two simple 
cases in which the same type of method leads to
a resummation of the perturbative series in the same manner: 
the first is the simple case of a weakly damped
harmonic oscillator with a small damping 
coefficient and the second, closer to the usual renormalization group
ideas is the scattering amplitude in a four dimensional scalar theory. 

\subsection{The weakly damped harmonic oscillator}

Consider the equation of motion for a weakly damped harmonic oscillator:
$$
\ddot{y}+y=-\epsilon \dot{y}~, \quad \epsilon \ll 1 \label{damposc}~.
$$
Attempting to solve this equation in a perturbative expansion in
$\epsilon$ leads to the lowest order solution  
$$
y(t)= A\; e^{it}\left[1-\frac{\epsilon}{2}\; t\right] + 
\mbox{c.c.}+\mbox{non-secular terms}~,
$$
where the term that grows in time, i.e., the linear secular term
leads to the breakdown of the perturbative expansion at time scales
$t_{\rm break} \propto 1/\epsilon$. The dynamical renormalization group
introduces a renormalization of the complex amplitude at a time scale
$\tau$ in the form  
$ A=Z(\tau)\,A(\tau)$ with $ Z(\tau)=1+ z_1(\tau)\,\epsilon +\cdots $.
Choosing $z_1$ to cancel the secular term at this time scale leads to
$$ 
y(t)= A(\tau)\; e^{it}\left[1-\frac{\epsilon}{2}\; (t-\tau)\right] + \mbox{c.c.}~.
$$
The solution $y(t)$ cannot depend on the arbitrary scale at which the
secular term 
(divergence) has been subtracted, and this independence $\partial
y(t)/\partial \tau =0$ 
 leads to the following renormalization group equation to lowest order
in $ \epsilon $ 
$$
\frac{d A(\tau)}{d \tau}+\frac{\epsilon}{2} \;A(\tau)=0~.
$$
Now choosing $ t=\tau $, the renormalization group  improved solution
is given by 
$$ 
y(t)=e^{-\epsilon t/2}\left[A(0)\;e^{it}+\mbox{c.c.}\right]~.
$$ 
This is obviously the correct solution to ${\cal O}(\epsilon)$.
The interpretation of the renormalization group resummation is very clear
in this simple example: the perturbative expansion is carried out to a time
scale $\tau \ll 1/\epsilon$ within which perturbation theory is
valid. The correction is recognized as a change in the 
amplitude, so at this time scale the correction is absorbed in a 
renormalization of the amplitude and the
perturbative expansion is carried out to a longer time but in terms of
the {\em amplitude at the renormalization scale}. The dynamical
renormalization group equation is the differential form of this
procedure of evolving in time, absorbing the corrections into the
amplitude (and phases) and continuing the evolution in terms of the
renormalized amplitudes and phases. As we will see with the next 
example this is akin to the renormalization group in field theory.

\subsection{Scattering amplitude in scalar field theory}

Consider the scalar field theory described by the Lagrangian density
(\ref{scalar:barelagrangian}) defined 
as a  field theory in four dimensions with an upper momentum
cutoff $ \Lambda $ and consider for  
simplicity the massless case. The 1PI four 
point function (two particles to two particles scattering amplitude)
at the off-shell symmetric point is given to one-loop at zero
temperature in Euclidean space by 
\begin{equation}
\Gamma^{(4)}(p,p,p,p)= \lambda_0
-\frac{3}{2}\lambda_0^2\ln\left(\frac{\Lambda}{p}\right) +
{\cal O}(\lambda_0^3)~,\label{gamma4}
\end{equation}
where $\lambda_0$ is the bare coupling, 
$p$ is the Euclidean four-momentum. 
Clearly perturbation theory breaks down for $\Lambda/p \gtrsim e^{1/\lambda_0^2} $.  

Let us introduce the renormalized coupling constant at a scale 
$\kappa$ as usual as 
$$
\lambda_0 = \; {\cal Z}_{\lambda}(\kappa)\,\lambda(\kappa)~,\quad 
{\cal Z}_{\lambda}(\kappa) = 1 + z_1(\kappa)\,\lambda(\kappa)
+{\cal O}(\lambda^3)~,
$$
and choose $z_1(\kappa)$ to cancel the logarithmic divergence 
at a arbitrary renormalization scale  $ \kappa $.
Then in terms of $ \lambda(\kappa) $ the scattering amplitude becomes 
\begin{eqnarray}
\Gamma^{(4)}(p,p,p,p)
&=&\lambda(\kappa)+\frac{3}{2}\lambda^2(\kappa)\ln{p \over \kappa}  
+ {\cal O}(\lambda^3)\label{RGFT}~,
\end{eqnarray}
with $ \Gamma^{(4)}(\kappa,\kappa,\kappa,\kappa)= \lambda(\kappa)$.
The scattering amplitude does not depend on the arbitrary
renormalization scale $ \kappa $ and this independence 
implies $ \kappa\,\partial\Gamma^{(4)}(p,p,p,p) /\partial \kappa =0 $, 
which to lowest order leads to the
{\em renormalization group equation} 
\begin{equation}
\kappa\; \frac{d\lambda(\kappa)}{d\kappa} =
\frac{3}{2}\lambda^2(\kappa) + {\cal O}(\lambda^3)~,
\label{RGEFT} 
\end{equation} 
where $ \beta_{\lambda} \equiv \frac{3}{2}\lambda^2(\kappa) + 
{\cal O}(\lambda^3) $ is recognized as the renormalization group
beta function. Solving this renormalization 
group equation with an initial condition $\lambda(\bar{p})=\bar{\lambda}$ 
that determines the
scattering amplitude at some value of the momentum 
and choosing $ \kappa=p $ in eq.~(\ref{RGFT}), 
one obtains the renormalization
group improved scattering amplitude (at an off-shell point)
\begin{equation}
\Gamma^{(4)}(p,p,p,p;\bar{p},\bar{\lambda})= \lambda(p)~,
\label{RGimproved} 
\end{equation}
with $ \lambda(p) $ the solution of the renormalization group equation (\ref{RGEFT})
$$
\lambda(p) = \frac{\bar{\lambda}}{1 - (3\bar{\lambda}/2) \ln(p/\bar{p})}~.
$$ 

The connection between the renormalization group in momentum space and 
the dynamical renormalization group in real time (resummation of secular terms) used
in previous sections is immediate through the identification
$$
t_0 \Leftrightarrow \ln(\Lambda/\bar{p})~,\quad
t\Leftrightarrow\ln(p/\bar{p})~,\quad
\tau\Leftrightarrow\ln(\kappa/\bar{p})~, 
$$
which when replaced into (\ref{gamma4}) illuminates the equivalence with secular terms. 

This simple analysis highlights how the {\em dynamical renormalization
group} does precisely the same in the real-time formulation of kinetics 
as the renormalization group in Euclidean or zero temperature field theory. 
Much in the same manner that the renormalization group improved scattering
amplitude (\ref{RGimproved}) is a {\em resummation} of the
perturbative expansion, the kinetic equations obtained from the
dynamical renormalization group improvement represent a 
resummation of the perturbative expansion. The lowest order renormalization group
equation (\ref{RGEFT}) resums the leading logarithms, while the lowest
order {\em dynamical} renormalization group equation resums 
the leading secular terms.  

We can establish a closer relationship to the usual renormalization
program of field theory in its momentum shell version with
the following alternative interpretation of the secular terms and
their resummation~\cite{boyrgir}. 

The initial distribution at a time $ t_0 $ is evolved in 
time perturbatively up to a time scale $t_0+\Delta t$ such that the
perturbative expansion is still valid, 
i.e., $ t_{\rm rel}\gg \Delta t $ with $t_{\rm rel}$ the relaxational time scale. 
Secular terms begin to dominate the perturbative
expansion at a time scale $ \Delta t\gg t_{\rm micro} $ with $t_{\rm micro}$ 
the microscopic time scale. Thus if there is
a separation of time scales such that $ t_{\rm rel}\gg \Delta
t\gg t_{\rm micro}$, then in this intermediate asymptotic regime
perturbation theory is reliable but secular terms appear and can be isolated.  
A renormalization of the distribution function absorbs the contribution 
from the secular terms. 
The ``renormalized'' distribution function is used as an initial
condition at $t_0+\Delta t$ to iterate forward in time to $t_0+2 \Delta t$ 
using perturbation theory but with 
{\em the propagators in terms of the distribution function at the time scale
$t_0+\Delta t$}. 
This procedure can be carried out ``infinitesimally'' 
(in the sense compared with the relaxational time scale) 
and the differential equation that describes the changes of 
the distribution function under the intermediate asymptotic time evolution
is the dynamical renormalization group equation. 

This has an obvious similarity to the renormalization in terms of
integrating in momentum shells, the 
result of integrating out degrees of freedom in a momentum shell are
absorbed in a renormalization of 
the couplings and an effective theory at a lower scale but in terms of
the effective couplings. This 
procedure is carried out infinitesimally and the differential equation
that describes the changes of 
the couplings under the integration of degrees of freedom in these
momentum shells is the renormalization 
group equation. For other
examples of the dynamical renormalization group and its  relation to
the Euclidean renormalization program see Ref.~\cite{boyrgir}.  

An important aspect of this procedure of evolving in time and
``resetting'' the distribution functions, is that in this process 
it is implicitly assumed that the density matrix is diagonal in the
basis of free quasiparticles. Clearly if at the initial 
time the density matrix was diagonal in this basis, because the
interaction Hamiltonian does not commute with the density matrix, 
off-diagonal density matrix elements will be generated upon time
evolution. In resetting the distribution functions and using the 
propagators in terms of these updated distribution functions we have
neglected off-diagonal correlations, for example, in terms of 
the creation and annihilation of quasiparticles $a^{\dagger}({\vec k})$
and $ a({\vec k})$ upon time evolution new correlations 
of the form $\langle a({\vec k})a({\vec k}) \rangle$ and its hermitian
conjugate will be generated. In neglecting these terms 
we are introducing a {\em coarse graining}~\cite{boyanrk}, thus several
stages of coarse-graining had been introduced: (i) integrating 
in time up to an intermediate asymptotics and resumming the secular
terms neglect transient phenomena, i.e., averages  
over the microscopic time scales and (ii) 
off diagonal matrix elements (in the basis of free quasiparticles)
had been neglected. This coarse graining also has an equivalent in the
language of Euclidean renormalization: these are the irrelevant 
couplings that are generated upon integrating out shorter
scales. Keeping {\em all} of the correlations in the density matrix
would be equivalent to a Wilsonian renormalization in which all
possible couplings are included in the Lagrangian and all of 
them are maintained in the renormalization on the same footing. 

\subsection{Quantum Kinetics}

Having provided a method to obtain kinetic equations by implementing
the dynamical renormalization group resummation and  
compared this method to the improvement of asymptotic solutions of
differential equations as well as with the more familiar 
renormalization group of Euclidean quantum field theory we are now in
position to provide a simple recipe to obtain  
kinetic equations from the microscopic theory in the general case:

\begin{itemize}
\item{{\bf (1)} The first step requires the proper identification of
the quasiparticle degrees of freedom and their 
dispersion relations that is  frequency vs. momentum which is
determined from the real part of the self-energies on shell.  
The damping of these excitations will arise as a result of their
interactions and will be accounted for by the kinetic  
description. Define the number operator $N_{\vec k}(t)$ that counts these
quasiparticles in phase space and split the Hamiltonian into a part 
that commutes with this number operator (non-interacting) and a part
that changes the particle number (interacting). It is important 
that these particles or quasiparticles are defined in terms of the
correct microscopic time scales by including the proper frequencies 
in their definition. In the case of scalar $\phi^4$ near equilibrium
at high temperature the renormalized mass is the hard 
thermal loop resummed, such would also be the case in a gauge theory
in thermal equilibrium in the HTL limit. This is important to
determine the regime 
of validity of the perturbative expansion within which the secular
terms can be identified unambiguously, i.e., the intermediate 
asymptotics. It is here where the assumption of wide separation of
time scales enters. 
Although in most circumstances the non-interacting part is simply the free
field Hamiltonian (in terms of renormalized masses and fields) there 
could be other circumstances in which the non-interacting part is more
complicated, for example in the case of collective modes. 
The initial density matrix is usually assumed to be diagonal 
in the basis of this number operator but with nonequilibrium
distribution functions at the initial time. 
The real-time propagators are then given by (\ref{fqppropagator}).}

\item{{\bf (2)} Use the Heisenberg equations of motion to obtain a
general equation for $ \dot{n}_{\vec k}(t)$ with 
$n_{\vec k}(t)=\langle N_{\vec k}(t) \rangle $. Perform 
a perturbative expansion of this equation to the desired order in
perturbation theory, using the Feynman rules of real-time perturbation 
theory and the propagators (\ref{fqppropagator}). The resulting
expression is a functional of the distribution functions {\bf at the
initial time}. The only time dependence arises from the explicit
time dependence of the free propagators
(\ref{fqppropagator}). Integrate this expression in time and {\em
recognize the secular terms}.}

\item{{\bf (3)} Introduce the renormalization of the distribution
functions as in (\ref{scalar:renormal}) with the renormalization
constant $ {\cal Z}(\tau) $ expanded consistently in perturbation theory as in
(\ref{scalar:renormal}). Fix the coefficients $z^{(n)}(\tau)$ to  
cancel the secular terms consistently at the time scale $ \tau $. Obtain
the renormalization group equation from the $\tau$ independence 
of the distribution function, i.e., $d n_k /d \tau =0$. This dynamical
renormalization group equation {\bf is the quantum kinetic equation}.}
\end{itemize}

\noindent{\bf Corollary}: The similarity with the renormalization of
couplings explored in the previous section, suggests 
that the collisional terms of the quantum kinetic  
equation can be interpreted as beta
functions of the dynamical renormalization group and
that the space of distribution functions can be interpreted as a
coupling constant space. The dynamical renormalization group
trajectories determine the flow in this space, therefore fixed points
of the dynamical renormalization group describe 
stationary solutions with given distribution functions. 
Thermal equilibrium distributions are thus fixed points of 
the dynamical renormalization group. Furthermore, there can be
{\bf other} stationary solutions with non-thermal distribution functions,
for example describing turbulent behavior~\cite{zak}. 

Linearizing around these fixed points corresponds to 
linearizing the kinetic equation and the 
linear eigenvalues are related to the {\em relaxation rates}, 
i.e., linearization around the fixed points of the dynamical 
renormalization group corresponds to the {\em relaxation time approximation}. 

We now implement the program described by steps (1-3) in several
relevant cases in scalar and gauge field theories.  

\section{O(4) linear sigma model: Cool Pions and Sigma mesons}

In this section we consider an $ O(4) $ linear sigma model in the
strict chiral limit, i.e., without an explicit chiral symmetry breaking term:
\begin{equation}
{\cal L}[{\bbox\pi},\sigma] = \frac{1}{2} 
\left(\partial_{\mu}{\bbox\pi}\right)^2+
\frac{1}{2} \left(\partial_{\mu}\sigma\right)^2
-\frac{\lambda}{4}\left({\bbox\pi}^2+\sigma^2-f_{\pi}^2\right)^2~,
\end{equation}
where $ {\bbox\pi}=(\pi^1,\pi^2,\pi^3)$
and $ f_{\pi}\sim 93 $~MeV is the pion decay constant.  
At high temperature $T>T_c$, 
where $ T_c\sim{\cal O}(f_{\pi}) $~\cite{Bochkarev:1996gi} 
is the critical temperature, the $ O(4) $ symmetry is restored by a 
second order phase transition. 

In the symmetric phase pions and sigma mesons are 
degenerate, and the linear sigma model reduces to a self-interacting scalar 
theory, analogous to that discussed in Sec.~\ref{section:scalar}.
Thus, we limit our discussion here to the low temperature 
broken symmetry phase in which the temperature $ T\ll f_{\pi} $.
Since at low temperature the $O(4)$ symmetry is spontaneously broken via the 
sigma meson condensate, we shift the sigma field 
$ \sigma({\vec x},t)=\bar{\sigma}({\vec x},t)+v $, 
where $ v $ is temperature dependent and yet to be determined.
In equilibrium  $ v $ is fixed by requiring that 
$ \left\langle\bar{\sigma}({\vec x},t)\right\rangle=0 $
to all orders in perturbation theory for temperature $ T<T_c $. 
In the real-time formulation of nonequilibrium quantum field theory,
this split must be performed on both branches of the path
integral. Along the forward $(+)$ and backward 
$(-)$ branches the sigma field $\sigma^{\pm}({\vec x},t)$ is written as
$$
\sigma^{\pm}({\vec x},t)=\bar{\sigma}^{\pm}({\vec x},t)+v~,
$$
with $ \langle\bar{\sigma}^{\pm}({\vec x},t)\rangle = 0 $.
The expectation value $ v $ is obtained by requiring that the 
expectation value of $ \bar{\sigma}({\vec x},t) $ vanishes in
equilibrium to all orders in perturbation theory. Using the tadpole
method~\cite{tadpole} to one-loop order the equation that determines
$ v $ is given by 
\begin{equation}
v\left[v^2-f_{\pi}^2+\langle{\bbox\pi}^2\rangle+
3\langle\bar{\sigma}^2\rangle\right]=0~,\label{vev} 
\end{equation}
Once the solution of this equation for $v$ is used in the perturbative
expansion up to one loop, the tadpole diagrams that arise from the
shift in the field cancel. This feature of cancelation of 
tadpole diagrams that would result in an expectation value of
$\bar{\sigma}$ by the consistent use of the tadpole equation persists
to all orders in perturbation theory. Furthermore, away from equilibrium, when 
$\langle {\bbox \pi}^2 \rangle$ and $\langle \sigma^2 \rangle$ depend on
time through the time dependence of the distribution 
function, the tadpole condition (\ref{vev}) implies that $v$ 
becomes implicitly time dependent.  

A solution of (\ref{vev}) with $ v\neq 0 $ signals broken symmetry and
massless pions (in the strict chiral limit).  Therefore once the
correct expectation value $ v $ is 
used,  the one-particle-reducible (1PR) tadpole diagrams do not contribute 
in the perturbative expansion of the kinetic equation. Up  to this
order the inverse pion propagator reads
$$
\Delta^{-1}_{\pi}(\omega,{\vec k})=\omega^2-k^2-\lambda\left[
 v^2-f_{\pi}^2 +\langle{\bbox\pi}^2\rangle+3\langle\bar{\sigma}^2
\rangle\right]~,
$$
which vanishes for vanishing energy and momentum whenever $ v\neq 0 $ by
the tadpole condition (\ref{vev}), hence 
Goldstone's theorem is satisfied and the pions are the Goldstone bosons.
The study of the relaxation of sigma  mesons (resonance) and pions
near  and below the chiral phase transition is an 
important phenomenological aspect of low energy chiral phenomenology
with relevance to heavy ion collisions. Furthermore, 
recent studies have revealed interesting features associated with the
dropping of the sigma mass near the chiral transition and the enhancement of
threshold effects with potential experimental
consequences~\cite{hatsuda}. The kinetic approach described here could
prove useful to further assess the contributions 
to the width of the sigma meson near the chiral phase transition,
this is an important study on its own right  
and we expect to report on these issues in the near future. 

With the purpose of comparing to recent results, we now focus on the
situation at low temperatures under the 
assumption that the distribution functions of sigma mesons and
pions are not too far from equilibrium, i.e., {\em cool} 
pions and sigma mesons. At low temperatures the relaxation of pions
and sigma mesons will be dominated by the one-loop 
contributions, and the scattering contributions will be
subleading. The scattering contributions are of the same form as those 
discussed for the scalar theory and involve at least two distribution functions 
and are subdominant in the low temperature limit as 
compared to the one loop contributions described below. 

Since the linear sigma model is renormalizable and we focus
on finite-temperature effects, 
we ignore the zero-temperature ultraviolet divergences which can be absorbed
into a renormalization of $f_{\pi}$.
For a small departure from thermal equilibrium, 
we can approximate 
$\langle{\bbox\pi}^2\rangle$ and $\langle\bar{\sigma}^2\rangle$
by their equilibrium values:
\begin{equation}
\langle{\bbox\pi}^2\rangle = \frac{T^2}{4}~,\quad
\langle\bar{\sigma}^2\rangle =
\int\frac{d^3q}{(2\pi)^3}\frac{n_B(\omega_{\vec q})}{\omega_{\vec q}}~,
\end{equation}
where $\omega_{\vec q}=\sqrt{q^2+m_{\sigma}^2}$. The sigma mass 
$m_{\sigma}^2=2\lambda v^2$ is to be determined self-consistently.
In the low temperature limit $T \ll f_{\pi}$, 
we find $v^2=f_{\pi}^2\,[1-{\cal O}(T^2/f_{\pi}^2)]$ and 
$m_{\sigma}=\sqrt{2\lambda}\,f_{\pi}[1-{\cal O}(T^2/f_{\pi}^2)]$.
Thus in the case of a cool linear sigma model where $T\ll f_{\pi}$, we can 
approximate $v$ and $m_{\sigma}$ by $f_{\pi}$ and 
$\sqrt{2\lambda}f_{\pi}$, respectively. 

The main reason
behind this analysis is to display the microscopic time scales for the mesons:
$t_{\rm micro,\sigma} \leq 1/m_{\sigma}$
and  $t_{\rm micro,\pi} = 1/k$ with $k$ being the momentum of the pion. 
The validity of a kinetic description will hinge upon the relaxation time
scales being much longer than these microscopic scales. 

Finally, the Lagrangian for cool linear sigma model reads to lowest order
\begin{equation}
{\cal L}[{\bbox\pi},\sigma]=
\frac{1}{2} \left(\partial_{\mu}{\bbox\pi}\right)^2+\frac{1}{2} 
\left(\partial_{\mu}\sigma\right)^2
-\frac{1}{2} m_{\sigma}^2 \sigma^2 - 
\lambda f_{\pi} \left(\sigma{\bbox\pi}^2+\sigma^3\right)
-\frac{\lambda}{4}\left({\bbox\pi}^2+\sigma^2\right)^2~,
\end{equation}
where  we have omitted the bar over the shifted 
sigma field for simplicity of notation. 

Our goal in this section is to derive the kinetic equations describing  
pion and sigma meson relaxation to lowest order. The unbroken $O(3)$ 
isospin symmetry ensures that all the pions have the same relaxation rate, 
and sigma meson relaxation rate 
is proportional to the number of pion species. 
Hence for notational simplicity the pion index will be suppressed. 
We now study the kinetic equations for the pion and sigma meson 
distribution functions. 

\subsection{Relaxation of cool pions}
 
Without loss of generality, in what follows we discuss the  
relaxation for one isospin component, say $\pi^3$, 
but we suppress the indices for simplicity of notation. 
As before, we consider the case in which at an initial time $t=t_0$,
the density matrix is diagonal in the basis of free quasiparticles, 
but with out of equilibrium initial
distribution functions $n^{\pi}_{\vec k}(t_0)$ and $n^{\sigma}_{\vec k}(t_0)$.
The field operators and the corresponding canonical momenta in the 
Heisenberg picture can be written as 
\begin{eqnarray}
\left\{\begin{array}{l}
\pi({\vec x},t) \\
P_{\pi}({\vec x},t)\\
\sigma({\vec x},t)\\
P_{\sigma}({\vec x},t)
\end{array}\right\}&=&
\int \frac{d^3k}{(2\pi)^{3/2}}\,
\left\{\begin{array}{l}
\pi({\vec k},t) \\
P_{\pi}({\vec k},t)\\
\sigma({\vec k},t)\\
P_{\sigma}({\vec k},t)
\end{array}\right\}\, 
e^{i{\vec k}\cdot {\vec x}}~,
\end{eqnarray}
where
\begin{mathletters}
\begin{eqnarray}
\pi({\vec k},t)&=&\frac{1}{\sqrt{2k}}
 \left[a_{\pi}({\vec k},t)+a^{\dagger}_{\pi}({-\vec k},t)\right]~,\\
P_{\pi}({\vec k},t)&=&-i\sqrt{\frac{k}{2}}
 \left[a_{\pi}({\vec k},t)-a^{\dagger}_{\pi}({-\vec k},t)\right]~,\\
\sigma({\vec k},t)&=&\frac{1}{\sqrt{2\omega_{\vec k}}}
 \left[a_{\sigma}({\vec k},t)+a^{\dagger}_{\sigma}({-\vec k},t)\right]~,\\
P_{\sigma}({\vec k},t)&=&-i\sqrt{\frac{\omega_{\vec k}}{2}}
 \left[a_{\sigma}({\vec k},t)-a^{\dagger}_{\sigma}({-\vec k},t)\right]~,
\end{eqnarray}
\end{mathletters}
with $\omega_{\vec k}=\sqrt{k^2+m_{\sigma}^2}$.
The expectation value of pion number operator can be 
expressed in terms of $\pi({\vec k},t)$ and $P_{\pi}({\vec k},t)$ as
\begin{eqnarray*}
n^{\pi}_{\vec k}(t)&=&
\left\langle a^{\dagger}_{\pi}({\vec k},t)a_{\pi}({\vec k},t)\right\rangle\\
&=& \frac{1}{2k}\biggl\{\left\langle P_{\pi}({\vec k},t)P_{\pi}({-\vec k},t)
\right\rangle 
+ k^2 \left\langle\pi({\vec k},t)\pi({-\vec k},t)\right\rangle\\ 
&& +\,ik\Bigl[\left\langle\pi({\vec k},t)P_{\pi}({-\vec k},t)\right\rangle
- \left\langle P_{\pi}({\vec k},t)\pi({-\vec k},t)\right\rangle\Bigr]
\biggr\}~. 
\end{eqnarray*}
Use the Heisenberg equations of motion, to leading order in $\lambda$, 
we obtain (no tadpole diagrams are included since these are canceled by the choice of $v$)
\begin{equation}
\dot{n}^{\pi}_{\vec k}(t)= -\frac{2\lambda f_{\pi}}{k} 
\int\frac{d^3q}{(2\pi)^{3/2}}
\left.\left(\frac{\partial}{\partial t'}\right)
\Bigl\langle\sigma^{+}({\vec k}-{\vec q},t)\pi^{+}({\vec q},t)
\pi^{-}({-\vec k},t')\Bigr\rangle\right|_{t'=t}~.
\label{pion:ndot}
\end{equation}
The expectation values can be calculated perturbatively in terms of 
nonequilibrium vertices and Green's functions. To ${\cal O}(\lambda)$ the right
hand side of eq.~(\ref{pion:ndot}) vanishes identically.
Fig.~2a shows the Feynman diagrams that contribute to order
$ \lambda^2 $. It is now  straightforward to show that 
$\dot{n}^{\pi}_{\vec k}(t)$ reads
\begin{eqnarray*}
\dot{n}^{\pi}_{\vec k}(t)&=& \frac{\lambda^2 f_{\pi}^2}{k}
\int \frac{d^3 q}{(2 \pi)^3}
\frac{1}{q\,\omega_{\vec k+q}}
\int^t_{t_0} dt''
\Bigl\{{\cal N}_1(t_0)\cos[(k+q+\omega_{\vec k+q})(t-t'')]
+{\cal N}_2(t_0)\cos[(k-q-\omega_{\vec k+q})(t-t'')]\\
&&+\,{\cal N}_3(t_0)\cos[(k-q+\omega_{\vec k+q})(t-t'')]
+{\cal N}_4(t_0)\cos[(k+q-\omega_{\vec k+q})(t-t'')]\Big\}~,
\end{eqnarray*}
where
\begin{mathletters}
\begin{eqnarray}
{\cal N}_1(t) &=&[1+ n^{\pi}_{\vec k}(t)][1+ n^{\pi}_{\vec q}(t)] 
[1+ n^{\sigma}_{\vec k+q}(t)]
-n^{\pi}_{\vec k}(t) \; n^{\pi}_{\vec q}(t)\; n^{\sigma}_{\vec k+q}(t)~, \\
{\cal N}_2(t) &=&[1+ n^{\pi}_{\vec k}(t)]\; n^{\pi}_{{\vec q}}(t) \;
n^{\sigma}_{\vec k+q}(t) -n^{\pi}_{\vec k}(t)\; [1+n^{\pi}_{\vec q}(t)] 
[1+n^{\sigma}_{\vec k+q}(t)]~, \\ {\cal N}_3(t) &=&[1+ n^{\pi}_{\vec k}(t)] 
\; n^{\pi}_{{\vec q}}(t) [1+ n^{\sigma}_{\vec k+q}(t)]
-n^{\pi}_{\vec k}(t)\; [1+n^{\pi}_{\vec q}(t)] \; n^{\sigma}_{\vec k+q}(t)~,\\
{\cal N}_4(t) &=&[1+ n^{\pi}_{\vec k}(t)] 
[1+ n^{\pi}_{{\vec q}}(t)]\; n^{\sigma}_{\vec k+q}(t)
 -n^{\pi}_{\vec k}(t)\;  n^{\pi}_{\vec q}(t)\;[1+n^{\sigma}_{\vec k+q}(t)]~.
\end{eqnarray}\label{pion:ocupanumber}
\end{mathletters}
The different contributions have a very
natural interpretation in terms of `gain minus loss' processes. 
The first term in brackets corresponds to the process 
$0\rightarrow \sigma+\pi+\pi$ minus the process $\sigma+\pi+\pi\rightarrow 0$, 
the second and third terms correspond to the scattering  
$\pi+\sigma\rightarrow\pi$ minus $\pi\rightarrow\pi+\sigma$, 
and the last term corresponds to the decay of the sigma meson 
$\sigma\rightarrow\pi+\pi$ minus the inverse process 
$\pi+\pi\rightarrow\sigma$.

Just as in the scalar case, since the propagators entering in 
the perturbative expansion of the kinetic equation are in terms of the
distribution functions at the initial time, the time integration can 
be done straightforwardly leading to the following equation:
\begin{equation}
\dot{n}^{\pi}_{\vec k}(t)= \lambda^2\int d\omega\,
{\cal R}_{\pi}[\omega,{\vec k};{\cal N}_i(t_0)] \;
\frac{\sin[(\omega-k)(t-t_0)]}{\pi(\omega-k)}~~,\label{pion:ke}
\end{equation}
where $ {\cal R}_{\pi}[\omega,{\vec k};{\cal N}_i(t_0)] $ is given by
\begin{eqnarray}
{\cal R}_{\pi}[\omega,{\vec k};{\cal N}_i(t_0)]&=&\frac{f_{\pi}^2}{k} \,
\int \frac{d^3 q}{(2 \pi)^3} \frac{1}{q\,\omega_{\vec k+q}}
\Bigl[\delta(\omega+q+\omega_{\vec k+q})\;{\cal N}_1(t_0)
+\delta(\omega-q-\omega_{\vec k+q})\;{\cal N}_2(t_0)\nonumber\\
&&+\,\delta(\omega-q+\omega_{\vec k+q})\;{\cal N}_3(t_0)
+\delta(\omega+q-\omega_{\vec k+q})\;{\cal N}_4(t_0)
\Bigr]~.
\end{eqnarray} 
Eq.(\ref{pion:ke}) can be solved by direct integration over $t$ with 
the given initial condition at $ t_0 $, thus leading to
\begin{equation}
n^{\pi}_{\vec k}(t)=n^{\pi}_{\vec k}(t_0)+\lambda^2 \int d\omega\;
{\cal R}_{\pi}[\omega,{\vec k};{\cal N}_i(t_0)] \;
\frac{1-\cos[(\omega-k)(t-t_0)]}{\pi(\omega-k)^2}~~.
\label{ndot:pion} 
\end{equation}  
Potential secular term arises at large times when the resonant
denominator in (\ref{ndot:pion}) vanishes, i.e., 
$\omega\approx k$. A detailed analysis reveals that
${\cal R}_{\pi}[\omega,{\vec k};{\cal N}_i(t_0)]$ is regular at
$\omega=k$, hence using (\ref{FGR}) and (\ref{FGR2}) 
we find that at intermediate asymptotic time $ k(t-t_0)\gg 1 $,
the time evolution of the pion distribution function reads
\begin{eqnarray}
n^{\pi}_{\vec k}(t)&=& n^{\pi}_{\vec k}(t_0)+\lambda^2\,
{\cal R}_{\pi}[ k ,{\vec k};{\cal N}_i(t_0)]\,(t-t_0) +
\text{non-secular terms}~, 
\label{pion:perturb}
\end{eqnarray}
where ${\cal R}_{\pi}[ k ,{\vec k};{\cal N}_i(t_0)]$ 
does not depend on $t_0$ explicitly.

At this point we would be tempted to follow the same steps as in the
scalar case and introduce the dynamical renormalization of the 
pion distribution function. However, much in the same manner as the
renormalization program in a theory with several coupling constants,
in the case under consideration the $\bbox\pi$ field and the $\sigma$ field are
coupled. Therefore one must renormalize {\em all} of the distribution
functions on the same footing. Hence our next task is to obtain the
kinetic equations for the sigma meson distribution functions. 

\subsection{Relaxation of cool sigma mesons}

As before, we consider the case in which at an initial time $t=t_0$,
the density matrix is diagonal in the basis of free quasiparticles, 
but with initial out of equilibrium distribution functions 
$n^{\pi}_{\vec k}(t_0)$ and $n^{\sigma}_{\vec k}(t_0)$.
Again, for notational simplicity we suppress the pion isospin index. 
The expectation value of sigma meson number operator can be 
expressed in terms of $\sigma({\vec k},t)$ and $P_{\sigma}({\vec k},t)$ as
\begin{eqnarray*}
n^{\sigma}_{\vec k}(t)&=&
\left\langle a^{\dagger}_{\sigma}({\vec k},t)a_{\sigma}({\vec k},t)
\right\rangle\\
&=& \frac{1}{2k}\biggl\{\left\langle P_{\sigma}
({\vec k},t)P_{\sigma}({-\vec k},t)
\right\rangle 
+ k^2 \left\langle\sigma({\vec k},t)\sigma({-\vec k},t)\right\rangle \\ 
&& +\,ik\Bigl[\left\langle\sigma({\vec k},t)
P_{\sigma}({-\vec k},t)\right\rangle
- \left\langle P_{\sigma}({\vec k},t)\sigma({-\vec k},t)\right\rangle\Bigr]
\biggr\}~. 
\end{eqnarray*}
Using the Heisenberg equations of motion to leading order in $\lambda$,
and requiring again that the tadpole diagrams are canceled by the proper
choice of $ v $,  we obtain 
\begin{eqnarray}
\dot{n}^{\sigma}_{\vec k}(t)&=&
-\frac{3\lambda f_{\pi}}{\omega_{\vec k}} 
\int\frac{d^3q}{(2\pi)^{3/2}}
\left(\frac{\partial}{\partial t'}\right)
\Bigl[\left\langle\pi^{+}({\vec k}-{\vec q},t)\pi^{+}({\vec q},t)
\sigma^{-}({-\vec k},t')\right\rangle 
+3\left\langle\sigma^{+}({\vec k}-{\vec q},t)\sigma^{+}({\vec q},t)
\sigma^{-}({-\vec k},t')\right\rangle\Bigr]
\Bigl|_{t'=t} ~,\label{sigma:ndot}
\end{eqnarray}
where the factor $ 3 $ accounts for three isospin components of the
pion field. The expectation values can be calculated perturbatively in
terms of nonequilibrium vertices and Green's functions. 
To $ {\cal O}(\lambda) $ the right hand side of 
eq.~(\ref{sigma:ndot}) vanishes identically.
Fig.~3a depicts the one-loop Feynman diagrams that enter in the
kinetic equation for the sigma meson to order $ \lambda^2 $. To the
same order there will be the same type of two loops diagrams as in the
self-interacting scalar 
theory studied in the previous section, but in the low temperature
limit the two-loop diagrams will be suppressed with 
respect to the one-loop diagrams. Furthermore, in the low temperature
limit, the focus of our attention here, only the 
pion loops will be important in the relaxation of the sigma mesons. 
A straightforward calculation leads to the following expression 
\begin{eqnarray}
\dot{n}^{\sigma}_{\vec k}(t)&=&
\frac{3\lambda^2 f_{\pi}^2}{2\omega_{\vec k}}\,
\int \frac{d^3 q}{(2 \pi)^3}
\frac{1}{q\,|{\vec k+q}|}\int^t_{t_0}dt''\nonumber \\
&&\times\,
\biggl\{\Bigl[{\cal N}^{\pi}_1(t_0)
\cos[(\omega_{\vec k}+q+|{\vec k+q}|)(t-t'')]
+ {\cal N}^{\pi}_2(t_0)
\cos[(\omega_{\vec k}+q-|{\vec k+q}|)(t-t'')] \nonumber \\
&&+\,{\cal N}^{\pi}_3(t_0)
\cos[(\omega_{\vec k}-q+|{\vec k+q}|)(t-t'')]
+ {\cal N}^{\pi}_4(t_0)
\cos[(\omega_{\vec k}-q-|{\vec k+q}|)(t-t'')]\Bigr]\nonumber \\
&&+\,\frac{9}{\omega_{\vec q}\,\omega_{\vec k+q}}
\Bigl[{\cal N}^{\sigma}_1(t_0) 
\cos[(\omega_{\vec k}+\omega_{\vec q}
 +\omega_{\vec k+q})(t-t'')]
+ {\cal N}^{\sigma}_2(t_0)
\cos[(\omega_{\vec k}+\omega_{\vec q}
-\omega_{\vec k+q})(t-t'')\nonumber \\
&& +\,{\cal N}^{\sigma}_3(t_0)
\cos[(\omega_{\vec k}-\omega_{\vec q}
+\omega_{\vec k+q})(t-t'')]
+{\cal N}^{\sigma}_4(t_0)
\cos[(\omega_{\vec k}-\omega_{\vec q}
-\omega_{\vec k+q})(t-t'')]
\Bigr]\biggr\}~, \label{sigmadot0}
\end{eqnarray}
where
\begin{mathletters}
\begin{eqnarray}
{\cal N}^{\pi}_1(t) &=&[1+ n^{\sigma}_{\vec k}(t)][1+ n^{\pi}_{\vec q}(t)] 
[1+ n^{\pi}_{\vec k+q}(t)]
-n^{\sigma}_{\vec k}(t)\; n^{\pi}_{\vec q}(t)\; n^{\pi}_{\vec k+q}(t)~, \\
{\cal N}^{\pi}_2(t) &=&[1+ n^{\sigma}_{\vec k}(t)] 
[1+ n^{\pi}_{{\vec q}}(t)] n^{\pi}_{\vec k+q}(t)
 -n^{\sigma}_{\vec k}(t)\;  n^{\pi}_{\vec q}(t)\;[1+n^{\pi}_{\vec k+q}(t)]~,\\
{\cal N}^{\pi}_3(t) &=&[1+ n^{\sigma}_{\vec k}(t)] \;
 n^{\pi}_{{\vec q}}(t)\; [1+ n^{\pi}_{\vec k+q}(t)]
-n^{\sigma}_{\vec k}(t)\; [1+n^{\pi}_{\vec q}(t)]\;  n^{\pi}_{\vec k+q}(t)~,\\
{\cal N}^{\pi}_4(t) &=&[1+ n^{\sigma}_{\vec k}(t)]\; n^{\pi}_{{\vec q}}(t) 
\;n^{\pi}_{\vec k+q}(t)
 -n^{\sigma}_{\vec k}(t)\; [1+n^{\pi}_{\vec q}(t)] [1+n^{\pi}_{\vec k+q}(t)]~,
\end{eqnarray}\label{sigma:ocupanumber1}
\end{mathletters}
and
\begin{mathletters}
\begin{eqnarray}
{\cal N}^{\sigma}_1(t)&=&[1+ n^{\sigma}_{\vec k}(t)]
[1+ n^{\sigma}_{\vec q}(t)] [1+ n^{\sigma}_{\vec k+q}(t)]
-n^{\sigma}_{\vec k}(t)\; n^{\sigma}_{\vec q}(t)\; n^{\sigma}_{\vec
k+q}(t)~, \\ {\cal N}^{\sigma}_2(t) &=&[1+ n^{\sigma}_{\vec k}(t)] 
[1+ n^{\sigma}_{{\vec q}}(t)]\; n^{\sigma}_{\vec k+q}(t)
 -n^{\sigma}_{\vec k}(t)\; n^{\sigma}_{\vec q}(t)\;[1+n^{\sigma}_{\vec
k+q}(t)]~,\\ {\cal N}^{\sigma}_3(t) &=&[1+ n^{\sigma}_{\vec k}(t)] \;
 n^{\sigma}_{{\vec q}}(t)\; [1+ n^{\sigma}_{\vec
k+q}(t)]-n^{\sigma}_{\vec k}(t)\; [1+n^{\sigma}_{\vec q}(t)]\;
n^{\sigma}_{\vec k+q}(t)~,\\ {\cal N}^{\sigma}_4(t) &=&[1+
n^{\sigma}_{\vec k}(t)]\; n^{\sigma}_{{\vec q}}(t)\;  n^{\sigma}_{\vec k+q}(t)
 -n^{\sigma}_{\vec k}(t)\; [1+n^{\sigma}_{\vec q}(t)]
[1+n^{\sigma}_{\vec k+q}(t)]~. 
\end{eqnarray}\label{sigma:ocupanumber2}
\end{mathletters}
Although the above expression is somewhat unwieldy,
the different contributions have a very
natural interpretation in terms of `gain minus loss' processes. 
In the first brackets (i.e., the pion contribution) 
the first term corresponds to the process 
$0\rightarrow \sigma+\pi+\pi$ minus the process $\sigma+\pi+\pi\rightarrow 0$, 
the second and third terms correspond to the scattering  
$\pi\rightarrow\pi+\sigma$ minus $\pi+\sigma\rightarrow\pi$, 
and the last term corresponds to the decay of the sigma meson 
$\sigma\rightarrow\pi+\pi$ minus the inverse 
process $\pi+\pi\rightarrow\sigma$.
Similarly, in the second brackets (i.e., the sigma meson contribution) 
the first term corresponds to the process 
$0\rightarrow\sigma+\sigma+\sigma$ minus the process 
$\sigma+\sigma+\sigma\rightarrow 0$, 
the second and third terms correspond to annihilation of two sigma mesons and
creation of one sigma meson minus the inverse process, 
and the last term corresponds to annihilation of a sigma meson and
creation of two sigma mesons minus the inverse process.

Since the propagators entering in 
the perturbative expansion of the kinetic equation are in terms of the
distribution functions at the initial time, the time integration can 
be done straightforwardly leading to the following equation:
\begin{equation}
\dot{n}^{\sigma}_{\vec k}(t)= \lambda^2\int d\omega\,
{\cal R}_{\sigma}[\omega,{\vec k};{\cal N}_i(t_0)] \;
\frac{\sin[(\omega-\omega_{\vec k})(t-t_0)]}
{\pi(\omega-\omega_{\vec k})}~~,\label{sigma:ke}
\end{equation}
where
\begin{eqnarray}
{\cal R}_{\sigma}[\omega,{\vec k};{\cal N}_i(t_0)]&=&
\frac{3f_{\pi}^2}{2\omega_{\vec k}}\,
\int \frac{d^3 q}{(2 \pi)^3} 
\Biggl\{\frac{1}{q\,|{\vec k+q}|}
\biggl[\delta(\omega+q+|{\vec k+q}|){\cal N}^{\pi}_1(t_0) 
+ \delta(\omega+q-|{\vec k+q}|){\cal N}^{\pi}_2(t_0) \nonumber\\
&& +\,\delta(\omega-q+|{\vec k+q}|){\cal N}^{\pi}_3(t_0)
+\delta(\omega-q-|{\vec k+q}|){\cal N}^{\pi}_4(t_0)
\biggr]\nonumber\\
&&+\,\frac{9}{\omega_{\vec q}\,\omega_{\vec k+q}}
\bigg[\delta(\omega+\omega_{\vec q}+
\omega_{\vec k+q}){\cal N}^{\sigma}_1(t_0) 
+ \delta(\omega+\omega_{\vec q}-\omega_{\vec k+q})
{\cal N}^{\sigma}_2(t_0) \nonumber\\
&& +\,\delta(\omega-\omega_{\vec q}+\omega_{\vec k+q})
{\cal N}^{\sigma}_3(t_0)
+\delta(\omega-\omega_{\vec q}-\omega_{\vec k+q})
{\cal N}^{\sigma}_4(t_0)
\bigg]\Biggr\}~. \label{Rsigma}
\end{eqnarray}

Just as before ${\cal R}_{\sigma}[\omega,{\vec k};{\cal N}_i(t_0)]$ 
is fixed at initial time $t_0$, 
eq.~(\ref{sigma:ke}) can be integrated over $t$ 
with the given initial condition at $t_0$, thus leading to
\begin{equation}
n^{\sigma}_{\vec k}(t)=n^{\sigma}_{\vec k}(t_0)+\lambda^2 \int d\omega\, 
{\cal R}_{\sigma}[\omega,{\vec k};{\cal
N}_i(t_0)]\frac{1-\cos[(\omega-\omega_{\vec k})(t-t_0)]} 
{\pi(\omega-\omega_{\vec k})^2}~.\label{sigma:solofke}
\end{equation} 
At intermediate asymptotic times $m_{\sigma}(t-t_0)\gg 1$,
potential secular term arises when 
$\omega\sim\omega_{\vec k}$ in
eq.~(\ref{sigma:solofke}). 
We notice that, although ${\cal R}_{\sigma}[\omega,{\vec k};{\cal N}_i(t_0)]$ has 
{\em threshold (infrared) singularities} at $\omega=\pm k$, 
it is regular on the sigma meson mass-shell. 
This observation will allow us to explore a crossover behavior 
for very large momentum later. 

Since the spectral density is regular near the resonance region  
$\omega=\pm \omega_{\vec k}$, 
the behavior at intermediate asymptotic times is given by
\begin{equation}
n^{\sigma}_{\vec k}(t)=n^{\sigma}_{\vec k}(t_0)+\lambda^2 
{\cal R}_{\sigma}[\omega_{\vec k},{\vec k};{\cal N}_i(t_0)](t-t_0)+\mbox{non-secular terms}.\label{sigmasecular}
\end{equation} 
We note that the perturbative expansions for the pion and sigma meson 
distribution functions contain secular terms that grow linearly in time,  
{\em unless} the system is initially prepared in thermal equilibrium. 
We must now renormalize both equations
(\ref{pion:perturb}) and (\ref{sigmasecular}) 
{\em simultaneously}, since it is a field theory with two coupled fields.

Introduce the renormalized initial distribution functions $
n^{\pi}_{\vec p}(\tau) $ and $ n^{\sigma}_{\vec p}(\tau) $, which are
related to the bare initial distribution functions $ n^{\pi}_{\vec
p}(t_0) $ and $ n^{\sigma}_{\vec p}(t_0) $ via respective
renormalization constants ${\cal Z}^{\pi}_{\vec p}(\tau,t_0)$ and
${\cal Z}^{\sigma}_{\vec p}(\tau,t_0)$ by  
\begin{mathletters}
\begin{eqnarray}
n^{\pi}_{\vec p}(t_0)={\cal Z}^{\pi}_{\vec p}(\tau,t_0)\,
n^{\pi}_{\vec p}(\tau)~,\quad
{\cal Z}^{\pi}_{\vec p}(\tau,t_0)= 1 + 
 \lambda^2 \; z^{\pi(1)}_{\vec p}(\tau,t_0)+\cdots~,\\
n^{\sigma}_{\vec p}(t_0)= {\cal Z}^{\sigma}_{\vec p}(\tau,t_0)\,
 n^{\sigma}_{\vec p}(\tau)~,\quad
{\cal Z}^{\sigma}_{\vec p}(\tau,t_0)= 1 + \lambda^2\,
z^{\sigma(1)}_{{\vec p}}(\tau,t_0)+\cdots~,
\end{eqnarray}\label{pion:renormal}
\end{mathletters}
where $ \tau $ is an arbitrary renormalization scale at which the
secular terms will be canceled.  
The renormalization constants $ z^{\pi(1)}_{\vec p}(\tau,t_0) $
and  $ z^{\sigma(1)}_{\vec p}(\tau,t_0) $
are chosen so as to cancel the secular term at the arbitrary scale
$ \tau $ consistently in perturbation theory. Substitute
eq.~(\ref{pion:renormal}) into eq.~(\ref{pion:perturb}), consistently
up to $ {\cal O}(\lambda^2) $ we obtain
\begin{eqnarray*}
n^{\pi}_{\vec k}(t)&=&n^{\pi}_{\vec k}(\tau)+\lambda^2 
\left\{ z^{\pi(1)}_{\vec k}(\tau,t_0)\;n^{\pi}_{\vec k}(\tau)+(t-t_0)\;
{\cal R}_{\pi}[k,{\vec k};{\cal N}_i(\tau)]\right\}+{\cal O}(\lambda^4)~,\\
n^{\sigma}_{\vec k}(t)&=&n^{\sigma}_{\vec k}(\tau)+\lambda^2 
\left\{z^{\sigma(1)}_{\vec k}(\tau,t_0)\; n^{\sigma}_{\vec k}(\tau)+
(t-t_0)\;{\cal R}_{\sigma} [\omega_{\vec k},{\vec k};{\cal
N}_i(\tau)] \right\}+{\cal O}(\lambda^4)~.
\end{eqnarray*}
To this order, the choices 
\begin{eqnarray*}
z^{\pi(1)}_{\vec k}(\tau,t_0)&=&-(\tau-t_0)
{\cal R}_{\pi}[k,{\vec k};{\cal N}_i(\tau)] / n^{\pi}_{\vec k}(\tau) 
\\
z^{\sigma(1)}_{\vec k}(\tau,t_0)&=&-(\tau-t_0)
{\cal R}_{\sigma}[\omega_{\vec k},{\vec k};{\cal N}_i(\tau)] / n^{\sigma}_{\vec k}(\tau) 
\end{eqnarray*} 
lead to
\begin{eqnarray*}
&& n^{\pi}_{\vec k}(t)=n^{\pi}_{\vec k}(\tau)+\lambda^2\;(t-\tau)\;
{\cal R}_{\pi}[k,{\vec k};{\cal N}_i(\tau)]+{\cal O}(\lambda^4)~,\\
&& n^{\sigma}_{\vec k}(t)=n^{\sigma}_{\vec k}(\tau)+\lambda^2\;(t-\tau)\;{\cal
R}_{\sigma}[\omega_{\vec k},{\vec k};{\cal N}_i(\tau)]+{\cal O}(\lambda^4)~.
\end{eqnarray*} 
The independence of $n^{\pi}_{\vec k}(t)$ and $n^{\sigma}_{\vec k}(t)$ on the arbitrary
renormalization scale $\tau$ leads to the simultaneous set of
dynamical renormalization group equations to lowest order: 
\begin{eqnarray*}
\frac{d}{d\tau}n^{\pi}_{\vec k}(\tau)&=&\lambda^2\,
{\cal R}_{\pi}[k,{\vec k};{\cal N}_i(\tau)]~, 
\\
\frac{d}{d\tau}n^{\sigma}_{\vec k}(\tau)&=&\lambda^2\,
{\cal R}_{\sigma}[\omega_{\vec k},{\vec k};{\cal N}_i(\tau)]~.
\end{eqnarray*}
These equations have an obvious resemblance to a set of
renormalization group equations for ``couplings'' $ n^{\pi}_{\vec k}$ and $n^{\sigma}_{\vec k}$ 
where the right hand sides are the corresponding beta functions.  

As before, choosing the arbitrary scale $\tau$ to coincide with the
time $t$ and keeping only the terms whose delta functions have support
on the mass shells we obtain the kinetic equations describing pion and
sigma relaxation: 
\begin{eqnarray}
&&\dot{n}^{\pi}_{\vec k}(t) =  \frac{\pi\lambda^2 f_{\pi}^2}{k} \,
\int \frac{d^3 q}{(2 \pi)^3}\frac{\delta(k+q-\omega_{\vec
k+q})}{q\,\omega_{\vec k+q}} \left\{ [1+ n^{\pi}_{\vec k}(t)] 
[1+ n^{\pi}_{{\vec q}}(t)] \; n^{\sigma}_{\vec k+q}(t)
 -n^{\pi}_{\vec k}(t) \;  n^{\pi}_{\vec q}(t)\; [1+n^{\sigma}_{\vec
k+q}(t)]\right\}~~.\label{pionboltz}\\ 
&&\dot{n}^{\sigma}_{\vec k}(t) =  \frac{3\pi\lambda^2 
f_{\pi}^2}{2\omega_{\vec k}} \, 
\int \frac{d^3 q}{(2\pi)^3}\frac{\delta(\omega_{\vec
k}-q-|{\vec k+q}|)}{q\,|{\vec k+q}|} 
\left\{[1+ n^{\sigma}_{\vec k}(t)] \; n^{\pi}_{{\vec q}}(t) \; 
n^{\pi}_{\vec k+q}(t) -n^{\sigma}_{\vec k}(t)\;  [1+n^{\pi}_{\vec
q}(t)][1+n^{\pi}_{\vec k+ \vec q}(t)]\right\}~~, \label{sigmaboltz} 
\end{eqnarray}
The processes that contribute to (\ref{pionboltz}) are depicted in
Fig.~2b and those that contribute to (\ref{sigmaboltz}) are depicted
in Fig.~3b.  

\subsection{Relaxation time approximation}

Thermal equilibrium is a {\em fixed point} of the 
dynamical renormalization group equations
(\ref{pionboltz}) and (\ref{sigmaboltz}), i.e., 
a stationary solution of the kinetic equations. 

A linearized kinetic equation can be obtained in 
relaxation time approximation, in which only the mode with
momentum $ {\vec k} $ is slightly out of equilibrium whereas all the other
modes are in equilibrium: 
\begin{eqnarray*}
&&\delta \dot{n}^{\pi}_{\vec k}(t) = -\gamma_{\pi}({\vec k})\,\delta
n^{\pi}_{\vec k}(t)~,
\\
&&\delta \dot{n}^{\sigma}_{\vec k}(t) = -\gamma_{\sigma}({\vec k})\,
\delta n^{\sigma}_{\vec k}(t)~,
\end{eqnarray*}
where $ \gamma_{\pi}({\vec k}) $ and $ \gamma_{\sigma}({\vec k}) $
are respectively the cool pion and sigma meson relaxation rates which 
are identified with twice the damping rates of
the corresponding field amplitudes. 
Linearizing (\ref{pionboltz}) we obtain
\begin{eqnarray}
\gamma_{\pi}({\vec k})&=&\frac{\pi\lambda^2 f_{\pi}^2}{k} \,
\int \frac{d^3 q}{(2\pi)^3}\frac{\left[n_B(q)-n_B(\omega_{\vec k+q})\right]}
{q\,\omega_{\vec k+q}}\;
\delta(k+q-\omega_{\vec k+q})\nonumber\\
&=&\frac{\lambda^2 f_{\pi}^2 T}{4\pi k^2} 
\ln\left[\frac{1-e^{-\beta(m_{\sigma}^2/4k+k)}}
{1-e^{-\beta m_{\sigma}^2/4k}}\right]~.\label{relrate:pion} 
\end{eqnarray}
This is a remarkable expression because it reveals that the physical
processes that contribute to cool pion relaxation are the {\em decay}
of sigma meson $\sigma\rightarrow\pi+\pi$ and its inverse process
$\pi+\pi\rightarrow\sigma$. The form of (\ref{relrate:pion}) is
reminiscent of the Landau damping contribution to 
the pion self-energy and in fact a simple calculation reveals this to
be correct. The sigma particles present in 
the medium can decay into pions and this increases the number of
pions, but at the same time pions recombine into 
sigma particles, and since there are more pions in the medium because
they are lighter the loss part of the 
process prevails producing a non-zero relaxation rate. This is an
induced phenomenon in the medium in the very definitive sense that the
decay of the heavier sigma meson induces the decay of the pion distribution
function, it is a non-collisional process. 

Such relaxation of cool pions is analogous to the induced relaxation
of fermions in a fermion-scalar plasma induced by the decay of a
massive scalar into fermion pairs~\cite{fermiondamping}.  

For soft, cool pion mode $(k\ll T\ll f_{\pi})$, the pion relaxation rate reads
\begin{equation}
\gamma_{\pi}(k\ll T)\approx\frac{\lambda^2 f_{\pi}^2}{4\pi k}\exp\left(
-\frac{m_{\sigma}^2}{4kT}\right)~.
\end{equation}
The exponential suppression in the soft, cool pion relaxation rate is a 
consequence of the heavy sigma mass.
Our results of the pion relaxation rate are in agreement with the
pion damping rate found in Ref.~\cite{Pisarski:1996mt}. These results
(accounting for the factor 2 necessary to 
relate the relaxation rate to the damping rate) also agree with those reported 
recently in Ref.~\cite{patkos} wherein a related and clear analysis of pion
and sigma meson {\em damping rates} was presented.  

For the relaxation rate of the sigma mesons, we find
\begin{eqnarray}
\gamma_{\sigma}({\vec k})&=&
\frac{3\pi\lambda^2 f_{\pi}^2}{2\omega_{\vec k}}\,
\int \frac{d^3 q}{(2\pi)^3}
\frac{\left[1+n_B(q)+n_B(|{\vec k+q}|)\right]}{q\,|{\vec k+q}|}\;
\delta(\omega_{\vec k}-q-|{\vec k+q}|)\nonumber\\
&=&\frac{3\lambda^2 f_{\pi}^2}{8\pi\omega_{\vec k}}
\left[1 + \frac{2T}{k} 
\ln\left(\frac{1-e^{-\beta(\omega_{\vec k}+k)/2}}
{1-e^{-\beta(\omega_{\vec k}-k)/2}}\right)\right]~.
\label{relrate:sigma}
\end{eqnarray}
The first temperature-independent term in $\gamma_{\sigma}({\vec k})$
is the usual zero-temperature sigma meson decay rate~\cite{Serot:1986ey}, whereas the finite temperature factors
result from the {\em same processes} that determine the pion relaxation rate, i.e., 
$\sigma \leftrightarrow \pi+\pi$.

For soft sigma meson ($k\ll T\ll f_{\pi}$), we obtain
$$
\gamma_{\sigma}({k\approx 0})\approx 
\frac{3\lambda^2 f_{\pi}^2}{8\pi m_{\sigma}}
\coth\left(\frac{m_{\sigma}}{4 T}\right)~.
$$
It agrees with the decay rate for a sigma meson at rest 
found in Refs.~\cite{patkos,sigmadecay,csernai}. 

On the other hand consider the theoretical 
high temperature and large momentum limit  
$ k\gg m_{\sigma}\gtrsim T$ such that $ \omega_{\vec k}-k \ll T $. 
In this limit the sigma meson relaxation rate (\ref{relrate:sigma}) 
becomes logarithmic (infrared) divergent. 
The reason for this divergence is that as was mentioned below 
eq.~(\ref{sigma:solofke}), 
$ {\cal R}_{\sigma}(\omega,{\vec k}; {\cal N}_i) $ has an infrared
threshold singularity at $ \omega=k $ arising from the 
contribution proportional to $ {\cal N}^{\pi}_4 $ in eq.~(\ref{Rsigma}). 
In the presence of this threshold singularity, 
we can no longer apply eqs.~(\ref{FGR}) and (\ref{FGR2}) 
and instead we must study the long time limit in (\ref{sigmadot0}) more carefully. 
Understanding the influence of threshold behavior of the sigma meson 
on its relaxation could be important in view of the recent proposal
by Hatsuda and collaborators~\cite{hatsuda} that near the chiral phase
transition the mass of the sigma meson drops and threshold
effects become enhanced with distinct phenomenological consequences. 
We expect to report on a more detailed study of threshold effects 
near the critical temperature in the near future.   
 
\subsection{Threshold singularities and crossover}

As mentioned above, in the discussion following
eq.(\ref{sigma:solofke}), $ {\cal R}_{\sigma}[\omega,{\vec k};{\cal
N}_i(t_0)] $ in (\ref{sigma:solofke}) has threshold singularities at
$\omega=\pm k$ arising from the emission and 
absorption of collinear massless pions. For $ k \sim m_{\sigma} $, the
point at which the resonant denominator in (\ref{sigma:solofke})
vanishes (i.e., $\omega = \omega_{\vec k}$)
is far away from threshold and  
$ {\cal R}_{\sigma}[\omega,{\vec k};{\cal N}_i(t_0)] $ is regular at
this point (on-shell), hence Fermi's Golden Rule (\ref{FGR}) is applicable. 
However in the large momentum limit, when 
$\omega_{\vec k} \rightarrow k $ the point at which the resonant
denominator vanishes becomes closer to threshold and such singular
point begins to influence the long-time behavior.  

That this is the case can be seen in the expression for the relaxation
rate (\ref{relrate:sigma}) which displays a logarithmic (infrared)
divergence as $ \omega_{\vec k} \rightarrow k $.  
A close inspection at the terms that contribute to $ {\cal
R}_{\sigma}[\omega,{\vec k};{\cal N}_i(t_0)] $ in 
eq.(\ref{Rsigma}) reveals that the threshold divergence arising  as
$ \omega \approx \omega_{\vec k} \rightarrow k $ originates 
in the term proportional to $ {\cal N}^{\pi}_4(t_0) $ which accounts for
the emission and absorption of collinear massless pions.   

In order to understand how this threshold
divergence modifies the long-time behavior, let us focus on 
the mode of sigma mesons with momentum 
$k\gg m_{\sigma}\gtrsim T$. 
This situation is not relevant to the phenomenology
of the cool pion-sigma meson system 
for which relevant temperatures are $ T\ll m_{\sigma}$. 
However studying this limiting case 
will yield to important insight into how threshold divergences
invalidate the simple Fermi's Golden Rule analysis leading 
to on-shell delta functions in the intermediate asymptotic regime. 
This issue will become more pressing in the case of gauge theories studied below.  

To present this case in the simplest and clearest manner, 
we will study the relaxation time approximation, by 
assuming that only one mode of sigma mesons, with momentum ${\vec k}$,
is slightly displaced from equilibrium such that
$n^{\sigma}_{\vec k} = 
n_B(\omega_{\vec k}) + \delta n^{\sigma}_{\vec k}(t_0)$, whereas  
all other pion and sigma meson modes are in equilibrium,  
i.e., $n^{\pi}_{\vec q}(t_0) = n_B(q)$ for all ${\vec q}$ and 
$n^{\sigma}_{\vec q}(t_0) = n_B(\omega_{\vec q})$ for all ${\vec q}\ne{\vec k}$. 
In this approximation and keeping 
the only term that contributes to 
${\cal R}_{\sigma}[\omega,{\vec k};{\cal N}_i(t_0)]$ for  
$ \omega \approx \omega_{\vec k} $, 
[i.e., the one proportional to $ {\cal N}^{\pi}_4(t_0)$], we find that
eq.~(\ref{sigma:solofke}) simplifies to 
\begin{equation}
\delta n^{\sigma}_{\vec k}(t)=\delta n^{\sigma}_{\vec k}(t_0)
\left[ 1- \int d\omega\, 
\gamma_{\sigma}(\omega,{\vec k}) \;
\frac{1-\cos[(\omega-\omega_{\vec k})(t-t_0)]}  
{\pi(\omega-\omega_{\vec k})^2}\right]~,\label{linearsigma}
\end{equation} 
with
\begin{eqnarray}
\gamma_{\sigma}(\omega,{\vec k})&=&
\frac{3\pi\lambda^2 f_{\pi}^2}{2\omega_{\vec k}}\,
\int \frac{d^3 q}{(2\pi)^3}
\frac{1+n_B(q)+n_B(|{\vec k+q}|)}{q\,|{\vec k+q}|}
\delta(\omega-q-|{\vec k+q}|)\nonumber\\
&=&\frac{3\lambda^2 f_{\pi}^2}{8\pi\omega_{\vec k}}
\left[1 + \frac{2T}{k} 
\ln\left(\frac{1-e^{-\beta(\omega+k)/2}}
{1-e^{-\beta(\omega-k)/2}}\right)\right]~.
\label{gammasigma}
\end{eqnarray}

At intermediate asymptotic times $m_{\sigma}(t-t_0)\gg 1$, 
the region $ \omega \approx \omega_{\vec k}\approx k $ 
dominates the integral and in the limit $ k\gg m_{\sigma}\gtrsim T$ 
we can further approximate
\begin{equation}
\gamma_{\sigma}(\omega,{\vec k}) \buildrel{ \omega\to k}\over=
\frac{3\lambda^2 f_{\pi}^2 T}{4\pi  k^2}  \ln\left[\frac{\bar{T}}
{\omega-k}\right] + {\cal O}(\omega-k)~, 
\label{gammathreshold} 
\end{equation}
where $\bar{T} = 2\,T\,[1-\exp(-k/T))]\approx 2T$.
The integral over $\omega$ in eq.(\ref{linearsigma}) can be performed
when $ \gamma_{\sigma}(\omega,{\vec k}) $ is given by the first term
in eq.(\ref{gammathreshold}) and we obtain  
$$
\int d\omega\, \gamma_{\sigma}(\omega,{\vec
k})\frac{1-\cos[(\omega-\omega_{\vec k})(t-t_0)]} 
{\pi(\omega-\omega_{\vec k})^2}  
\approx
\frac{3\lambda^2 f_{\pi}^2 T}{4\pi  k^2}{\cal F}(t-t_0,{\vec k})
$$
for $m_{\sigma}(t-t_0)\gg 1$, where
\begin{equation}
 {\cal F}(t-t_0,{\vec k}) = (t-t_0)\left\{
\ln\left[\frac{\bar{T}}{\omega_{\vec k}-k}\right] +  
\text{ci} \left[(\omega_{\vec k}-k)(t-t_0)\right]-
\frac{\sin\left[(\omega_{\vec k}-k)(t-t_0)\right]}{(\omega_{\vec
k}-k)(t-t_0)}\right\}~,  
\end{equation}
with ${\rm ci}(x)$ being the cosine integral function
$$
{\rm ci}(x) \equiv - \int_x^{+\infty}dt\;{\cos t \over t}~. 
$$
For fixed ${\vec k}$, ${\cal F}(t-t_0,{\vec k})$ has the following limiting behaviors
\begin{eqnarray}
{\cal F}(t-t_0,{\vec k}) 
&=& (t-t_0)\left\{ \ln\left[(t-t_0)\bar{T} 
e^{\gamma-1}\right]+{\cal O}\left((\omega_{\vec
k}-k)^2(t-t_0)^2\right) \right\}\quad\quad
\text{for}~(\omega_{\vec k}-k)(t-t_0) \ll 1~, 
\label{muchsmaller} \\
{\cal F}(t-t_0,{\vec k}) 
&=&(t-t_0)\left\{ \ln\left[\frac{\bar{T}}{\omega_{\vec k}-k}\right] + 
{\cal O}\left(\frac{1}{(\omega_{\vec k}-k)^2(t-t_0)^2}\right)\right \}
\quad \quad \text{for}~(\omega_{\vec k}-k)(t-t_0) \gg 1~, 
\label{muchgreater}  
\end{eqnarray}
where $ \gamma = 0.5772157 \ldots $ is the Euler-Mascheroni constant.
Thus, we see that there is a {\em crossover}
time scale 
$ t_{\rm c} \approx (\omega_{\vec k}-k)^{-1}$
at which the time dependence of the function $ {\cal F}(t-t_0,{\vec k}) $ 
changes from $\sim t\ln t $ for $ t-t_0\lesssim t_{\rm c}$ 
to linear in $t$ for $ t-t_0\gtrsim t_{\rm c}$. 
In the large momentum limit, as the sigma meson mass-shell approaches
threshold, this crossover time scale becomes longer such that an ``anomalous''
(non-linear) secular term of the form $t\ln t$ 
dominates during most of the time whereas 
the usual secular term linear in $t$ ensues at very large times.

We can now proceed with the dynamical renormalization
group to resum the secular terms. Introducing the renormalization constant
 ${\cal Z}^{\sigma}_{\vec k}$ by
\begin{eqnarray}
\delta n^{\sigma}_{\vec k}(t_0)= {\cal Z}^{\sigma}_{\vec k}(\tau,t_0)\,
 \delta n^{\sigma}_{\vec k}(\tau)~,\quad
{\cal Z}^{\sigma}_{\vec k}(\tau,t_0)= 1 + \lambda^2 
z^{\sigma(1)}_{{\vec k}}(\tau,t_0)+\cdots~, \label{linsigren}
\end{eqnarray}
and choosing
\begin{equation}
z^{\sigma(1)}_{{\vec k}}(\tau,t_0)= \frac{3 f_{\pi}^2 T}{4\pi  k^2} \;
{\cal F}(\tau-t_0,{\vec k}) 
\end{equation}
to cancel the secular divergences at the time scale $\tau$, we find
that dynamical renormalization group equation 
\begin{equation}
\frac{d \delta n^{\sigma}_{\vec k}(\tau)}{d \tau}+
\frac{3 \lambda^2 f_{\pi}^2 T}{4\pi  k^2}   \;
\frac{d {\cal F}(\tau-t_0,{\vec k})}{d \tau}= 0 \label{RGlogi}
\end{equation}
\noindent leads to the following solution in relaxation time approximation
\begin{equation}
\delta n^{\sigma}_{\vec k}(t)= \delta n^{\sigma}_{\vec k}(t_0) 
\exp\left[- \frac{3 \lambda^2 f_{\pi}^2 T}{4\pi
k^2} {\cal F}(t-t_0,{\vec k})\right]~. 
\end{equation}  
In the large momentum limit, using (\ref{muchsmaller}) and (\ref{muchgreater})
we find that the crossover in the form of the secular terms results in a 
crossover in the sigma meson relaxation:
an ``anomalous'' (non-exponential) relaxation will dominate 
the relaxation during most of the time and usual exponential relaxation ensues 
at very large times. 

This simple exercise has revealed several important features
highlighted by a consistent resummation via the dynamical renormalization group:
\begin{itemize}
\item{Threshold infrared divergences result in a breakdown of Fermi's
Golden Rule. The secular terms of the 
perturbative expansion are no longer linear in time but include
logarithmic contributions arising from these infrared divergences. }

\item{The concept of the damping rate is directly tied to exponential
relaxation. The infrared divergences of the damping rate reflect the
breakdown of Fermi's Golden Rule and signal a 
very different relaxation from a simple exponential. }

\item{Whereas the usual calculation of damping rates will lead to a
divergent result arising from the infrared threshold divergences,
the dynamical renormalization group approach recognizes that these 
threshold divergences result in secular terms that are non-linear 
in time as discussed above. 
While in relaxation time approximation linear secular terms lead to
exponential relaxation and therefore to an unambiguous definition of
the damping rate, non-linear secular terms lead to novel non-exponential
relaxational phenomena for which the concept of a damping rate may not
be appropriate.} 

\end{itemize}

This discussion of threshold singularities and anomalous 
relaxation has paved the way to studying the case of gauge theories, wherein
the emission and absorption of (transverse) photons that are 
only dynamically screened lead to a similar anomalous relaxation~\cite{boyrgir}. 

\section{Hot Scalar QED}

In this section we  study the relaxation of the distribution function of 
charged scalars in hot SQED as a prelude to studying 
the more technically involved cases of hot QED and QCD~\cite{qedqcd}. 
Hot SQED shares many of the important features of hot 
QED and QCD in leading order in the hard thermal loop (HTL) resummation~\cite{rebhan,thoma,boyhtl,thoma2}. 
Furthermore, the infrared physics in hot QED captured in the eikonal 
(Bloch-Nordsieck) approximation~\cite{iancu} has been reproduced recently 
via the dynamical renormalization group in hot SQED~\cite{boyrgir}, 
thus lending more support to the similarities of both theories at least 
in leading HTL order. However, unlike hot QED and QCD there are 
two simplifications~\cite{rebhan,thoma} in this theory 
that allows a more clear presentation of the relevant results: 
(i) there are {\em no HTL corrections to the vertex} and (ii) 
the HTL resummed scalar self-energy is momentum independent~\cite{rebhan,thoma}. 
These features of hot SQED enable us to probe the relaxation of charged scalars 
with arbitrary momentum within a simplified setting that 
nevertheless captures important features that are relevant to QED and QCD. 
This study is different from those in 
Ref.~\cite{boyrgir} in that we here include the contribution from the
longitudinal, Debye screened photons and discuss in detail the crossover between the 
relaxational time scales associated with
the transverse and longitudinal photons for arbitrary momentum of the charged scalar. 
Furthermore, in order to provide an unambiguous definition of the distribution
function, our study is done directly in a gauge invariant formulation. This formulation has several advantages, in that gauge
invariance is built in from the outset and the distribution functions are defined for gauge invariant objects. 

In the Abelian theory under consideration, it is rather
straightforward to implement a gauge invariant formulation 
by projecting the Hilbert space on states
annihilated by the two primary first class constraints: Gauss' law and vanishing canonical momentum conjugate to the temporal component of the gauge field. 
Gauge invariant operators are those that commute with both constraints and are obtained
systematically, finally the Hamiltonian and Lagrangian can be written in terms of these gauge invariant operators~\cite{gaugeinv},
details are presented in  Appendix~A. 
The resulting Lagrangian is exactly the same as that in Coulomb gauge~\cite{gaugeinv}
and is given by (see Appendix~A)
\begin{eqnarray}  
{\cal L}=&&\partial_\mu\Phi^\dagger\,\partial^\mu\Phi -m^2 \Phi^{\dagger}\Phi  
+\frac{1}{2}\partial_\mu {\vec A}_T\cdot\partial^\mu{\vec A}_T 
-e{\vec A}_T\cdot{\vec j}_T 
-e^2{\vec A}_T\cdot{\vec A}_T\, \Phi^\dagger\Phi \nonumber \\ 
&&+\frac{1}{2}\left({{\vec \nabla}} A_0 \right)^2+ {e^2}A^2_0 \, 
\Phi^{\dagger}\Phi+ e A_0 j_0~, \nonumber \\ 
{\vec j}_T=&&i[\Phi^\dagger{{\vec \nabla}}\Phi-({\vec \nabla}
\Phi^\dagger)\Phi]~, \quad \
j_0 = -i(\Phi\dot{\Phi}^{\dagger}-{\Phi}^{\dagger}\dot{\Phi})~,\nonumber 
\end{eqnarray} 
where $e$ is the gauge coupling, 
${\vec A}_T$ is the transverse component of the gauge field satisfying
${\vec \nabla}\cdot {\vec A}_T({\vec x},t)=0$, $\Phi$ and 
$\Phi^{\dagger}$ are charged but {\em gauge invariant} fields, 
and we have traded the instantaneous Coulomb interaction for a
{\em gauge invariant} auxiliary field $A_0$ 
which should not be confused with the time component of the gauge field.
Since we are only interested in obtaining the relaxation behavior arising
from finite temperature effects we do not introduce the renormalization
counterterms to facilitate the study, although these can be systematically 
included in our formulation~\cite{boyrgir}. 
Furthermore we will consider a neutral system with vanishing chemical 
potential.

Medium effects are included via the {\em equilibrium} 
hard thermal loop resummation, hence we will restrict 
our study to the relaxation time approximation 
in which only one mode of the scalar field, with momentum ${\vec k}$ 
is perturbed off equilibrium while all other scalar modes and 
the gauge fields will be taken to be in equilibrium. 
Considering the full nonequilibrium quantum kinetic equation 
will require an extrapolation of the hard thermal loop program 
to situations far from equilibrium, clearly a task beyond the scope of this article. 
Hence the propagators to be used in the calculation for the modes and fields in equilibrium will be hard thermal loop resummed.   

Since for hot SQED the leading one-loop contributions to 
scalar self-energy is momentum independent and $\sim{\cal O}(e^2T^2)$~\cite{rebhan}, 
the leading order HTL resummed inverse scalar propagator reads 
(here and henceforth, we neglect the zero-temperature scalar mass $m$)
\begin{equation}
\Delta_{s}^{-1}(\omega,{\vec k})=\omega^2-k^2-m^2_s~,
\end{equation}
where $m_s=eT/2$ is the thermal mass of the charged scalar. 
The dispersion relation of scalar quasiparticles 
to leading order in HTL is given by $\omega_{\vec k}=\sqrt{k^2+m^2_s}$. Just as in the scalar
case studied in section~\ref{section:scalar}, 
the mass $m_s$ is included in the Hamiltonian and a counterterm is
considered as part of the interaction to cancel the tadpole contributions. 

In terms of the free scalar quasiparticles of mass $m_s$,
the field operators in the Heisenberg picture 
are  written as 
\begin{eqnarray*}
\Phi({\vec x},t) & = & \int {d^3k\over (2\pi)^{3/2}}\,\phi({\vec k},t)\, 
e^{i{\vec k}\cdot {\vec x}}~,\\
\Pi({\vec x},t) & = & \int {d^3k\over (2\pi)^{3/2}}\,
\pi({\vec k},t)\, e^{i{\vec k} \cdot {\vec x}}~, 
\end{eqnarray*}
where
\begin{eqnarray*}
\phi({\vec k},t) & = & \frac{1}{\sqrt{2\omega_{\vec k}}}
\left[a({\vec k},t)+b^{\dagger}({-\vec k},t)\right]~,\\
\pi({\vec k},t) & = & i\sqrt{\omega_{\vec k} \over 2}
\left[a^{\dagger}({-\vec k},t) - b({\vec k},t)\right]~.
\end{eqnarray*} 
The number of positively charged scalars (which at zero chemical
potential is equal to the number of negatively charged scalars) 
is then given by 
\begin{eqnarray*}
n_{\vec k}(t)&=&\left\langle 
a^{\dagger}({\vec k},t)a({\vec k},t)\right\rangle\\
&=& \frac{1}{2\omega_{\vec k}}
\biggl\{\left\langle\pi({-\vec k},t)\pi^{\dagger}({\vec k},t)\right\rangle+
\omega^2_{\vec k}\,
\left\langle\phi^{\dagger}({-\vec k},t)\phi({\vec k},t)\right\rangle\\ 
&& +\,i\omega_{\vec k} \Bigl[\left\langle\phi^{\dagger}({-\vec k},t)
\pi^{\dagger}({\vec k},t)\right\rangle-
\left\langle \pi({-\vec k},t)\phi({\vec k},t)\right\rangle\Bigr]\biggr\}~.  
\end{eqnarray*}
We emphasize that this number operator, is a gauge invariant quantity
by construction. Using the Heisenberg equations of motion, to lowest
order in $e$, we obtain 
$$
\dot{n}_{\vec k}(t) = \dot{n}_{L,{\vec k}}(t) + \dot{n}_{T,{\vec k}}(t)~,
$$
where $\dot{n}_{L,{\vec k}}(t)$ and $\dot{n}_{T,{\vec k}}(t)$  
correspond to the longitudinal photon (plasmon) and transverse photon  
contributions respectively:
\begin{eqnarray}
\dot{n}_{L,{\vec k}}(t) &=&  \frac{e}{2\omega_{\vec k}} 
\int {d^3q\over (2\pi)^{3/2}}
\left(\frac{\partial}{\partial t'}+i\omega_{\vec k}\right)\nonumber\\
&&\times
\left[\Bigl\langle\phi^{\dagger,+}({\vec -k},t')\phi^-({\vec k-q},t)
{\cal A}^{-}_0({\vec q},t)\Bigr\rangle\right. \nonumber\\
&& +\left.\left. i
\left\langle\phi^{\dagger,+}({\vec -k},t')\dot{\phi}^-({\vec k-q},t)
{\cal A}^{-}_0({\vec q},t)\right\rangle\right]
\right|_{t'=t} +\;\text{c.c.}~.\label{ndotl:sqed}\\
\dot{n}_{T,{\vec k}}(t) & = & \frac{e}{\omega_{\vec k}} 
\int {d^3q\over (2\pi)^{3/2}} k^i_T({\vec q})
\left(\frac{\partial}{\partial t'}\right) \nonumber\\
&&\times\left[
\left\langle\phi^+({\vec k},t')
\phi^{\dagger,-}({\vec -k-q},t) {\cal A}^{i,-}_T({\vec q},t) 
\right\rangle\right.\nonumber\\ 
&& +\left.\left.\left\langle{\cal A}^{i,+}_T({\vec q},t)
\phi^{+}({\vec k-\vec q},t)\phi^{\dagger,-}({-\vec k},t')
\right\rangle\right]\right|_{t'=t}~,
\label{ndott:sqed}
\end{eqnarray}
Here
${\vec k}_T({\vec q})={\vec k}-({\vec k}\cdot{\vec{\hat q}}){\vec {\hat q}}$, 
$\bbox{{\cal A}}_T({\vec k},t)$ and ${\cal A}_0({\vec k},t)$ are the spatial
Fourier transforms of the gauge fields:
\begin{equation}
{\vec A}_T({\vec x},t)=\int\frac{d^3k}{(2\pi)^{3/2}}\,
{\bbox{\cal A}}_T({\vec k},t)\,e^{i{\vec k}\cdot{\vec x}}~, \quad A_0({\vec x},t)=\int\frac{d^3k}{(2\pi)^{3/2}}\,
{\cal A}_0({\vec k},t)\,e^{i{\vec k}\cdot{\vec x}}~.
\end{equation}
As usual the expectation values are computed in nonequilibrium perturbation 
theory in terms of the real-time propagators and vertices.
A detailed study of this scalar theory has revealed that there are no
HTL vertex corrections in SQED~\cite{rebhan,thoma} and 
this facilitates the analysis of the time evolution of the
distribution function for soft quasiparticles.  

Since in SQED the leading order HTL contribution to the scalar
propagator is a mass shift, 
the real-time HTL effective scalar propagator is given in 
eqs.~(\ref{fqppropagator}) in terms of the quasiparticle frequency
$\omega_{\vec k} = \sqrt{k^2+m^2_s}$. When the 
internal photon lines in the Feynman diagrams for the kinetic equation
are soft, an HTL resummation of these 
photon lines is required~\cite{htl,rob2,rob3,lebellac}. It is important
to note that  the HTL resummed  
photon propagators are only valid in {\em thermal equilibrium} since
the KMS condition that relates the 
advanced and retarded Green's functions has been used to write these
in terms of the spectral density. Therefore 
an analysis of the kinetic equation for the distribution function that
uses the HTL resummation for the soft degrees of 
freedom will be restricted to the linearized, i.e. relaxation time
approximation.  A truly nonequilibrium description of 
the kinetic equations for charged or gauge fields will require an
extension of the hard thermal loop program to situations 
far away from equilibrium, clearly such extension is beyond the scope of this article.
Therefore the derivation of the kinetic equation for 
the charged scalar fields assumes that the photons are in equilibrium and the distribution function of the charged scalars
has been displaced slightly off equilibrium. 
Fig.~4a shows the lowest order $ {\cal O}(e^2) $ contribution to the  
kinetic equation from longitudinal photons and Fig.~4b shows the
contributions from transverse photons.  

\subsection{Longitudinal photon contribution}

In this gauge invariant formulation, the longitudinal photon is
associated with the auxiliary field $ A_0(\vec x,t) $ which is 
the Lagrange multiplier associated with Gauss' law constraint. Since
this is not a propagating field (no canonical momentum conjugate 
exists), proper care must be taken in obtaining the Green's functions
for this field. In appendix~B we provide the details to 
obtain the HTL resummed real-time Green's function for this auxiliary field. 

The HTL effective propagators of the longitudinal photons are given by
(see Appendix~B)
\begin{mathletters}
\begin{eqnarray} 
&&{\cal G}_{L,{\vec q}}^{>}(t,t')= -i \int d^3x
\, e^{-i{\vec q}\cdot{\vec x}} \,
\Big\langle A_0({{\vec x}},t) A_0({{\vec 0}},t') \Big\rangle ~, 
\label{sqed:greaterl}\\
&&{\cal G}_{L,{\vec q}}^{<}(t,t')= -i \int d^3x 
\, e^{-i{\vec q}\cdot{\vec x}} \,
\Big\langle A_0({\vec 0},t') A_0({\vec x},t)\Big\rangle~,
\label{sqed:gsmallerl}\\
&&{\cal G}_{L,{\vec q}}^{++}(t,t^\prime)= \frac{1}{ q^2}\delta(t-t')+
{\cal G}_{L,{\vec q}}^{>}(t,t^{\prime})\theta(t-t^{\prime}) 
+{\cal G}_{L,{\vec q}}^{<}(t,t^{\prime})\theta(t^{\prime}-t)~, \\ 
&&{\cal G}_{L,{\vec q}}^{--}(t,t^\prime)= -\frac{1}{ q^2}\delta(t-t')+
{\cal G}_{L,{\vec q}}^{>}(t,t^{\prime})\theta(t^{\prime}-t) 
+{\cal G}_{L,{\vec q}}^{<}(t,t^{\prime})\theta(t-t^{\prime})~,\\
&&{\cal G}_{L,{\vec q}}^{\pm\mp}(t,t^\prime)=
{\cal G}_{L,{\vec q}}^{<(>)}(t,t^{\prime})~,  
\end{eqnarray}
\end{mathletters}
where $q= |{\vec q}|$ and 
\begin{mathletters}
\begin{eqnarray}
{\cal G}_{L,{\vec q}}^{>}(t,t^{\prime})&=& -i\int dq_0
\,\tilde{\rho}_L(q_0,{\vec q})\,[1+n_B(q_0)]\,e^{-iq_0(t-t^\prime)}~,\\ 
{\cal G}_{L,{\vec q}}^{<}(t,t^{\prime})&=& -i\int dq_0\,
\tilde{\rho}_L(q_0,{\vec q})\, n_B(q_0)\,e^{-iq_0(t-t^\prime)}~.
\end{eqnarray}
\end{mathletters}
The HTL spectral density $\tilde{\rho}_L(q_0,{\vec q})$
is given by~\cite{rebhan,boyhtl}
\begin{mathletters}
\begin{eqnarray}
\tilde{\rho}_L(q_0,{\vec q}) & = &
\frac{1}{\pi}\frac{{\rm Im}\Sigma_{L}(q_0,{\vec q})\,\theta(q^2-q^2_0)}
{\left[ q^2+{\rm Re}\Sigma_{L}(q_0,{\vec q})\right]^2
+\left[{\rm Im}\Sigma_{L}(q_0,{\vec q})\right]^2} +\,\mbox{sgn}(q_0)\,
Z_{L}({\vec q})\,\delta(q^2_0-\omega^2_L({\vec q}))~, \label{rhophotonl} \\ 
{\rm Im}\Sigma_{L}(q_0,{\vec q}) & = & \frac{\pi e^2 T^2}{6}
\frac{q_0}{q}~,\label{landaucutl} \\ 
{\rm Re}\Sigma_{L}(q_0,{\vec q}) & = & \frac{e^2 T^2}{6}
\left[2-\frac{q_0}{q}\ln\left|\frac{q_0+q}{q_0-q}\right|\right]~, 
\end{eqnarray}\label{rholall}
\end{mathletters}
where $\omega_L({\vec q})$ is the longitudinal photon pole 
and $Z_L({\vec q})$ is the corresponding (momentum dependent)
residue, which will not be relevant for the following discussion.

Using the above expressions for the nonequilibrium propagators, and after
some tedious but straightforward algebra, we find 
$\dot{n}_{L,{\vec k}}(t)$ to lowest order in perturbation theory
${\cal O}(e^2)$ is given by 
\begin{eqnarray}
\dot{n}_{L,{\vec k}}(t)& = & \frac{e^2}{2\omega_{\vec k}}
\int \frac{d^3q}{(2\pi)^3\omega_{\vec k+\vec q}}
\int_{-\infty}^{\infty} dq_0~\tilde{\rho}_L(q_0,{\vec q}) 
\int^t_{t_0}dt''\nonumber\\
&& \times \Bigl[(\omega_{\vec k}-\omega_{\vec k+\vec q})^2\,
{\cal N}_1(t_0)\cos[(\omega_{\vec k}+
\omega_{\vec k+\vec q}+q_0)(t-t'')]\nonumber\\
&& +\,(\omega_{\vec k}+\omega_{\vec k+\vec q})^2\,{\cal N}_2(t_0)
\cos[(\omega_{\vec k}-\omega_{\vec k+q}-q_0)(t-t'')]\Bigr]~,
\label{finalndotl} 
\end{eqnarray} 
where  
\begin{mathletters}
\begin{eqnarray}
{\cal N}_1(t)&=&[1+n_{\vec k}(t)]\,[1+n_{\vec k+q}(t)]\,
[1+n_B(q_0)]-n_{\vec k}(t)\,
n_{\vec k+\vec q}(t)\,n_B(q_0)~,\\
{\cal N}_2(t)&=&[1+n_{\vec k}(t)]\,n_{\vec k+q}(t)\,n_B(q_0) -
n_{\vec k}(t)\,[1+n_{\vec k+q}(t)]\,[1+n_B(q_0)]~.
\end{eqnarray}\label{statfactors:sqed}
\end{mathletters}
To obtain eq.~(\ref{finalndotl}), we have used the following  
properties~\cite{lebellac} (see also Appendix~B)
\begin{equation}
\tilde{\rho}_L(-q_0,{\vec q}) = -\tilde{\rho}_L(q_0,{\vec q})~,\quad
n_B(-q_0)= -[1+n_B(q_0)]~. \label{properties2:sqed}  
\end{equation} 
The different contributions have a very
natural interpretation in terms of gain minus loss processes. 
The first term in brackets corresponds to the process 
$0 \rightarrow \gamma^{\ast}_L+s+\bar{s}$ 
minus the process $\gamma^{\ast}_L+s+\bar{s}\rightarrow 0$,
and the second term corresponds to the scattering in the medium  
$\gamma^{\ast}_L + s \rightarrow s$ minus the inverse process $s \rightarrow
\gamma^{\ast}_L +s$, where $\gamma^{\ast}_L$ refers to the HTL-dressed 
longitudinal photon and $s$, $\bar{s}$ refer to the charged quanta 
of the scalar field $\Phi$.   

As mentioned above the HTL resummation of the 
internal photon and scalar lines assume that these degrees of freedom are in
thermal equilibrium and that the kinetic equation is valid in relaxation time 
approximation which will be assumed henceforth.
Namely, we assume that at time $t=t_0$ the distribution
function for a fixed mode with momentum ${\vec k}$ is disturbed 
slightly off equilibrium such that $n_{L,{\vec k}}(t_0)=n_B(\omega_{\vec k})
+\delta n_{L,{\vec k}}(t_0)$,
while the rest of the modes
remain in equilibrium, i.e., 
$n_{L,{\vec k}+{\vec q}}(t_0)=n_B(\omega_{\vec k+q})$ 
for ${\vec q}\neq{\vec 0}$ and linearize the kinetic equation in $\delta n_{L,{\vec k}}$. 

Since the propagators entering in 
the perturbative expansion of the kinetic equation are in terms of the
distribution functions at the initial time, the time integration can 
be done straightforwardly leading to a linearized equation
in relaxation time approximation. In terms of the spectral density 
\begin{eqnarray}
\rho_L(\omega,{\vec k}) &=& \frac{2\pi^2}{\omega_{\vec k}} \int
\frac{d^3q}{(2\pi)^3\omega_{\vec k+q}}\,
\int_{-\infty}^{\infty} dq_0\,\tilde{\rho}_L(q_0,{\vec q})
\left[1+n_B(q_0)+n_B(\omega_{\vec k+q})\right]\nonumber\\
&&\times\left[(\omega_{\vec k}+\omega_{\vec k+\vec q})^2\,
\delta(\omega-\omega_{\vec k+\vec q}-q_0)-
(\omega_{\vec k}-\omega_{\vec k+\vec q})^2\,
\delta(\omega+\omega_{\vec k+\vec q}+q_0)\right]~,\label{sqed:specfuncl} 
\end{eqnarray}
we obtain the time derivative of distribution function in the form
\begin{equation}
\delta{\dot n}_{L,{\vec k}}(t)= -\alpha\,\Gamma_{L,{\vec k}}(t)\,
\delta n_{L,{\vec k}}(t_0)~, 
\end{equation}
where $ \alpha = e^2/4\pi $ and 
\begin{equation}
\Gamma_{L,{\vec k}}(t)= \int d\omega\,\rho_L(\omega,{\vec k})
\; \frac{\sin[(\omega-\omega_{\vec
k})(t-t_0)]}{\pi(\omega-\omega_{\vec k})}\;\;.\label{gammaldef} 
\end{equation}

Upon integrating over $t$ with the given initial condition at $t_0$ leads to the form
\begin{equation}
\delta n_{L,{\vec k}}(t)= \delta n_{L,{\vec k}}(t_0)\left[1- \alpha
\int^t_{t_0}\Gamma_{L,{\vec k}}(t')\,dt'\right]~.\label{relaxtimeappxl}
\end{equation}
As a consequence of the HTL resummation, the long-range instantaneous
Coulomb interaction is screened with a Debye screening 
length of ${\cal O}(1/eT)$. This results in that there are no
threshold or mass shell singularities in the spectral density
$\rho_L(\omega,{\vec k})$ which after HTL resummation is  a regular
function of $\omega$ both at threshold and on the mass shell
$\omega=\omega_{\vec k}$. Therefore the analysis leading to Fermi's Golden
Rule (\ref{FGR}) is valid and at intermediate asymptotic times
$ m_s(t-t_0)\gg 1$  we find a secular term that grows linear in time:  
\begin{eqnarray*}
\int^t_{t_0}\Gamma_{L,{\vec k}}(t')\,dt'&=& (t-t_0)\;
\rho_L(\omega_{\vec k},{\vec k})+\;\text{non-secular term}~.
\end{eqnarray*}
As before applying the dynamical renormalization group to resum
the secular term, one obtains the dynamical renormalization group 
(kinetic) equation
\begin{equation}
\delta{\dot n}_{L,{\vec k}}(t)= -\gamma_{L}({\vec k})\,
\delta n_{L,{\vec k}}(t)~, 
\end{equation}
where $ \gamma_{L}({\vec k}) $ is the scalar relaxation rate corresponding
to exchange of a longitudinal photon
\begin{eqnarray}
\gamma_{L}({\vec k})&=&\frac{2\pi^2\alpha}{\omega_{\vec k}}
\int\frac{d^3q}{(2\pi)^3}\;
\frac{(\omega_{\vec k}+\omega_{\vec k+\vec q})^2}{\omega_{\vec k+q}}\;
\tilde{\rho}_L(\omega_{\vec k}-\omega_{\vec k+\vec q},{\vec q})\nonumber\\
&&\times \left[1+n_B(\omega_{\vec k+q})+n_B(\omega_{\vec k}
-\omega_{\vec k+\vec q})\right]~.
\end{eqnarray}
Note that in obtaining $\gamma_{L}({\vec k})$ 
we have discarded  the second term in $\rho_L(\omega_{\vec k},{\vec k})$ which
vanishes due to kinematics.
With the initial condition 
$\delta n_{L,{\vec k}}(t=t_0) = \delta n_{L,{\vec k}}(t_0)$,
we find that the distribution function evolves in time as
\begin{equation}\label{evolulong}
\delta n_{L,{\vec k}}(t)=\delta n_{L,{\vec k}}(t_0) \;
e^{-\gamma_{L}({\vec k})(t-t_0)}~.
\end{equation}
Numerically, we find that $\gamma_L(k)$ is a rather smooth function of $k$
and approaches a constant value for $k\gtrsim T$. The numerical values 
of $\gamma_L(k)$ for static and hard scalars are, respectively, 
$\gamma_L(k\approx 0)\sim 0.721~\alpha T$ and $\gamma_L(k\gtrsim T)
\sim 1.10~\alpha T$ and interpolates monotonically 
in this range~\cite{thoma}. 
Our results of the scalar relaxation rate due to longitudinal photon
contribution are in agreement with the corresponding scalar damping rate 
found in Ref.~\cite{thoma}. For further comparison with the transverse photon
contribution, we write $\gamma_{L}({\vec k})$ in the form
\begin{equation}
\gamma_{L}({\vec k})= \alpha T f(k)~,\quad 0.721 \leq f(k) \leq 1.10~,\label{longirate}
\end{equation}
with $f(k)$ a smooth function of $k$. 

\subsection{Transverse photon contribution}

We anticipate that the transverse photon contribution will lead to infrared 
divergences because the transverse photons are only dynamically
screened through Landau damping  in the HTL
approximation~\cite{boyrgir,iancu,robinfra,boyhtl}. Since the scalar
is massive ($ m_s \sim eT $) the infrared region in the internal loop
momenta comes solely from soft transverse photons with $q_0,\,q \approx 0$.  
In terms of the spectral density, the HTL effective nonequilibrium
transverse photon propagators read~\cite{boyrgir,lebellac}: 
\begin{mathletters}
\begin{eqnarray} 
&&{\cal P}^{ij}({\vec q})\,
{\cal G}_{T,{\vec q}}^{>}(t,t')=i \int d^3x \, e^{-i{\vec q}\cdot{\vec x}} \,
\Big\langle A^i_T({{\vec x}},t) A^j_T({{\vec 0}},t') \Big\rangle ~, 
\label{sqed:greatert}\\
&&{\cal P}^{ij}({\vec q})\,
{\cal G}_{T,{\vec q}}^{<}(t,t')=i \int d^3x \, e^{-i{\vec q}\cdot{\vec x}} \,
\Big\langle A^j_T({\vec 0},t') A^i_T({\vec x},t)\Big\rangle~,
\label{sqed:gsmallert}\\
&&{\cal G}_{T,{\vec q}}^{++}(t,t^\prime)=
{\cal G}_{T,{\vec q}}^{>}(t,t^{\prime})\theta(t-t^{\prime}) 
+{\cal G}_{T,{\vec q}}^{<}(t,t^{\prime})\theta(t^{\prime}-t)~, \\ 
&&{\cal G}_{T,{\vec q}}^{--}(t,t^\prime)= 
{\cal G}_{T,{\vec q}}^{>}(t,t^{\prime})\theta(t^{\prime}-t) 
+{\cal G}_{T,{\vec q}}^{<}(t,t^{\prime})\theta(t-t^{\prime})~,\\
&&{\cal G}_{T,{\vec q}}^{\pm\mp}(t,t^\prime)=
{\cal G}_{T,{\vec q}}^{<(>)}(t,t^{\prime})~,  
\end{eqnarray}
\end{mathletters}
where
\begin{mathletters}
\begin{eqnarray}
{\cal G}_{T,{\vec q}}^{>}(t,t^{\prime})&=&i\int dq_0
\,\tilde{\rho}_T(q_0,{\vec q})\,[1+n_B(q_0)]\,e^{-iq_0(t-t^\prime)}~,\\ 
{\cal G}_{T,{\vec q}}^{<}(t,t^{\prime})&=&i\int dq_0\,
\tilde{\rho}_T(q_0,{\vec q})\, n_B(q_0)\,e^{-iq_0(t-t^\prime)}~,
\end{eqnarray}
\end{mathletters}
and ${\cal P}^{ij}({\vec q})=\delta^{ij}- q^i q^j/q^2$ is the transverse 
projector. Here, the HTL spectral density $\tilde{\rho}_T(q_0,{\vec q})$
is given by~\cite{boyrgir,rebhan,boyhtl}
\begin{eqnarray}
\tilde{\rho}_T(q_0,{\vec q}) & = &
\frac{1}{\pi}\frac{{\rm Im}\Sigma_{T}(q_0,{\vec q})\,\theta(q^2-q^2_0)}
{\left[q^2_0-q^2-{\rm Re}\Sigma_{T}(q_0,{\vec q})\right]^2
+\left[{\rm Im}\Sigma_{T}(q_0,{\vec q})\right]^2}+\,\mbox{sgn}(q_0)\,
Z_{T}({\vec q})\,\delta(q^2_0-\omega^2_T({\vec q}))~, \label{rhophotont} \\ 
{\rm Im}\Sigma_{T}(q_0,{\vec q}) & = & \frac{\pi e^2 T^2}{12}
\frac{q_0}{q}\left(1-\frac{q^2_0}{q^2}\right)~,\label{landaucutt} \\ 
{\rm Re}\Sigma_{T}(q_0,{\vec q}) & = & \frac{e^2 T^2}{12}
\left[2\,\frac{q^2_0}{q^2}+\frac{q_0}{q}
\left(1-\frac{q^2_0}{q^2}\right)\ln\left|\frac{q_0+q}{q_0-q}\right|\right]~, 
\end{eqnarray}
where $\omega_T({\vec q})$ is the transverse photon pole 
and $Z_T({\vec q})$ is the corresponding (momentum dependent)
residue, which will not be relevant for the following discussion. 
The important feature of this HTL spectral density is 
its support below the light cone. 
That is, for $q^2>q^2_0$ the imaginary part of the HTL resummed photon
self-energy, ${\rm Im}\Sigma_{T}(q_0,{\vec q})$, 
originates in the process of Landau
damping~\cite{robinfra,boyhtl} from scattering of quanta in the medium.

Using the above expressions for the nonequilibrium propagators and after
some tedious but straightforward algebra, we find that 
$\dot{n}_{T,{\vec k}}(t)$ to lowest order in perturbation theory
${\cal O}(e^2)$ is given by 
\begin{eqnarray}
\dot{n}_{T,{\vec k}}(t)& = & \frac{2 e^2}{\omega_{\vec k}}
\int \frac{d^3q}{(2\pi)^3}
\frac{{\vec k}_T^2({\vec q})}{\omega_{\vec k+\vec q}} 
\int_{-\infty}^{\infty} dq_0~\tilde{\rho}_T(q_0,{\vec q}) 
\int^t_{t_0}dt''\nonumber \\
&& \times \Bigl[{\cal N}_1(t_0)\cos[(\omega_{\vec k}+
\omega_{\vec k+\vec q}+q_0)(t-t'')]\nonumber\\
&& +\,{\cal N}_2(t_0)
\cos[(\omega_{\vec k}-\omega_{\vec k+q}-q_0)(t-t'')]\Bigr]~,
\label{finalndott} 
\end{eqnarray} 
where ${\cal N}_{1}(t)$ and ${\cal N}_{2}(t)$ are the same as that in 
eq.~(\ref{statfactors:sqed}).
To obtain eq.~(\ref{finalndott}), we have used the following  
properties~\cite{lebellac,boyrgir}
\begin{equation}
\tilde{\rho}_T(-q_0,{\vec q}) = -\tilde{\rho}_T(q_0,{\vec q})~,\quad
n_B(-q_0)= -[1+n_B(q_0)]~. \label{properties1:sqed}  
\end{equation} 
The different contributions have a very
natural interpretation in terms of gain minus loss processes. 
The first term in brackets corresponds to the process 
$0 \rightarrow \gamma^{\ast}_T+s+\bar{s}$ 
minus the process $\gamma^{\ast}_T+s+\bar{s}\rightarrow 0$,
and the second term corresponds to the scattering in the medium  
$\gamma^{\ast}_T + s \rightarrow s$ minus the inverse process 
$s \rightarrow
\gamma^{\ast}_T + s$, where $\gamma^{\ast}_T$ refers to the HTL-dressed 
transverse photon and $s$, $\bar{s}$ refer to the charged quanta of 
the scalar field $\Phi$.   

As mentioned above the HTL resummation of the internal 
photon and scalar lines assume that these degrees of freedom are in
thermal equilibrium and that the kinetic equation 
is valid in relaxation time approximation.
Hence we assume that at time $t=t_0$ the distribution
function for a fixed mode with momentum ${\vec k}$ is disturbed 
slightly off equilibrium such that $n_{T,{\vec k}}(t_0)=n_B(\omega_{\vec k})
+\delta n_{T,{\vec k}}(t_0)$, while the rest of the modes
remain in equilibrium, i.e., 
$n_{T,{\vec k}+{\vec q}}(t_0)=n_B(\omega_{\vec k+q})$ 
for ${\vec q}\neq{\vec 0}$ and linearize the kinetic equation in $\delta n_{T,{\vec k}}$.  
  
Since the propagators entering in 
the perturbative expansion of the kinetic equation are in terms of the
distribution functions at the initial time, the time integration can be
done straightforwardly.
In terms of the spectral density 
\begin{eqnarray}
\rho_T(\omega,{\vec k}) &=& \frac{8\pi^2}{\omega_{\vec k}} \int
\frac{d^3q}{(2\pi)^3}\frac{{\vec k}^2_T({\vec q})}{\omega_{\vec k+q}}\,
\int_{-\infty}^{\infty} dq_0\,
\tilde{\rho}_T(q_0,{\vec q})\nonumber \\ 
&& \times\left[1+n_B(q_0)+n_B(\omega_{\vec k+q})\right]\,
\delta(\omega-\omega_{\vec k+\vec q}-q_0)~,\label{sqed:specfunc} 
\end{eqnarray}
we obtain the time derivative of distribution function in the form
\begin{equation}
\delta{\dot n}_{T,{\vec k}}(t)= -\alpha\,\Gamma_{\vec k}(t)\,
\delta n_{T,{\vec k}}(t_0)~, 
\end{equation}
where
\begin{equation}
\Gamma_{T,{\vec k}}(t)= \int d\omega\,\rho_T(\omega,{\vec k})
\left[
\frac{\sin[(\omega-\omega_{\vec k})(t-t_0)]}{\pi(\omega-\omega_{\vec k})}
-(\omega \rightarrow -\omega)\right]~,\label{gammadef}
\end{equation}
It is to be noted that the spectral density in eq.~(\ref{sqed:specfunc}),
up to a prefactor, is the same as that studied within the
context of the relaxation of the amplitude of a mean-field in SQED~\cite{boyrgir}
and in the eikonal approximation~\cite{iancu} in QED.

Upon integrating over $t$ with the given initial condition at $t_0$ leads to the form
\begin{equation}
\delta n_{T,{\vec k}}(t)= \delta n_{T,{\vec k}}(t_0)\left[1- \alpha
\int^t_{t_0}\Gamma_{\vec k}(t')\,dt'\right]~.\label{relaxtimeappx}
\end{equation}
As before at intermediate asymptotic times $m_s(t-t_0)\gg 1$
if there are no singularities arising from the spectral density as 
$\omega\rightarrow\pm\omega_{\vec k}$, one finds a secular term 
linear in time. 
This is a perturbative signal of pure exponential relaxation at large times 
as we have discussed thoroughly in the previous sections. 
However, in the case under consideration the spectral density 
has an infrared singularity~\cite{boyrgir,iancu} 
and the long-time limit must be studied carefully.  

Potential secular terms (growing in time) could arise in the long time
limit $ t\gg t_0 $ whenever the denominators in
(\ref{gammadef}) vanish, i.e., for  the region of frequencies 
$\omega\approx \pm\omega_{\vec k}$. For $\omega \approx \omega_{\vec k}$ 
we see that the argument of the delta function in eq.~(\ref{sqed:specfunc}) 
vanishes in the region of the Landau damping cut of the exchanged transverse 
photon $q^2_0 < q^2$ and contributes to the infrared behavior. 
On the other hand, for $\omega\approx -\omega_{\vec k}$
the delta function in eq.~(\ref{sqed:specfunc}) is satisfied for 
$q_0 \approx -2\omega_{\vec k}$, and this
region gives a negligible contribution to the long time dynamics. 
Therefore, only the first term in (\ref{gammadef}) 
(with $\omega-\omega_{\vec k}$) contributes in the long time limit. 

This term is dominated by the Landau damping 
region of the spectral density 
of the exchanged soft photon given by (\ref{rhophotont}), since for 
$\omega \approx \omega_{\vec k}$ the argument of the delta function is 
$q_0+kq\cos\theta/\omega_{\vec k}$ and 
this is the region where the imaginary 
part of the HTL photon self-energy, ${\rm Im}\Sigma_T(q_0,{\vec q})$, 
has support. 
The second contribution (with $\omega + \omega_{\vec k}$)
oscillates in time and is always bound and perturbatively small.   

To extract the infrared behavior of the spectral density,
we focus on the infrared region of the loop momenta with 
$ q_0,\,q \ll eT $ in eq.~(\ref{sqed:specfunc})~\cite{boyrgir,iancu}. 
This is the region dominated by the exchange of very soft 
(HTL resummed) transverse photons~\cite{iancu,robinfra} 
and that dominates the long time evolution of the distribution function. 
For $ q_0\ll q\ll eT $, the contributions to the spectral density 
from zero temperature and massive scalars 
are sub-leading, therefore the term $ 1+n_B(\omega_{\vec k+\vec q}) $
can be neglected. The only dominant contribution is from very soft quasistatic 
($ q_0\sim 0 $) transverse photons for which  $ n_B(q_0) \approx T/q_0 $.
 
For $ q\ll eT $ the function $ \tilde{\rho}_T(q_0,{\vec q})/q_0 $ is 
strongly peaked at $ q_0=0 $ and is well approximated by~\cite{boyrgir,iancu}
\begin{equation}
\left.\frac{\tilde{\rho}_T(q_0,{\vec q})}{q_0}\right|_{q_0\ll q}
=\frac{1}{\pi q^2}\frac{d}{q_0^2+d^2}\approx\frac{\delta(q_0)}{q^2}
\label{quasistaticrho}
\end{equation}
as $ q\rightarrow 0 $, where $ d=12q^3/\pi e^2 T^2 $.
The remaining delta function $\delta(\omega -\omega_{\vec k + \vec
q})$ is satisfied in the kinematical region
$q_1\leq q \leq q_2$, with
$$
q_1 = |k-\sqrt{\omega^2-m^2_s}|~,\quad
q_2 =k+\sqrt{\omega^2-m^2_s}~. 
$$
The secular terms arise in the limit $ \omega \rightarrow \omega_{\vec k} $,
in this limit $ q_1 \rightarrow |\omega -\omega_{\vec k}|/v_{\vec
k} $ with $ v_{\vec k}= d\omega_{\vec k}/dk $ being the group velocity 
of the scalar quasiparticle, and  $q_2 \rightarrow 2k $. However the
region in which the above quasistatic approximation
(\ref{quasistaticrho}) is valid corresponds to $ q \leq eT $, therefore
the upper momentum cutoff $ q_2 $ in the integration region for $ q $
should be the {\em minimum} between $ 2k$  or $ eT $. Thus for  
momenta $k \geq eT$ the upper limit should be taken as 
$ q_2 \approx eT $ whereas for $ k \ll  eT $ the upper limit is $ q_2=2k $. 

Hence, we find that the spectral density diverges logarithmically as 
$ \omega\rightarrow \omega_{\bf k} $: 
$$
\rho_T(\omega,{\vec k}) \approx -2 v_{\vec k}T
\ln\frac{|\omega-\omega_{\vec k}|}{\mu_{\vec k}v_{\vec k}}
\left[1+{\cal O}(\omega-\omega_{\vec k})\right]~,
$$
where $\mu_{\vec k}\sim{\rm min}(\omega_{\rm pl},k)$
with $\omega_{\rm pl}\sim eT$ being the plasma frequency.

As it will be seen shortly, the external momentum dependence of the
upper momentum cutoff is crucial 
to determine the relaxational time scale of hard and soft scalars. 
At intermediate asymptotic times $\omega_{\rm pl}(t-t_0)\gg 1$ 
(recall that $\omega_{\rm pl}\sim m_s$), we find~\cite{boyrgir}
\begin{eqnarray}
\int^t_{t_0}\Gamma_{T,{\vec k}}(t')\;dt'&\approx&2v_{\vec k}T(t-t_0)
\,\ln[\bar{\mu}_{\vec k}v_{\vec k}(t-t_0)]+\;\text{non-secular terms}~,
\label{secular} 
\end{eqnarray} 
where $ \bar{\mu}_{\vec k}=\mu_{\vec k}\exp(\gamma -1)$ with 
$ \gamma = 0.5772157\dots $ being Euler-Mascheroni constant.
In lowest order in perturbation theory, the distribution functions that
enters in the loops are those at the initial time.
Obviously perturbation theory breaks down at time scales 
\begin{equation}
t-t_0\approx {1 \over 2\alpha v_{\vec k}T\ln[\bar{\mu}_{\vec k}
v_{\vec k}(t-t_0)]} \approx {1 \over 2\alpha v_{\vec
k}T\ln(\bar{\mu}_{\vec k} / 2\alpha T)}~.
\end{equation}

Now we apply the dynamical renormalization group to resum
the anomalous secular term, 
$(t-t_0)\ln[\bar{\mu}_{\vec k}v_{\vec k}(t-t_0)]$, 
in the perturbative expansion.
To achieve this purpose we introduce a renormalization constant for
the distribution function that absorbs the secular divergences at a fixed
time scale $\tau$ and write 
\begin{equation}
\delta n_{T,{\vec k}}(t_0) ={\cal Z}(\tau,t_0)\,
\delta n_{T,{\vec k}}(\tau)~,
\quad {\cal Z}(\tau,t_0)= 1 + \alpha \, z_1(\tau,t_0) + \cdots~, 
\label{renor}
\end{equation} 
and request that the coefficients $z_n$ cancel the secular divergences
proportional to $\alpha^n$ at a given time scale $\tau$. To lowest
order the choice
\begin{equation}
z_1(\tau,t_0) = \int^{\tau}_{t_0}\Gamma_{T,{\vec k}}(t')\,dt' 
\label{zeta1}
\end{equation}
\noindent leads to the renormalized distribution function at time $t$ 
in terms of the updated distribution function at the time scale $\tau$
$$
\delta n_{T,{\vec k}}(t) = \delta n_{T,{\vec k}}(\tau)
\left[1- \alpha\int^t_{\tau}\Gamma_{T,{\vec k}}(t')\,dt'\right]~.
$$
However, the distribution function $\delta n_{T,{\vec k}}(t)$ 
cannot depend on the
arbitrary renormalization scale $\tau$, this independence on the
renormalization scale leads to the renormalization group equation to
lowest order: 
\begin{equation}
\frac{d}{d\tau} \delta n_{T,{\vec k}}(\tau) + \alpha\,
\Gamma_{T,{\vec k}}(\tau)\, \delta n_{T,{\vec k}}(\tau) =0~. 
\label{RGeqn} 
\end{equation}
This renormalization group equation is now clearly of the form of a
kinetic equation in relaxation time approximation with a time dependent rate. 

Now choosing the renormalization scale to coincide with the time $t$ in
the solution of (\ref{RGeqn}) 
as is usually done in the scaling analysis of the solutions to the
renormalization group equations,  we find that the distribution
function in the linearized approximation evolves in time in the 
following manner
\begin{equation}
\delta n_{T,{\vec k}}(t)= \delta n_{T,{\vec k}}(t_0)\, 
\exp\left[-\alpha \int^t_{t_0}\Gamma_{T,{\vec k}}(t')\,dt'\right]
\label{expsol:sqed}
\end{equation}
with the initial condition
$\delta n_{T,{\vec k}}(t=t_0) = \delta n_{T,{\vec k}}(t_0)$.
In the long time limit $\omega_{\rm pl}(t-t_0) \gg 1$, using 
(\ref{secular}) 
we find that the distribution function relaxes towards equilibrium as 
\begin{equation}
\delta n_{T,{\vec k}}(t)\approx\delta n_{\vec k}(t_0)
\exp\Bigl[-2\alpha v_{\vec k}T(t-t_0)\ln[\bar{\mu}_{\vec k}
v_{\vec k}(t-t_0)]\Bigr]~. 
\label{linearrg}
\end{equation}

Furthermore, (\ref{linearrg}) reveals a time scale for the relaxation of the
charged scalar distribution function due to exchange of transverse photons  
$t_{{\rm rel},T} = \gamma^{-1}_{T}(\vec k) $,  with 
\begin{equation}
\gamma_T({\bf k})\approx 
\left\{
\begin{array}{ll}
\alpha v_{\vec k}T\left[\ln(1/\alpha)+{\cal O}(1)\right]&\quad 
\text{for}~k\gtrsim \alpha \, T~,\\
2k^2/m_s&\quad\text{for}~k\lesssim \alpha \, T~.
\end{array}\right. \label{gammatrans}
\end{equation}
Note that the transverse photon contribution to the scalar relaxation rate 
vanishes at zero momentum and $ \gamma_T(k)\ll\gamma_L(k) $ for
$ k\lesssim \alpha T $, this is very similar to the 
behavior of the damping rate of fermions in QCD found in Ref.~\cite{robinfra}. 

\subsection{Relaxational crossover in real time}

The real-time description of charged scalar relaxation discussed above allows 
us to study the crossover between exponential and anomalous relaxation. 
Combining the longitudinal and transverse photon contributions, 
we obtain in relaxation time approximation the following time 
evolution of the charged scalar distribution function 
\begin{equation}
\delta n_{\vec k}(t)\approx\delta n_{\vec k}(t_0)
\exp\left\{-\left[\gamma_{L}({\vec k})+2\alpha v_{\vec k}T 
\ln[\bar{\mu}_{\vec k}v_{\vec k}(t-t_0)]\right](t-t_0)\right\}~. 
\end{equation}

From the expression for $\gamma_L(\vec k)$ given by
eq.(\ref{longirate}) with $f(k) \approx 1$ we find that 
plain  exponential relaxation holds for 
$ 2 v_{\vec k}\ln[\bar{\mu}_{\vec k}v_{\vec k}(t-t_0)]\ll 1$ 
and $\omega_{\rm pl}(t-t_0)\gg 1 $, whereas 
anomalous exponential relaxation with an exponent $\sim t\ln t$
dominates for very long times. Hence there is a crossover in the form
of relaxation for the charged scalar distribution function at a time scale
$(t-t_0)\approx t_{\rm c}$, with
\begin{equation}
t_{\rm c} \approx \frac{\exp(1/2v_{\vec k})}
{\bar{\mu}_{\vec k}v_{\vec k}}~.
\end{equation}

For $ k\ll eT$ we have $\bar{\mu}_{\vec k} \sim k \ll eT$ and 
$v_{\vec k}\ll 1$, hence the crossover time scale is exceedingly long 
and the relaxation of the distribution function is dominated by (HTL
resummed) longitudinal photon exchange and is purely exponential 
in the asymptotic regime. On the other hand for $k\gtrsim eT$ then
$\bar{\mu}_{\vec k} \sim eT$ and $v_{\vec k} \sim {\cal O}(1)$ and 
$t_{\rm c} \sim \omega^{-1}_{\rm pl}$ in which case the relaxation is
dominated by (HTL resummed) transverse photon exchange and is 
anomalous with an exponent $ t\ln t$, hence faster than exponential and
with a relaxational time scale  
$t_{\rm rel} = \alpha v_{\vec k}T\ln(1/\alpha)$.

\section{Secular terms vs. pinch singularities}\label{section:pinch}

An important difference between the approach to nonequilibrium
evolution described by quantum kinetic 
equations advocated in this work and that often presented in the
literature is that we work directly in {\em real time}  
not taking Fourier transforms in time. This must be contrasted with
the real-time formulation (RTF) of finite temperature 
quantum field theory in which there are also four propagators and a
closed-time-path contour but the propagators and quantities 
computed therefrom are all in terms of temporal Fourier transforms. 
In thermal equilibrium the Fourier 
representation of these four propagators for a scalar field are given
by~\cite{chou,landsmann,lebellac}  
\begin{mathletters}
\begin{eqnarray}
G^{++}(K)&= &-[G^{--}(K)]^{\ast}\nonumber\\
&=&-\frac{1}{K^2-m^2_{\rm eff}+i\epsilon}
+2\pi i n_B(|k_0|) \delta(K^2-m^2_{\rm eff})~,\\
G^{+-}(K)&=& 2\pi i\,[\theta(-k_0)+n_B(|k_0|)] 
\delta(K^2-m^2_{\rm eff})~,\\
G^{-+}(K)&=& 2\pi i\,[\theta(k_0)+n_B(|k_0|)] 
\delta(K^2-m^2_{\rm eff})~,
\end{eqnarray}
\label{propagator1-ft}
\end{mathletters}
where $K=(k_0,{\vec k})$ is
the four-momentum and $ K^2=k_0^2-k^2 $,
whereas out of equilibrium the distribution functions are
simply replaced by non-thermal ones, i.e., 
$ n_B(|k_0|) \rightarrow n_{\vec k}(t_0)$.

Using the integral representation of the step function
$$
\theta(t)=\frac{i}{2\pi}\int\frac{d\omega}{\omega+i\epsilon}\;
e^{-i\omega t}~,
$$
one can easily show that eqs.~(\ref{propagator1-ft}) and the ones
obtained by replacing the thermal equilibrium 
distributions by the nonequilibrium ones are, 
respectively, the {\em temporal} Fourier transforms of
eqs.~(\ref{ffpropagator}) and (\ref{fqppropagator}).
The temporal Fourier transforms of the free retarded 
and advanced propagators are obtained similarly and read
\begin{equation}
G_{\rm R/A}(K)=-\frac{1}{K^2-m^2_{\rm eff}\pm i\,
\text{sgn}(k_0)\epsilon}~.
\end{equation}

Several authors have pointed out that the calculations using the CTP
formulation  in terms of the standard form of free propagators in
eqs.~(\ref{propagator1-ft}) or those obtained by the replacement of
the distribution functions by the nonequilibrium ones, lead to  pinch
singularities~\cite{landsmann,altherr,altherr2,dadic,bedaque,niegawa,niegawa2,greiner,carrington}.    

In a consistent  perturbative expansion both the retarded and advanced
propagators contribute and pinch singularities arise from the product
of these, for example for  a scalar field this product is of the form 
\begin{equation}
G_{\rm R}(K)G_{\rm A}(K)=\frac{1}{[K^2-m^2_{\rm eff}+i\,
\text{sgn}(k_0)\epsilon]
[K^2-m^2_{\rm eff}-i\,\text{sgn}(k_0)\epsilon]}~.\label{pinchterm}
\end{equation}
For finite $\epsilon$ this expression is regular, whereas
when $\epsilon\rightarrow 0^+$ it gives rise to singular products such
as $[\delta(K^2-m^2_{\rm eff})]^2$ as discussed
in Refs.~\cite{landsmann,altherr,altherr2,dadic,bedaque,niegawa,niegawa2,greiner,carrington}.
Singularities of this type are ubiquitous and are not particular to
scalar theories.  

A detailed analysis of these pinch terms reveals that they do not cancel each 
other in perturbation theory unless the system is in thermal 
equilibrium~\cite{landsmann,altherr,altherr2,dadic,bedaque,niegawa,niegawa2,greiner}. Indeed, this severe problem has cast 
doubt on the validity or usefulness of the  CTP formulation to describe nonequilibrium 
phenomena~\cite{altherr}. Although this singularities have been found in many circumstances and analyzed and discussed
in the literature often, a systematic and satisfactory treatment of these singularities is still lacking. In Ref.~\cite{carrington}
it was suggested that including an in-medium width of the quasiparticles to replace the Feynman's $\epsilon$ does provide a 
physically reasonable solution, however this clearly 
casts doubt on the consistency of any perturbative approach to describe
even weakly out of equilibrium phenomena. 

Recently some authors have conjectured that pinch 
singularities in perturbation theory might be attributed to a misuse
of Fourier transforms (for a detailed discussion
see~\cite{bedaque,niegawa,niegawa2,greiner}). 
As an illustrative and simple example of these type of pinch
singularities, these authors  discussed the elementary derivation of 
Fermi's Golden Rule in time-dependent perturbation theory in 
quantum mechanics. In calculating total transition probabilities there
appears  the square of energy conserving  
$\delta$-function, which arises due to taking the infinite time limit of 
scattering probabilities. In this setting, such terms are interpreted
as the elapsed scattering time multiplied by the energy-conservation
constraint rather than a pathological singularity.  
A close look at eq.~(\ref{pinchterm}) reveals that the pinch term 
is the square of on-shell condition for the free quasiparticle, which implies
a temporal Fourier transform in the infinite time limit and of the
same form as the square of the energy conservation constraint 
for the transition probability obtained in time-dependent perturbation theory. 

By assuming that the interaction duration time is large but finite,
Ni\'egawa~\cite{niegawa} and Greiner and Leupold~\cite{greiner} showed
that for a self-interacting scalar field 
the pinch part of the distribution function can be regularized by the 
interaction duration time 
as\footnote{See eqs.~(14) and (29) of Ni\'egawa~\cite{niegawa} and 
eqs.~(14) and (22) of Greiner and Leupold~\cite{greiner}.
Note that eq.~(14) in Ref.~\cite{greiner} contains a typographic error.} 
\begin{equation}
n^{\rm pinch}_{\vec k}(t)\simeq (t-t_0)\,\Gamma^{\rm net}_{\vec k}~,
\label{pinchpart}
\end{equation}
where ``$\simeq$'' denotes that only the pinch singularity
contribution is included, $ t-t_0 $ is the interaction duration time
and $ \Gamma^{\rm net}_{\vec k} $ is the net gain rate 
of the quasiparticle distribution function per unit time
\begin{equation}
\Gamma^{\rm net}_{\vec k}=
\frac{-i}{2\omega_{\vec k}}\Big[[1+n_{\vec k}(t_0)]
\Sigma^{<}(\omega_{\vec k},{\vec k})-n_{\vec k}(t_0) 
\Sigma^{>}(\omega_{\vec k},{\vec k})\Big]~. \label{gammapinch}
\end{equation}
Here $\Sigma^{>}(\omega_{\vec k},{\vec k})-
\Sigma^{<}(\omega_{\vec k},{\vec k})=
2\, i\, \mbox{Im}\Sigma_R(\omega_{\vec k},{\vec k})$
with  $ \Sigma_R(\omega_{\vec k},{\vec k}) $ 
being the retarded scalar self-energy on mass-shell.
Comparing (\ref{pinchpart}) with (\ref{scalar:perturb}) and
(\ref{gammapinch}) with (\ref{secularpart}), we clearly see the
{\em equivalence} between the linear secular terms in the perturbative
expansion and the presence of pinch singularities in the usual CTP
description. In the discussion following eq.(\ref{scalar:perturb}) we 
have recognized that secular terms are not present if the system is in
equilibrium, much in the same manner as the case of pinch singularities
as discussed originally by Altherr~\cite{altherr,altherr2}. 
Thus our conclusion is that {\bf pinch singularities are a temporal
Fourier transform representation of linear secular terms}. 

The dynamical renormalization group provides a systematic
resummation of these secular terms and provides the consistent
formulation to implement the renormalization of the distribution 
function suggested in Refs.~\cite{niegawa,niegawa2}. 

Hence we emphasize that the dynamical renormalization group advocated
in this article, explains the physical origin of 
the pinch singularities in terms of secular terms and Fermi's Golden
Rule, and provides a consistent and systematic resummation 
of these secular terms that lead to the quantum kinetic equation as
a renormalization group equation that determines the 
time evolution of the distribution function. This result justifies in
a systematic manner the conclusions and interpretation obtained  
in Ref.~\cite{carrington} where a possible regularization of the pinch
singularities was achieved by including the width  
of the quasiparticle obtained via the resummation of hard thermal loops. 

Furthermore we emphasize that the dynamical renormalization group is
far more general in that it allows to treat situations where the long
time evolution is modified by 
threshold (infrared) singularities in spectral densities, thereby
providing a resolution of infrared singularities in damping rates and
a consistent resummation scheme to extract the asymptotic time
evolution of the distribution function. The infrared singularities in 
these damping rates is a reflection of anomalous (i.e.,
non-exponential) relaxation as a result of threshold effects.  

The pinch singularities signal the breakdown of perturbation theory,
just as the secular terms in real time, however, 
the advantage of working directly in real time is that the time scale
at which perturbation theory breaks down is  
recognized clearly from the real-time perturbative expansion and is
identified directly with the relaxational time scale.  
The dynamical renormalization group justifies this identification by
providing a resummation of the 
perturbative series that improves the solution beyond the
intermediate asymptotics.  

The resolution of pinch singularities via the dynamical
renormalization group is general. As originally pointed out
in~\cite{altherr,altherr2} the pinch singularities typically multiply
expressions of the form (\ref{gammapinch}) which 
vanish in equilibrium, just as the linear secular terms multiply
similar terms in the real-time perturbative expansion, 
as highlighted by eqs.~(\ref{secularpart}). These terms are of the
typical form gain minus loss, in equilibrium they vanish, but 
their non-vanishing simply indicates that the distribution functions
are evolving in time and it is precisely this time evolution that is
described consistently by the dynamical renormalization group.  

\section{Conclusions}

In this article we have introduced a novel method to obtain quantum
kinetic equations via a field theoretical and diagrammatic 
perturbative expansion improved via a dynamical renormalization group
resummation in {\em real time}. The first step of this method is to
use the microscopic equations of motion to obtain the evolution
equation of the quasiparticle distribution function, this is the
expectation value of the quasiparticle number operator in the initial
density matrix. This evolution equation can be solved in a consistent 
diagrammatic perturbative expansion and one finds that the solution
for the time evolution of the distribution function features 
{\em secular} terms, i.e., terms that grow in time. In perturbation
theory the microscopic and relaxational time scales are widely 
separated and there is a regime of intermediate asymptotics within
which (i) the secular terms dominate the time evolution of the 
distribution function and (ii) perturbation theory is valid. A
renormalization of the distribution function absorbs the contribution 
from the secular terms at a given renormalization time scale thereby
improving the perturbative expansion. The arbitrariness of this 
renormalization scale leads to the dynamical renormalization group
equation, which is recognized as the quantum kinetic  
equation. Linear secular terms are recognized to lead to usual
exponential relaxation (in relaxation time approximation), whereas 
non-linear secular terms lead to anomalous relaxation. The dynamical
renormalization group provides a consistent resummation of the 
secular terms.  There are many advantages in this formulation: 

\begin{itemize}
\item{It is based on straightforward quantum field theoretical
diagrammatic perturbation theory, hence it allows a systematic 
calculation to any arbitrary order. It allows to include resummations
of medium effects such as nonequilibrium generalizations of hard
thermal loop resummation in the quantum kinetic equation. This is
worked out in detail in a  scalar field theory.} 

\item{It allows a detailed understanding of crossover between
different relaxational phenomena directly in real time. This is 
important in the case of wide resonances where threshold effects may
lead to non-exponential relaxation on some time scales, and  
also near phase transitions where soft excitations dominate the dynamics.} 

\item{ It describes non-exponential relaxation directly in real time
whenever threshold effects are important thus providing  
a real-time interpretation of infrared divergent damping rates in
gauge theories. This we consider one of the most valuable 
features of the dynamical renormalization group which makes this
approach particularly suited to study relaxation in gauge field 
theories in a medium where the emission and absorption of soft gauge
fields typically lead to threshold infrared divergences. This  
important feature was highlighted in this article by studying the
quantum kinetic equation for the distribution function of 
charged quasiparticles in SQED.  } 

\item{This method provides a simple and natural resolution of pinch
singularities found often when the distribution functions 
are non-thermal. Pinch singularities are the temporal Fourier
transform manifestation of the real-time secular terms, and their 
resolution is via the resummation implemented by the dynamical
renormalization group.} 
\end{itemize}

We have tested this new method within the familiar setting of a scalar
field theory, thus reproducing previous results but with these 
new methods and moved on to apply the dynamical renormalization group
to describe the quantum kinetics of a cool gas of pions
and sigma mesons described by the $O(4)$ linear sigma
model in the chiral limit.  This particular example reveals 
a crossover behavior in the case of hard resonances because of
threshold singularities associated with the emission and absorption of 
massless pions. In relaxation time approximation we find a
crossover between purely exponential relaxation and anomalous 
relaxation with an exponent of the form $ t\ln t $ which is faster
than exponential, 
the crossover scale depends on the momentum of the resonance. The 
regime of exponential relaxation (in the relaxation time
approximation) is described by a relaxation rate which is simple
related to the damping rate found recently for the same
model~\cite{Pisarski:1996mt,patkos,csernai}. The (faster) anomalous
relaxation is a novel result and could be of phenomenological
relevance in view of recent suggestions of novel threshold effects of
the sigma resonance near the chiral phase transition~\cite{hatsuda},
this possibility is worthy of a deeper study and we are currently
generalizing these methods to reach the critical region.  

We consider that the most important aspect of this article is the
study of relaxation of charged quasiparticles in a gauge theory. 
As a prelude to studying quantum kinetics in QED and QCD~\cite{qedqcd}
in this article we studied the case of SQED.
In equilibrium, this theory shares many important features with QED and
QCD in leading order in the hard thermal loop resummation 
and is a relevant model to study kinetics and relaxation in the hot
Electroweak theory~\cite{fermiondamping,thoma2}.  
This Abelian theory allowed us to begin our study by providing 
a {\em gauge invariant} description of 
the distribution functions, thus bypassing potential 
ambiguities in the definition of gauge covariant Wigner transforms which
is the usual approach. The hard thermal loop resummation for both
longitudinal and transverse photons as well as for the 
scalar is included consistently in the derivation of the quantum
kinetic equation for the charged scalar quasiparticles in 
relaxation time approximation. The real-time solution of the kinetic
equation for the distribution function features linear and 
non-linear secular terms which are resummed consistently by the
dynamical renormalization group. The HTL longitudinal photons 
are Debye screened and do not lead to infrared divergences resulting
in purely exponential relaxation with a well defined 
relaxation rate. On the other hand, transverse photons are only
dynamically screened by Landau damping and the emission and 
absorption of photons at right angles leads to infrared threshold
divergences resulting in anomalous relaxation. We studied in 
detail the crossover between purely exponential and anomalous
relaxation. The crossover time scale depends on the momentum and 
for soft quasiparticles exponential relaxation dominates the dynamics
for a longer period of time, whereas for hard quasiparticles 
anomalous (with an exponent of the form $ t\ln t $) dominates the
relaxation. Recent approaches to quantum  kinetics including
HTL resummations have encountered infrared divergent relaxation
rates~\cite{Blaizot:1999xk}, the dynamical renormalization group
reveals very clearly that this is a manifestation of non-exponential
relaxation arising from threshold infrared effects that results in a violation 
of Fermi's Golden Rule. The time scales that can be extracted both from
the exponential and the non-exponential regimes agree with 
those obtained by Pisarski~\cite{robinfra} for QCD after
self-consistently including a width for the quasiparticle in the
calculation of the damping rate~\cite{robinfra}. Therefore, the study of
this Abelian model has indeed offered a novel method to study relaxation in
real time which is a useful arena for QCD and QED. 

We envisage several important applications of the dynamical
renormalization group method primarily to study transport phenomena 
and relaxation of collective modes in gauge theories where infrared
effects are important, as well as to study relaxational phenomena 
near critical points where soft fluctuations dominate the dynamics. An
important aspect of  this method is that it does not rely on a
quasiparticle approximation and allows a direct interpretation of
infrared phenomena directly in real time. Furthermore, we have 
established a very close relationship between the usual 
renormalization group and the dynamical renormalization 
group approach to kinetics. We have proved that the dynamical
renormalization group equation is the quantum kinetic equation, 
the collisional terms are the equivalent of the beta functions in
Euclidean renormalization group. 
Fixed points of the dynamical renormalization group are identified with 
stationary solutions of the kinetic equation and the exponents that
determine the stability of the fixed points are identified 
with the relaxation rates in the relaxation time
approximation. Furthermore we have suggested that in this language
coarse graining is the equivalent to neglecting irrelevant couplings
in the Euclidean renormalization program. This identification brings a
new and rather different perspective to kinetics and relaxation that
will hopefully lead to new insights.  
 
\acknowledgements

D.B. and S.-Y.W. thank the NSF for partial support through grants
PHY-9605186, INT-9815064 and INT-9905954. 
S.-Y.W. thanks the Andrew Mellon Predoctoral Fellowship for support. 
D.B. and H.J.d.V. acknowledge support from NATO.  

\appendix 

\section{Gauge-invariant formulation for Scalar QED}

In this appendix we summarize the gauge-invariant formulation~\cite{gaugeinv} 
for SQED with the Lagrangian density given by
$$
{\cal{L}} = D^{\mu}\phi^{\dagger} D_{\mu}\phi- m^2 \phi^{\dagger}\phi  
-\frac{1}{4}F^{\mu \nu} F_{\mu \nu}~,
$$
where
\begin{eqnarray*}
&& F_{\mu \nu} = \partial_{\mu} A_{\nu} - \partial_{\nu} A_{\mu}~,\\
&& D_{\mu}\phi = \partial_{\mu} \phi +i e A_{\mu} \phi~.
\end{eqnarray*}
The description in terms of gauge invariant states and operators is best
achieved within the canonical formulation, 
which begins with the identification of canonical field  
variables and constraints. These will determine the
classical physical phase space and, at the quantum level, 
the physical Hilbert space.

The canonical momenta conjugate to the gauge and scalar
fields are given by
\begin{eqnarray*}
\Pi^0 & = & 0 ~,\\
\Pi^i & = & \dot{A}^i+\nabla^i A^0 = -E^i ~,\\
\pi & = & \dot{\phi}^{\dagger}+i e A^0 \phi^{\dagger}~,\\
\pi^{\dagger} & = & \dot{\phi}-i e A^0 \phi~. 
\end{eqnarray*}
Hence, the Hamiltonian is
\begin{eqnarray*}
H &=& \int d^3x \bigg\{ \frac{1}{2}{\vec \Pi}\cdot{\vec \Pi}+\pi^{\dagger}\pi+
\left({\vec \nabla}\phi^{\dagger}+ie {\vec A}\phi^{\dagger}\right)
\cdot\left({{\vec \nabla}}\phi- ie {\vec A}\phi\right)+
\frac{1}{2}\left({\vec \nabla} \times {\vec A}\right)^2 + \nonumber \\ 
&& m^2\phi^{\dagger}\phi + A_0\left[{\vec \nabla}\cdot{\vec \Pi}-ie 
\left(\pi\phi-\pi^{\dagger}\phi^{\dagger}\right)\right]\bigg\}~.
\end{eqnarray*}

There are several different manners of quantizing a gauge theory, 
but the one
that exhibits the gauge invariant states and operators, 
originally due to
Dirac, begins by recognizing the first class constraints 
(with mutually vanishing Poisson brackets between constraints). 
From here there are two possibilities: 
(i) The constraints become operators in the quantum 
theory and are imposed onto the
physical states, thus defining the physical subspace 
of the Hilbert space and gauge invariant operators. 
(ii) Introduce a gauge, converting the first class
system of constraints into a second class 
(with non-zero Poisson brackets between constraints) 
and introducing Dirac brackets.  
This second possibility is the popular way of 
dealing with the constraints 
and leads to the usual gauge-fixed path
integral representation~\cite{slavnov} 
in terms of Faddeev-Popov determinants and ghosts.
We will instead proceed with the first possibility that leads to an 
unambiguous projection of the physical states and operators. 
Such a method has been previously used by James and Landshoff 
within a different context~\cite{landshoff}.

In Dirac's method of quantization~\cite{hat} there are two first class
constraints which are:
\begin{eqnarray}
&&\Pi^0= \frac{\delta {\cal{L}}}{\delta A^0} = 0~, \label{pi0const}\\
&&{\cal{G}}({\vec x},t)={\vec \nabla}\cdot{\vec \Pi}+e\rho=0~,
\label{gausslaw} 
\end{eqnarray}
with $\rho=-i(\phi\pi- \phi^{\dagger}\pi^{\dagger})$ being the scalar 
charge density. eq.~(\ref{gausslaw}) is Gauss's law, which
can be seen to be a constraint in two ways: 
either because it cannot be obtained as a Hamiltonian 
equation of motion, or because in Dirac's
formalism, it is the secondary (first class) constraint 
obtained by requiring that the primary constraint 
eq.~(\ref{pi0const}) remain constant in time. 
Quantization is now achieved by imposing the canonical equal-time
commutation relations
\begin{mathletters}   
\begin{eqnarray*}
&&\left[A^0({\vec x},t),\Pi^0({\vec y},t)\right]=
  i\delta^3({\vec x}-{\vec y})~, \\ 
&&\left[A^i({\vec x},t),\Pi^j({\vec y},t),\right] =
  i\delta^{ij} \delta^3({\vec x}-{\vec y})~, \\
&&\left[\phi({\vec x},t),\pi({\vec y},t),\right]= i
\delta^3({\vec x}-{\vec y})~,\\
&&\left[\phi^{\dagger}({\vec x},t),\pi^{\dagger}({\vec y},t)\right] =
  i\delta^3({\vec x}-{\vec y})~.
\end{eqnarray*}
\end{mathletters}   

In Dirac's formulation, the projection onto the gauge 
invariant subspace of the
full Hilbert space is achieved by imposing the first 
class constraints onto the
states. Physical operators are those that commute with the first class
constraints. With the above equal-time commutation relations it is
straightforward to see that the unitary operator
$$
U_{\Lambda}= \exp{\left[i\int\left(\Pi^0 
\dot{\Lambda}+{\cal{G}} \Lambda\right) d^3x\right]} 
$$
performs the local gauge transformations. Thus the first class constraints are
recognized as the generators of gauge transformations. In particular, Gauss's
law operator ${\cal{G}}$ is the generator of time independent gauge
transformations. Requiring that the physical states be annihilated by these
constraints is tantamount to selecting the gauge invariant states.
Consequently operators that commute with the first class constraints are gauge
invariant.

In the Schr\"odinger representation of field theory, 
in which the field operators are diagonal,
states are represented by wave functionals, 
and the canonical momenta conjugate to the field 
operators are represented by Hermitian functional differential operators.
The constraints applied onto the states become functional 
differential equations that the wave functionals must satisfy:
\begin{mathletters}   
\begin{eqnarray}
&& \frac{\delta}{\delta A_0({\vec x})} \Psi[{\vec A},\phi,\phi^{\dagger}] = 0
 \label{1stconst} \\ 
&& \left[ {{\vec \nabla}}_{\vec x} \cdot\frac{\delta}{\delta
 {\vec A}({\vec x})}-i e \left(\phi({\vec x})\frac{\delta}
{\delta\phi({\vec x})}-
 \phi^{\dagger}({\vec x})\frac{\delta}{\delta \phi^{\dagger}({\vec x})}\right)
 \right] \Psi[{{\vec A}},\phi,\phi^{\dagger}] = 0 \label{2ndconst}
\end{eqnarray} 
\end{mathletters}   
The first equation simply means that the wave functional does not
depend on $A_0$, whereas the second equation means that the wave functional is
only a functional of the combination of fields that is annihilated by the
Gauss's law functional differential operator.  
It is a simple calculation to prove that the fields
\begin{mathletters}   
\begin{eqnarray}
\Phi({\vec x})&=&\phi({\vec x})\exp\left[ie \int d^3 y {\vec A}({\vec y})\cdot
{{\vec \nabla}}_{\vec y} G({\vec y}-{\vec x})\right]\label{gauginvphi} \\
\Phi^{\dagger}({\vec x}) & = & \phi^{\dagger}({\vec x})\exp
\left[-ie \int d^3 y{\vec A}({\vec y})\cdot 
{\vec \nabla}_{\vec y} G({\vec y}-{\vec x})\right]
\label{gauginvphidag}
\end{eqnarray}
\end{mathletters}   
are annihilated by Gauss's law functional differential equation with 
$G({\vec y}-{\vec x})$ the Coulomb Green's function
\begin{equation}
{\vec \nabla}^2_{\vec y} G({\vec y}-{\vec x})=\delta^3({\vec y}-{\vec x})~. 
\label{coulombgf}
\end{equation}
Furthermore, writing the gauge field in terms of transverse and longitudinal
components as
\begin{equation}
{\vec A}({\vec x}) = {\vec A}_L({\vec x})+{\vec A}_T({\vec x})~, 
\label{fieldsplit}
\end{equation}
where
\begin{equation}
{{\vec \nabla}} \times {\vec A}_L({\vec x}) = 0 \label{long}~,\quad
{{\vec \nabla}} \cdot {\vec A}_T({\vec x}) = 0 \label{trans}~,
\end{equation}
one finds
\begin{equation}
{{\vec \nabla}}_{\vec x}\cdot\frac{\delta}{\delta {\vec A}({\vec x})} = 
{{\vec \nabla}}_{\vec x}\cdot\frac{\delta}{\delta{\vec A}_L({\vec x})}~.
\label{difflong}
\end{equation}
Therefore the transverse component ${\vec A}_T$ is also annihilated
by the Gauss's law operator, and ${\vec A}$ in the exponential 
in eqs.~(\ref{gauginvphi}) and (\ref{gauginvphidag}) 
can be replaced by ${\vec A}_L$. 
This analysis shows that the wave functional
solutions of the functional differential equations that represent the
constraints in the Schroedinger representation are of the form
\begin{equation}
\Psi[{\vec A},\phi,\phi^{\dagger}] = \Psi[{\vec A}_T,\Phi,\Phi^{\dagger}]~.
\label{gaugeinvfunc}
\end{equation}
The fields ${\vec A}_T$, $\Phi$ and $ \Phi^{\dagger}$ are {\em gauge invariant}
as they commute with the constraints. The canonical momenta conjugate to 
$\Phi$ and $\Phi^{\dagger}$ are found to be
\begin{mathletters}   
\begin{eqnarray}
\Pi({\vec x}) & = & \pi({\vec x}) \exp\left[-ie\int d^3 y{\vec A}_L({\vec y})
\cdot{{\vec \nabla}}_{\vec y} G({\vec y}-{\vec x})\right]~, \label{canophi} \\
\Pi^{\dagger}({\vec x}) & = & \pi^{\dagger}({\vec x}) \exp\left[ie\int d^3 y
{\vec A}_L({\vec y})\cdot {{\vec \nabla}}_{\vec y} G({\vec y}-{\vec x})
\right]~.
\label{canophidag}
\end{eqnarray}
\end{mathletters}   
The momentum ${\vec \Pi}$ canonical to ${\vec A}$ 
can also be written in terms of
longitudinal and transverse components
\begin{equation}
{\vec \Pi}({\vec x}) ={\vec \Pi}_L({\vec x})+{\vec \Pi}_T({\vec x})~. 
\label{PiA}
\end{equation}
It is straightforward to check that both components are gauge invariant.
In the physical subspace of gauge invariant wave functionals, matrix elements
of ${{\vec \nabla}}\cdot {\vec \Pi}$ can be replaced by matrix elements of the
charge density $\rho=-i(\Phi \Pi - \Phi^{\dagger}\Pi^{\dagger})$.
Therefore in all matrix elements between gauge
invariant states (or functionals) one can replace
\begin{equation}
{\vec \Pi}_L({\vec x})\rightarrow - e\,{{\vec \nabla}}_{\vec x} 
\int d^3y G({\vec x}-{\vec y})
\rho({\vec y})~. \label{coulomb}
\end{equation}

Finally in the gauge invariant subspace the Hamiltonian becomes
\begin{eqnarray}
H &=& \int d^3x \bigg\{ \frac{1}{2}{\vec \Pi}_T\cdot{\vec \Pi}_T+
\Pi^{\dagger}\Pi+\Big({{\vec \nabla}}\Phi^{\dagger}+ 
ie {\vec A}_T \Phi^{\dagger}\Big)
\cdot\Big({{\vec \nabla}}\Phi- ie {\vec A}_T \Phi\Big)+
\frac{1}{2}\left({{\vec \nabla}} \times {\vec A}_T\right)^2\nonumber \\ 
&& +\,m^2 \Phi^{\dagger}\Phi \bigg\}-\frac{e^2}{2}\int d^3x d^3y\,
\rho({\vec x})G({\vec x}-{\vec y})\rho({\vec y})~.
\label{gauginham}
\end{eqnarray}
Clearly the Hamiltonian is gauge invariant, and it manifestly has the global
$U(1)$ gauge symmetry under which $\Phi$ transforms with a constant phase, 
$\Pi$ transforms with the opposite phase, and
${\vec A}_T$ and ${\vec \Pi}_T$ is invariant.
This Hamiltonian is reminiscent of the Coulomb gauge Hamiltonian, but we
emphasize that we have not imposed any gauge fixing condition. The formulation
is fully gauge invariant, written in terms of operators that commute with the
generators of gauge transformations and states that are invariant under these
transformations.

To obtain a gauge invariant description in Lagrangian formalism, we switch
to the path integral representation for field theory in which the 
vacuum-to-vacuum amplitude is defined as
\begin{equation}
\int {\cal D}{\vec A}_T {\cal D}{\vec \Pi}_T 
{\cal D}\Phi{\cal D}\Pi{\cal D}\Phi^{\dagger}{\cal D}\Pi^{\dagger}
\exp\left[i\int d^4x 
\left(\Pi\dot{\Phi}+\Pi^{\dagger}\dot{\Phi}^{\dagger}
+{\vec \Pi}_T\cdot\dot{{\vec A}}_T\right)-i\int dt H\right]~.
\end{equation}
Note that the last term in the Hamiltonian, eq.~(\ref{gauginham}),
is the instantaneous Coulomb 
interaction, which  can be traded for a gauge invariant 
auxiliary field $A_0$, up to an overall factor,
the vacuum-to-vacuum amplitude becomes
\begin{equation}
\int{\cal D}A_0 {\cal D}{\vec A}_T {\cal D}{\vec \Pi}_T 
{\cal D}\Phi{\cal D}\Pi{\cal D}\Phi^{\dagger}{\cal D}\Pi^{\dagger}
\exp\left[i\int d^4x 
\left(\Pi\dot{\Phi}+\Pi^{\dagger}\dot{\Phi}^{\dagger}
+{\vec \Pi}_T\cdot\dot{{\vec A}}_T\right)-i\int dt \bar{H}\right]~,
\end{equation}
where
\begin{eqnarray}
\bar{H} &=& \int d^3x \bigg\{\frac{1}{2}{\vec \Pi}_T\cdot{\vec \Pi}_T+
\Pi^{\dagger}\Pi+\Big({\vec \nabla}\Phi^{\dagger}+ 
ie {\vec A}_T \Phi^{\dagger}\Big)
\cdot\Big({\vec \nabla}\Phi- ie {\vec A}_T \Phi\Big)+
\frac{1}{2}\left({\vec \nabla} \times {\vec A}_T\right)^2 + \nonumber \\ 
&& m^2 \Phi^{\dagger}\Phi -\frac{1}{2}
\left({\vec \nabla}A_0\right)^2- e A_0 \rho\bigg\}~.
\end{eqnarray}
Since the exponent is now quadratic in the conjugate momenta, 
we can complete the squares and evaluate the 
${\cal D}{\vec \Pi}_T$, ${\cal D}\Pi$ and ${\cal D}\Pi^{\dagger}$  
integrals to obtain
\begin{equation}
\int {\cal D}A_0{\cal D}{\vec A}_T{\cal D}\Phi{\cal D}\Phi^{\dagger}
\exp\left[i\int d^4x {\cal L}[A_0,{\vec A}_T,\Phi,\Phi^\dagger]\right]~,
\end{equation}
up to an overall factor,
where ${\cal L}[A_0,{{\vec A}}_{T},\Phi,\Phi^\dagger]$ is the gauge 
invariant Lagrangian,
\begin{eqnarray}
{\cal L}[A_0,{{\vec A}}_T,\Phi,\Phi^\dagger]
&=&\partial_\mu\Phi^\dagger\,\partial^\mu\Phi -m^2 \Phi^{\dagger}\Phi  
+\frac{1}{2}\partial_\mu {{\vec A}}_T\cdot\partial^\mu{{\vec A}}_T
-e {{\vec A}}_T\cdot{{\vec j}}_T \nonumber \\
&& - e^2{{\vec A}}_T\cdot{{\vec A}}_T \Phi^\dagger\Phi  
 +\frac{1}{2}\left({{\vec \nabla}}A_0\right)^2 +
 e^2 A_0^2 \Phi^\dagger\Phi  + e A_0 j_0~. 
\end{eqnarray}
with  
${\vec j}_T=i[\Phi^\dagger({\vec \nabla}_T\Phi)-({\vec \nabla}_T
\Phi^\dagger)\Phi]$ and
$j_0=-i(\Phi{\dot\Phi}^{\dagger}-\Phi^{\dagger}\dot{\Phi})$.
Note that $A_0$ satisfies an {\em algebraic} equation of motion 
${\vec \nabla}^2 A_0 = e \rho$.

\section{Full Propagators for Auxiliary Fields}

In this appendix we derive the {\em full} real-time CTP propagators
for the auxiliary field {\em in equilibrium}. 
We will consider as an example the longitudinal photon field
$A_0$ in SQED (the extension to other cases is straightforward). 
Since an auxiliary field is non-dynamical, 
it satisfies an {\em algebraic} equation of motion without time derivative. 
As a consequence, the free longitudinal photon propagators are local in
time and there is no mixture between fields on ``$+$'' and ``$-$''
branches of the CTP contour. Namely,
\begin{mathletters}
\begin{eqnarray} 
&&\Big\langle{\cal A}^{+}_0({\vec q},t){\cal A}^{+}_0({{\vec -q}},t') 
\Big\rangle_0 = \frac{i}{q^2}\delta(t-t')~, \\ 
&&\Big\langle{\cal A}^{-}_0({\vec q},t){\cal A}^{-}_0({{\vec -q}},t') 
\Big\rangle_0 = -\frac{i}{q^2}\delta(t-t')~,\\
&&\Big\langle{\cal A}^{+}_0({\vec q},t){\cal A}^{-}_0({{\vec -q}},t') 
\Big\rangle_0 = 
\Big\langle{\cal A}^{-}_0({\vec q},t){\cal A}^{+}_0({{\vec -q}},t') 
\Big\rangle_0 = 0~,
\end{eqnarray}
\end{mathletters}
where $\langle\cdots\rangle_0$ denotes expectation value of
free fields in equilibrium.

Now we consider the full longitudinal photon propagators. 
Neglecting the tadpole type term $e^2 A_0^2 \Phi^{\dagger}\Phi$
[which yields local (momentum independent) contribution and
higher order contribution, hence is irrelevant to the one-loop result 
that we are interested in], 
we have 
$$
{\cal L}_{\rm int}= e A_0 j_0~,
$$
where $j_0=-i(\Phi{\dot\Phi}^{\dagger}-\Phi^{\dagger}\dot{\Phi})$.
Straightforward diagrammatic expansions show that the following 
equalities 
hold to all orders in perturbation theory
\begin{mathletters}
\begin{eqnarray}
\Big\langle{\cal A}^{+}_0({\vec q},t){\cal A}^{+}_0({{\vec -q}},t') 
\Big\rangle &=& \frac{i}{q^2}\delta(t-t') + \frac{e^2}{q^4}
\Big\langle j^{+}_0({\vec q},t) j^{+}_0({{\vec -q}},t') 
\Big\rangle~,\\
\Big\langle{\cal A}^{+}_0({\vec q},t){\cal A}^{-}_0({{\vec -q}},t') 
\Big\rangle &=& \frac{e^2}{q^4}
\Big\langle j^{+}_0({\vec q},t) j^{-}_0({{\vec -q}},t') 
\Big\rangle~,
\end{eqnarray}
\end{mathletters}
where $\langle\cdots\rangle$ denotes the full equilibrium expectation value.
It is convenient to introduce the current-current 
spectral densities $\rho^{>}_j(q_0,{\vec q})$ and 
$\rho^{<}_j(q_0,{\vec q})$ defined by
\begin{mathletters}
\begin{eqnarray}
\Big\langle j^{+}_0({\vec q},t) j^{+}_0({{\vec -q}},t') 
\Big\rangle &=& \int dq_0 \Big[\rho^{>}_j(q_0,{\vec q}) \theta(t-t')
+\rho^{<}_j(q_0,{\vec q}) \theta(t'-t)\Big] e^{-iq_0(t-t')}~,\\
\Big\langle j^{+}_0({\vec q},t) j^{-}_0({{\vec -q}},t') 
\Big\rangle &=&\int dq_0\,\rho^{<}_j(q_0,{\vec q})\,e^{-iq_0(t-t')}~.
\end{eqnarray}
\end{mathletters}
Inserting a complete set of eigenstates of the full interacting 
Hamiltonian, one obtains the KMS condition
\begin{equation}
\rho^{<}_j(q_0,{\vec q})=e^{-\beta q_0}\rho^{>}_j(q_0,{\vec q})~.
\label{jjkms}
\end{equation}
In terms of the $\rho^{>}_j(q_0,{\vec q})$ the full {\em retarded} 
longitudinal photon propagator can be written as
\begin{eqnarray*}
\Big\langle{\cal A}_0({\vec q},t){\cal A}_0({{\vec -q}},t') 
\Big\rangle_{\rm R} &\equiv& \Big\langle
{\cal A}^{+}_0({\vec q},t){\cal A}^{+}_0({{\vec -q}},t')\Big\rangle -
\Big\langle{\cal A}^{+}_0({\vec q},t){\cal A}^{-}_0({{\vec -q}},t')
\Big\rangle\\
&=&
i\int \frac{dq_0}{2\pi}
\rho_0(q_0,{\vec q})e^{-iq_0(t-t')}~,
\end{eqnarray*}
where
$$
\rho_0(q_0,{\vec q})=\frac{1}{q^2}
+\frac{e^2}{q^4}\int d\omega
\frac{\rho^{>}_j(\omega,{\vec q})}{q_0-\omega+i\epsilon}
\left(1-e^{-\beta\omega}\right)~,
$$
and the KMS condition eq.~(\ref{jjkms}) is used. Thus, we obtain
$$
\text{Im}\rho_0(q_0,{\vec q})=-\pi\left(1-e^{-\beta q_0}\right)
\frac{e^2}{q^4}\rho^{>}_j(q_0,{\vec q})~.
$$
Again using the KMS condition eq.~(\ref{jjkms}), we can finally write 
the full longitudinal photon propagator as 
\begin{eqnarray*}
\Big\langle{\cal A}^{+}_0({\vec q},t){\cal A}^{+}_0({\vec -q},t') 
\Big\rangle &=& i\left[\frac{1}{q^2}\delta(t-t') 
 + {\cal G}^{>}_{L,{\vec q}}(t,t') \theta(t-t') 
 + {\cal G}^{<}_{L,{\vec q}}(t,t') \theta(t'-t)\right]~,\\
\Big\langle{\cal A}^{+}_0({\vec q},t){\cal A}^{-}_0({\vec -q},t') 
\Big\rangle &=& i\,{\cal G}^{<}_{L,{\vec q}}(t,t')~,
\end{eqnarray*}
where
\begin{eqnarray}
{\cal G}^{>}_{L,{\vec q}}(t,t') &=& 
\frac{i}{\pi}\int dq_0\,\text{Im}\rho_0(q_0,{\vec q})\,[1+n_B(q_0)]
\,e^{-iq_0(t-t')}~,\label{appb:glgrt}\\
{\cal G}^{<}_{L,{\vec q}}(t,t') &=& 
\frac{i}{\pi}\int dq_0\,\text{Im}\rho_0(q_0,{\vec q})\,n_B(q_0)
\,e^{-iq_0(t-t')}~.\label{appb:gllsr}
\end{eqnarray}
It is easy to check that the KMS condition 
${\cal G}^{>}_{L,{\vec q}}(t-i\beta,t')={\cal G}^{<}_{L,{\vec q}}(t,t')$ holds. 
By the same token, we obtain
\begin{eqnarray*}
\Big\langle{\cal A}^{-}_0({\vec q},t){\cal A}^{-}_0({\vec -q},t') 
\Big\rangle &=& i\left[-\frac{1}{q^2}\delta(t-t') 
 + {\cal G}^{>}_{L,{\vec q}}(t,t') \theta(t'-t) 
 + {\cal G}^{<}_{L,{\vec q}}(t,t') \theta(t-t')\right]~,\\
\Big\langle{\cal A}^{-}_0({\vec q},t){\cal A}^{+}_0({\vec -q},t') 
\Big\rangle &=& i\,{\cal G}^{>}_{L,{\vec q}}(t,t')~.
\end{eqnarray*}
The imaginary part of the full retarded longitudinal photon propagator,
$\text{Im}\rho_0(q_0,{\vec q})$, can be calculated in 
perturbation theory via the tadpole method~\cite{boyhtl}. 
To one-loop order and in HTL limit~\cite{boyhtl}, one finds
$\text{Im}\rho_0(q_0,{\vec q})= -\pi \tilde{\rho}_L(q_0,{\vec q})$,
where $\tilde{\rho}_L(q_0,{\vec q})$ is given in eq.~(\ref{rholall}).
Finally, using the condition 
${\cal G}^{>}_{L,{\vec q}}(t,t')={\cal G}^{<}_{L,{\vec q}}(t',t)$ 
[cf. eq.~(\ref{gfrelation})] and making change of variable 
$q_0\rightarrow -q_0$ in eq.~(\ref{appb:glgrt}), we find that
$\text{Im}\rho_0(-q_0,{\vec q})=-\text{Im}\rho_0(q_0,{\vec q})$,
hence $\tilde{\rho}_L(-q_0,{\vec q})=-\tilde{\rho}_L(q_0,{\vec q})$.




\newpage



\begin{center}
\begin{figure}
\epsfig{file=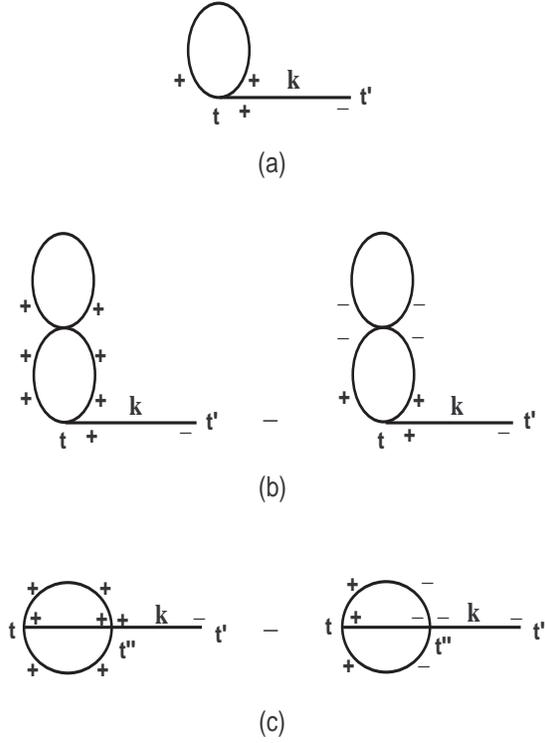,width=4in,height=4in}
\vspace{.1in}
\caption{The Feynman diagrams contribute to the quantum kinetic equation 
for a self-interacting scalar theory up to two loops order. 
The tadpole contributions, (a) and (b), are canceled by a proper choice of $\Delta$.
\label{fig:scalarloops}}
\end{figure}  
\end{center}



\begin{center}
\begin{figure}
\epsfig{file=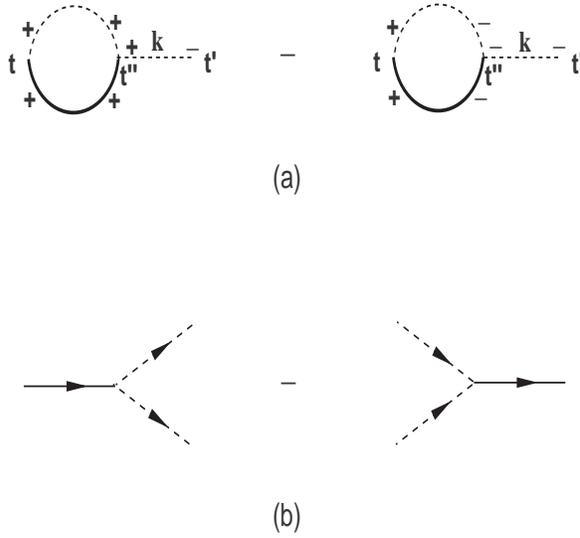,width=4in,height=3in}
\vspace{.1in}
\caption{(a) The Feynman diagrams that contribute to the quantum kinetic equation 
for the pion distribution function. 
The solid line is the sigma meson propagator and the dashed line is the pion propagator. 
(b) The only contribution on-shell is the decay of a
sigma meson into two pions minus the reverse process.
\label{fig:pionsigmaloop}}
\end{figure}  
\end{center}


\newpage

\begin{center}
\begin{figure}
\epsfig{file=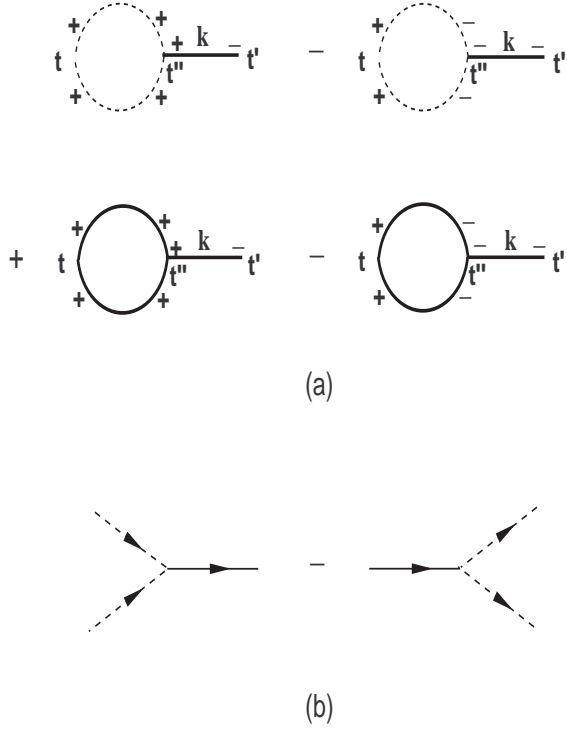,width=4in,height=4in}
\vspace{.1in}
\caption{(a) The Feynman diagrams that contribute to the quantum kinetic equation 
for the sigma meson distribution function. The solid line is the sigma meson 
propagator and the dashed line is the pion propagator. 
(b) The only contribution on-shell is recombination of two pions into a sigma meson
minus the decay of a sigma meson into two pions.
\label{fig:sigmapionloop}}
\end{figure}  
\end{center}



\begin{center}
\begin{figure}
\epsfig{file=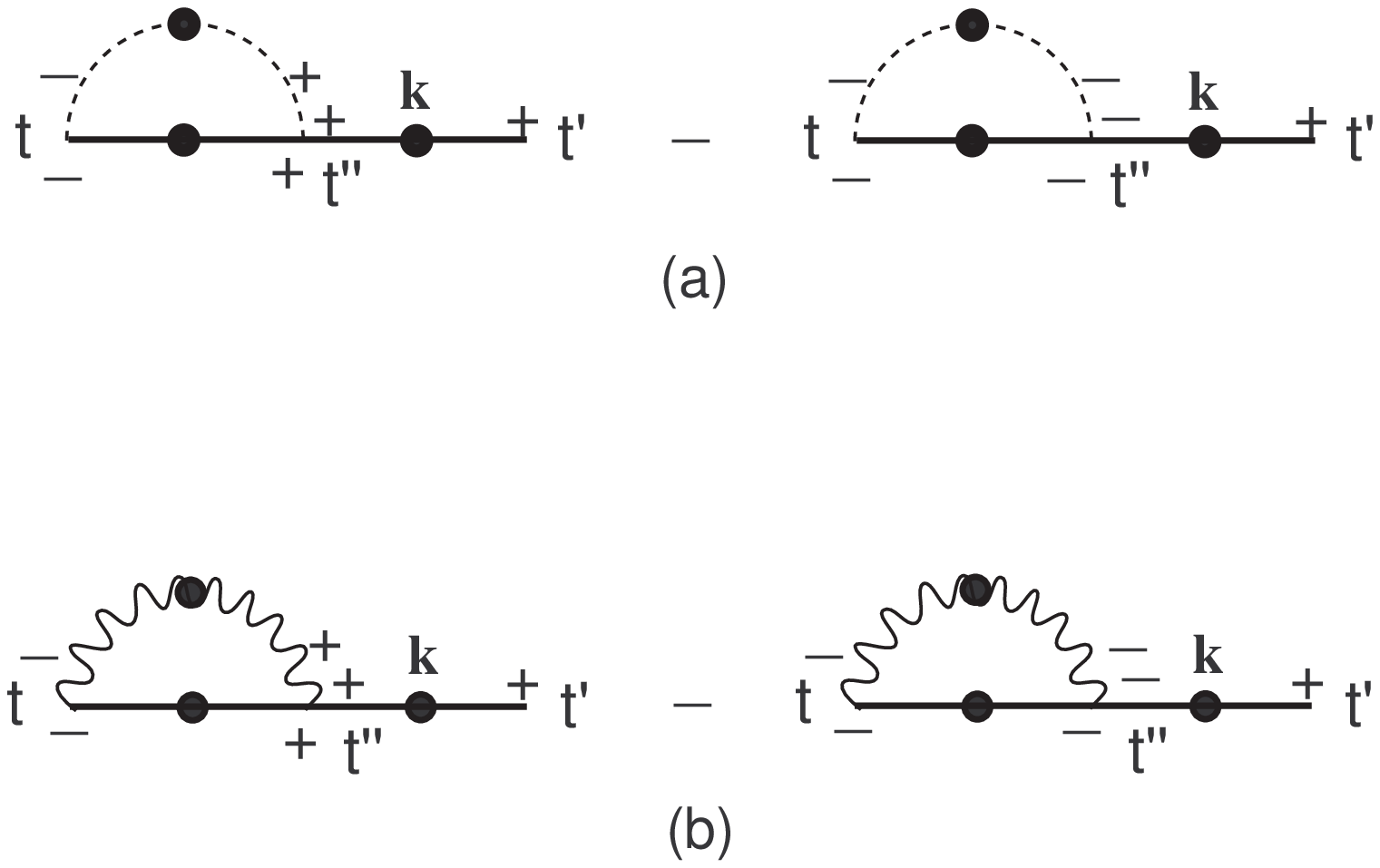,width=4in,height=3.5in}
\vspace{.1in}
\caption{ The Feynman diagrams that contribute to the quantum kinetic equation 
for the charged scalar distribution function to lowest order in $e^2$. 
The dashed and wavy lines are the HTL resummed longitudinal transverse  
photon propagators respectively, and the solid line is the HTL
resummed scalar propagator. 
(a) Contribution from longitudinal photons. 
(b) Contribution from transverse photons.
\label{fig:scalarkinetic}} 
\end{figure}  
\end{center}



\begin{thebibliography}{99}
%
%
\bibitem{qgp} J.W. Harris and B. Muller, Annu. Rev. Nucl. Part. Sci.
{\bf  46}, 71 (1996); 
B. Muller in {\it Particle Production in Highly Excited Matter}, 
edited by H.H. Gutbrod and J. Rafelski, NATO ASI series B, vol. 303 (1993); 
H. Elze and U. Heinz, Phys. Rep. 183, 81 (1989);
H. Meyer-Ortmanns, Rev. Mod. Phys. {68}, 473 (1996); 
H. Satz, in {\it Proceedings of the Large Hadron 
Collider Workshop}, edited by G. Jarlskog and D. Rein (CERN, Geneva), Vol.~1,
page 188; and in {\it Particle Production in Highly Excited Matter},
edited by H.H. Gutbrod and J. Rafelski, NATO ASI series B, Vol.~303 (1993).

\bibitem{books} B. Muller, {\it The Physics of the Quark-Gluon Plasma}, 
Lecture Notes in Physics, Vol.~225 
(Springer-Verlag, Berlin, Heidelberg, 1985); 
L.P. Csernai, {\it Introduction to Relativistic Heavy Ion Collisions}
(John Wiley and Sons, England, 1994); 
C.Y. Wong, {\it Introduction to High-Energy Heavy Ion
Collisions} (World Scientific, Singapore, 1994).

\bibitem{geiger} K. Geiger, Phys. Rep. {\bf 258}, 237 (1995);
Phys. Rev. {\bf D54}, 949 (1996); Phys. Rev. {\bf D56}, 2665 (1996);
Phys. Rev. {\bf D46}, 4965 (1992); Phys. Rev. {\bf D47}, 133 (1993); 
{\it Quark-Gluon Plasma 2}, edited by R.C. Hwa (World Scientific, Singapore, 1995). 

\bibitem{heinz} H.-T. Elze and U. Heinz, in {\em Quark-Gluon Plasma}
edited by R.C. Hwa (World Scientific, Singapore, 1990); 
H.-T. Elze and U. Heinz, Phys. Rep. {\bf 183}, 81 (1989).  

\bibitem{mrow} S. Mr\'owczy\'nski, in {\em Quark-Gluon Plasma} \cite{heinz}; 
S. Mr\'owczy\'nski and P. Danielewicz, Nucl. Phys. {\bf B342}, 345 (1990);
S. Mr\'owczy\'nski and U. Heinz, Ann. Phys. (N.Y.) {\bf 229}, 1 (1994);
S. Mr\'owczy\'nski, hep-ph/9805435.

\bibitem{bass} S.A. Bass {\em et. al.}, Prog. Part. Nucl. Phys. {\bf 41} 225 (1998). 

\bibitem{lawrie}  I.D. Lawrie and D.B. McKernan, Phys. Rev. {\bf D55} 2290 (1997);  
I.D. Lawrie, J. Phys. {\bf A21}, L823 (1988);
Phys. Rev. {\bf D40}, 3330 (1989); 
Can. J. Phys. {\bf 71}, 262 (1993). 

\bibitem{boyanrk}
D. Boyanovsky, I.D. Lawrie, and D.-S. Lee, Phys. Rev. {\bf D54}, 4013 (1996).

\bibitem{frenkel} F.T. Brandt, J. Frenkel, A. Guerra, Int. J. Mod. Phys. 
{\bf A13}, 4281 (1998). 

\bibitem{rau} J. Rau and B. Muller, Phys. Rept. {\bf 272}, 1 (1996). 

\bibitem{wang} X.-N. Wang, Phys. Rep. {\bf 280}, 287 (1997).

\bibitem{eskola}  K.J. Eskola, Prog. Theor. Phys. Suppl. {\bf 129}, 1 (1997); 
K.J. Eskola, K. Kajantie, P.V. Ruuskanen, Eur. Phys. J. {\bf C1} 627 (1998); 
K.J. Eskola, B. Müller, X.-N. Wang, Phys. Lett. {\bf B374}, 20 (1996); 
K. J. Eskola, Nucl. Phys. {\bf A590}, 383c (1995). 

\bibitem{smit} G. Aarts and J. Smit, Nucl. Phys. {\bf B555}, 355 (1999); 
hep-ph/9906538; hep-ph/9909040. 

\bibitem{venu} A. Krasnitz and R. Venugopalan, hep-ph/9909203; hep-ph/9905319; hep-ph/9809433; hep-ph/9808332; hep-ph/9706329; R. Venugopalan, hep-ph/9907209.  

\bibitem{raja} A. Rajantie and  M. Hindmarsh, hep-ph/9904270. 


\bibitem{mueller3} W. Poeschl and  B. Mueller, nucl-th/9812066; hep-ph/9811441. 

\bibitem{Jeon:1996zm} S. Jeon and L. Yaffe, Phys. Rev. {\bf D53}, 5799 (1996); 
S. Jeon, Phys. Rev. {\bf D52}, 3591 (1995).  

\bibitem{Bodeker:1999ud} D.~Bodeker,
hep-ph/9903478; Phys. Lett. {\bf B426}, 351 (1998).

\bibitem{Arnold:1999xk}
P.~Arnold, D.T.~Son and L.G.~Yaffe,
Phys. Rev. {\bf D59}, 105020 (1999).

\bibitem{Blaizot:1999xk} J.-P.~Blaizot and E.~Iancu, hep-ph/9903389; hep-ph/9906485. 

\bibitem{basagoiti} M.A. Valle Basagoiti, hep-ph/9903462.

\bibitem{Litim:1999id} D.F.~Litim and C.~Manuel, hep-ph/9906210.

\bibitem{daniel} P. Danielewicz, Ann. Phys. (N.Y.) {\bf 152}, 239 (1984). 

\bibitem{boyrgir}
D. Boyanovsky, H.J. de Vega, R. Holman, and M. Simionato, 
Phys. Rev. {\bf D60}, 065003 (1999);
D. Boyanovsky and H.J. de Vega, Phys. Rev. {\bf D59}, 105019 (1999). 

\bibitem{iancu} 
J.-P. Blaizot and E. Iancu, Phys. Rev. Lett. {\bf 76}, 3080 (1996);
Phys. Rev. {\bf D55}, 973 (1997); {\em ibid}. {\bf 56}, 7877 (1997). 
K. Takashiba, Int. J. Mod. Phys {\bf A 11}, 2309 (1996). 

\bibitem{robinfra} R.D. Pisarski, Phys. Rev. Lett. {\bf 63}, 1129 (1989);
Phys. Rev. {\bf D47},5589 (1993). 

\bibitem{htl} 
E. Braaten and R.D. Pisarski, Nucl. Phys. {\bf B337}, 569 (1990); 
{\em ibid}. {\bf B339}, 310 (1990). 

\bibitem{rob2} R.D. Pisarski, Physica A {\bf 158}, 146 (1989); 
Phys. Rev. Lett. {\bf 63}, 1129 (1989); 
Nucl. Phys. {\bf A525}, 175 (1991).  

\bibitem{rob3} R.D. Pisarski, Nucl. Phys. {\bf B309}, 476 (1988). 

\bibitem{landsmann}
N.P. Landsmann and C.G. van Weert, Phys. Rep. {\bf 145}, 141 (1987).

\bibitem{altherr}
T. Altherr and D. Seibert, Phys. Lett. B{\bf 333}, 149 (1994);
T.~Altherr, Phys. Lett. B{\bf 341}, 325 (1995).

\bibitem{hatsuda}  T. Hatsuda, T. Kunihiro, and H. Shimizu,
Phys. Rev. Lett. {\bf 82} 2840 (1999);  
S. Chiku and  T. Hatsuda, Phys. Rev. {\bf D58} 076001 (1998); 
{\em ibid}. Phys. Rev. {\bf D57}, 6  (1998).   

\bibitem{rebhan} U. Kraemmer, A. Rebhan, and H. Schulz, Ann. Phys. (N.Y.)
{\bf 238}, 286 (1995). 

\bibitem{thoma}M.H.~Thoma and C.T.~Traxler, Phys. Lett. {\bf B378}, 233 (1996).

\bibitem{boyhtl}  
D. Boyanovsky, H.J. de Vega, R. Holman, S.P. Kumar, and R.D. Pisarski, 
Phys. Rev. {\bf D58}, 125009 (1998).

\bibitem{thoma2} S. Leupold and M.H. Thoma, hep-ph/9908460.

\bibitem{schwinger} J.~Schwinger, J. Math. Phys. {\bf 2}, 407 (1961).
 
\bibitem{kt} K.T. Mahanthappa, Phys. Rev. {\bf 126}, 329 (1962); 
P.M. Bakshi and K.T. Mahanthappa, J. Math. Phys. {\bf 41}, 12 (1963).

\bibitem{keldysh} L.V. Keldysh, JETP {\bf 20}, 1018 (1965).

\bibitem{chou}
K.-C. Chou, Z.-B. Su, B.-L. Hao, and L. Yu, Phys. Rep. {\bf 118}, 1 (1985).

\bibitem{disip}
D. Boyanovsky, H.J. de Vega and R. Holman, in Proceedings of
the Second Paris Cosmology Colloquium, Observatoire de Paris, 1994,
edited by H.J. de Vega and N. S\'anchez 
(World Scientific, Singapore 1995), pp. 127-215;
in {\em Advances in Astrofundamental Physics}, 
Erice Chalonge School, edited by N. S\'anchez and
A. Zichichi (World Scientific, Singapore 1995); 
D. Boyanovsky, H.J. de Vega, R. Holman, D.-S. Lee and A. Singh, 
Phys. Rev. {\bf D51}, 4419 (1995); 
D. Boyanovsky, H.J. de Vega, R. Holman and J. Salgado, 
Phys. Rev. {\bf D54}, 7570 (1996).  

\bibitem{tadpole} 
D. Boyanovsky, H.J. de Vega and R. Holman, 
in {\em Current Topics in Astrofundamental Physics}, 
Vth Erice Chalonge School, edited by N. S\'anchez and A. Zichichi
(World Scientific, Singapore 1996), pp. 183-270; 
D. Boyanovsky, H.J. de Vega, R. Holman and D.-S. Lee, 
Phys. Rev. {\bf D52}, 6805 (1995); 
D. Boyanovsky, M. D'Attanasio, H.J. de Vega and R. Holman,
Phys. Rev. {\bf D54}, 1748 (1996).

\bibitem{lebellac} M. Le Bellac, {\em Thermal Field Theory} 
(Cambridge University Press, Cambridge, England, 1996). 

\bibitem{parwani} R.R. Parwani, Phys. Rev. {\bf D45}, 4695 (1992).

\bibitem{dolan} L. Dolan and R. Jackiw, Phys. Rev. {\bf D9}, 332 (1974).

\bibitem{elmfors} 
P. Elmfors, K. Enqvist, and I. Vilja, Nucl. Phys. {\bf B422[FS]}, 521 (1994).

\bibitem{goldenfeld} L.-Y. Chen, N. Goldenfeld and Y. Oono,
Phys. Rev. Lett. {\bf 73}, 1311 (1994); Phys. Rev.{\bf E54}, 376 (1996).

\bibitem{kuni} T. Kunihiro, Prog. Theor. Phys. {\bf 95}, 503 (1995); 
{\em ibid}. {\bf 97}, 179 (1997);
S.~Ei, K.~Fujii and T.~Kunihiro, hep-th/9905088.

\bibitem{salgado} 
H.J. de Vega and J.F.J. Salgado, Phys. Rev. {\bf D56} 6524, (1997). 

\bibitem{qm} 
I.L. Egusquiza and M.A. Valle-Basagoiti, hep-th/9611143; 
T.~Kunihiro, Prog. Theor. Phys. Suppl. 
{\bf 131}, 459 (1998).

\bibitem{zak} V.E. Zakharov, V.S. L\'vov, G. Falkovich, 
{\it Kolmogorov Spectra of Turbulence I} 
(Springer-Verlag, Berlin Heidelberg, 1992). 

\bibitem{Bochkarev:1996gi}
A.~Bochkarev and J.~Kapusta, Phys. Rev. {\bf D54}, 4066 (1996).

\bibitem{fermiondamping}
D. Boyanovsky, H.J. de Vega, D.-S. Lee, Y.J. Ng, and S.-Y. Wang,
Phys. Rev. {\bf D59}, 105001 (1999); 
S.-Y. Wang, D. Boyanovsky, H.J. de Vega, D.-S. Lee, and Y.J. Ng,
hep-ph/9902218.

\bibitem{Pisarski:1996mt}
R.D.~Pisarski and M.~Tytgat,
Phys. Rev. {\bf D54}, 2989 (1996).

\bibitem{patkos} A. Jakovac, A. Patkos, P. Petreczky, Zs. Szep, hep-ph/9905439.
The authors thank A.~Patkos for illuminating correspondence on these issues. 

\bibitem{Serot:1986ey}B.D.~Serot and J.D.~Walecka,
Adv. Nucl. Phys. {\bf 16}, 1 (1986). 

\bibitem{sigmadecay} D.H. Rischke, Phys. Rev.  {\bf C58}, 2331 (1998).

\bibitem{csernai}
L.P. Csernai, P.J. Ellis, S. Jeon, and J.I. Kapusta, nucl-th/9908020.

\bibitem{qedqcd}
D. Boyanovsky, H.J. de Vega, D.-S. Lee, and S.-Y. Wang (work in progress).

%
%

\bibitem{altherr2}
T.~Altherr, Phys. Lett. {\bf B341}, 325 (1995).

\bibitem{dadic}
I.~Dadic, Phys. Rev. {\bf D59}, 125012 (1999).

\bibitem{bedaque}
P.F.~Bedaque, Phys. Lett. {\bf B344}, 23 (1995).

\bibitem{niegawa} 
A. Ni\'egawa, Phys. Lett. B{\bf 416}, 137 (1998).

\bibitem{niegawa2}
A. Ni\'egawa, hep-th/9905090.

\bibitem{greiner}
C.~Greiner and S.~Leupold,
Eur. Phys. J. C{\bf 8}, 517 (1999).

\bibitem{carrington} M.E. Carrington, H. Defu and M. Thoma, 
Eur. Phys. J. {\bf C7}, 347 (1999).

%
%

\bibitem{gaugeinv} 
D. Boyanovsky, D. Brahm, R. Holman and D.-S. Lee, 
Phys. Rev. {\bf D54}, 1763 (1996).

\bibitem{slavnov} 
L. Faddeev and A. Slavnov, {\em Gauge Fields: An Introduction to 
Quantum Theory} (Addison-Wesley, Redwood City, CA, 1991).


\bibitem{landshoff} 
K. James and P.V. Landshoff, Phys. Lett. B{\bf 251}, 167 (1990).

\bibitem{hat} 
The original method is due to Dirac: P.A.M. Dirac,
{\em Lectures in Quantum Mechanics} 
(Yeshiva University, New York, 1964).
See also: B.~Hatfield, 
{\em Quantum Field Theory of Point Particles and Strings} 
(Addison Wesley, Redwood City, 1992);
A.~Hansson, T.~Regge, and C.~Teitelboim, 
{\em Constrained Hamiltonian Systems} 
(Academia Nazionale dei Lincei, Rome, 1976); 
M.~Henneaux and C.~Teitelboim, {\em Quantization of Gauge Systems}
(Princeton University Press, 1992), and references therein.


\end{thebibliography}
\end{document}